\documentclass[acmtog,nonacm]{acmart}
\usepackage[utf8]{inputenc}

\acmSubmissionID{papers\_127}
\usepackage[utf8]{inputenc}
\usepackage[ruled,vlined,linesnumbered]{algorithm2e}        % For pseudo-code
\usepackage{listings}                                       % For code
\usepackage{newtxmath}                                      % For blackboard letters
\usepackage {enumitem}
\usepackage{subcaption}
\usepackage{tikz}
\newlength{\unit}
\usepackage{overpic}
\usepackage{calc}
\newlength{\unitOther}
\newlength{\labelHeight}
\newlength{\gap}
\newcommand{\edited}[1]{{#1}}

\usepackage{tikz}
\usepackage{pgfplots}
\pgfplotsset{compat=newest}

\author{Abdalla G. M. Ahmed}
\orcid{0000-0002-2348-6897}
\affiliation{%
  \institution{KAUST}%
  \country{KSA}}
\email{abdalla_gafar@hotmail.com}

\author{Matt Pharr}
\orcid{0000-0002-0566-8291}
\affiliation{%
  \institution{NVIDIA}%
  \country{USA}}
\email{matt.pharr@gmail.com}

\author{Peter Wonka}
\orcid{0000-0003-0627-9746}
\affiliation{%
  \institution{KAUST}%
  \country{KSA}}
\email{pwonka@gmail.com}

\title{ART-Owen Scrambling}

\setcopyright{rightsretained}
\acmJournal{TOG}
\ccsdesc[500]{Computing methodologies~Ray tracing}

\keywords{sampling, nets, digital nets, dyadic nets, Sobol sequences, quasi-Monte Carlo, low-discrepancy sequences, Owen's scrambling}

\date{Jan 2023}

\begin{abstract}
We present a novel algorithm for implementing Owen-scrambling, combining the generation and distribution of the scrambling bits in a single self-contained compact process.
We employ a context-free grammar to build a binary tree of symbols, and equip each symbol with a scrambling code that affects all descendant nodes.
We nominate the grammar of adaptive regular tiles (ART) derived from the repetition-avoiding Thue-Morse word, and we discuss its potential advantages and shortcomings.
Our algorithm has many advantages, including random access to samples, fixed time complexity, GPU friendliness, and scalability to any memory budget.
Further, it provides two unique features over known methods: it admits optimization, and it is invertible, enabling screen-space scrambling of the high-dimensional Sobol sampler.
\end{abstract}
\keywords{sampling, Sobol sequences, quasi-Monte Carlo, dyadic nets, Owen scrambling}

\begin{teaserfigure}
  {\centering
  \scriptsize%
  \setlength{\unit}{0.4\textwidth}%
  \setlength{\unitOther}{0.316151203\unit}%
  \settoheight{\labelHeight}{Unscrambled}%
  \setlength{\gap}{(\unit - 3\unitOther)/2}%
  \begin{tikzpicture}[
          img/.style = {inner sep=0, anchor=south west},
          lbl0/.style = {inner sep=0, anchor=south,text=black},
          lbl1/.style = {inner sep=0, anchor=south,text=white,yshift=0.5pt,xshift=-0.5pt},
          lbl/.style = {align=center,yshift=-1.2\labelHeight},
          lblVertical/.style={rotate=90,inner sep= 0pt,yshift=-1.2\labelHeight}
      ]
      \node[img,anchor=south east] at (0,0) (all) {\includegraphics[height=1\unit]{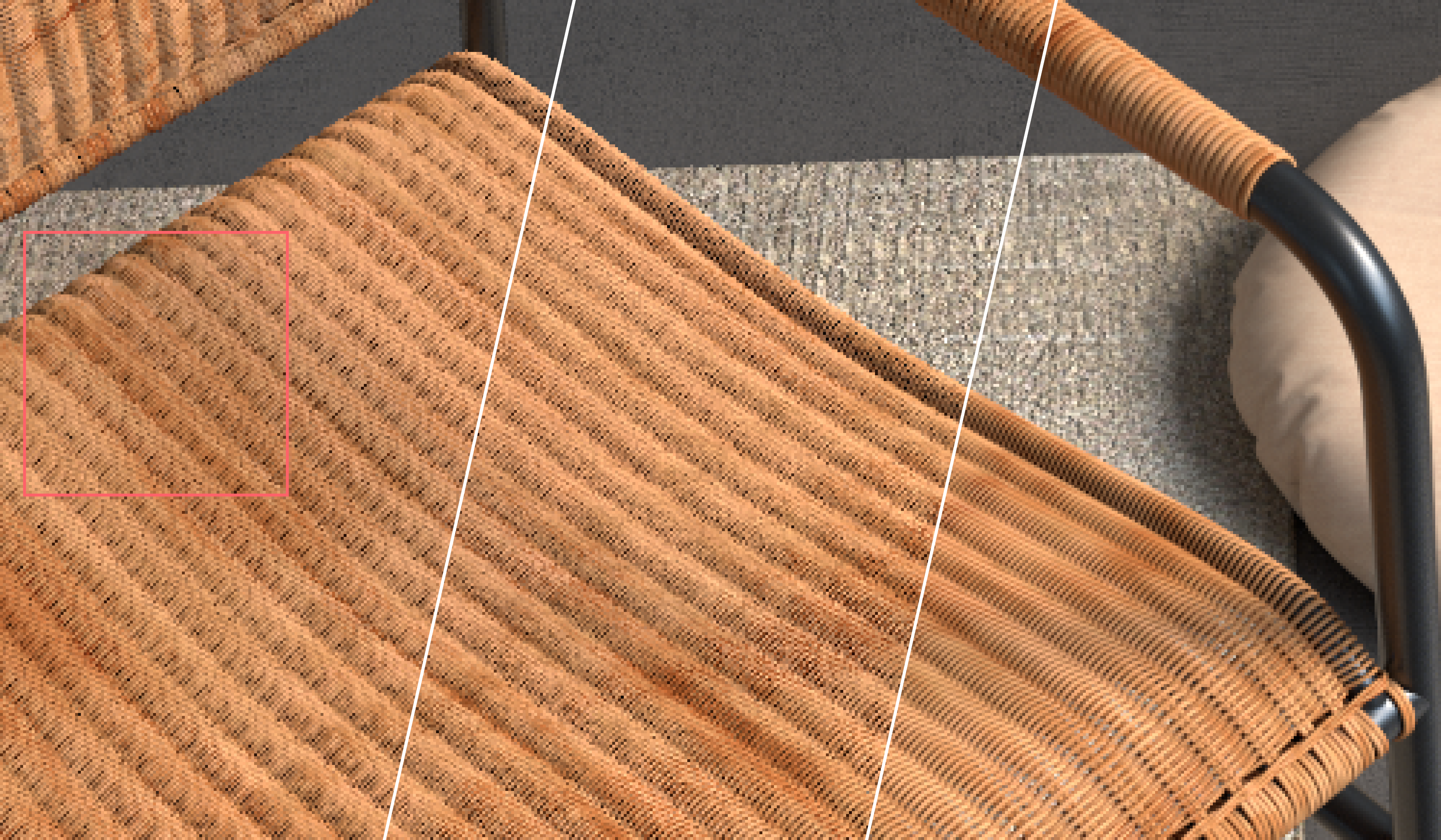}};
      \node[img] at (1cm,                0) (insetSobol) {\includegraphics[height=1\unitOther]{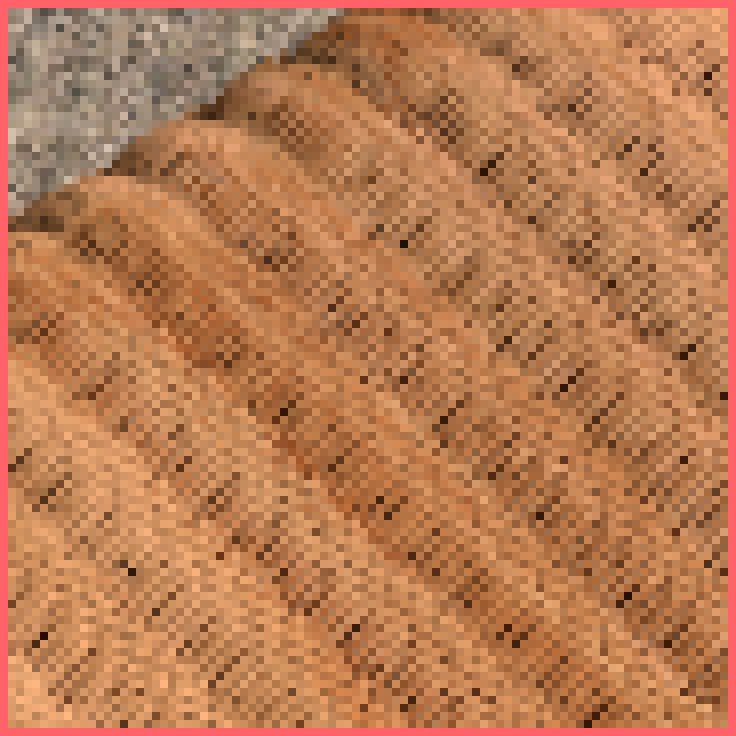}};
      \node[img] at (1cm,1\unitOther+1\gap) (insetOwen)  {\includegraphics[height=1\unitOther]{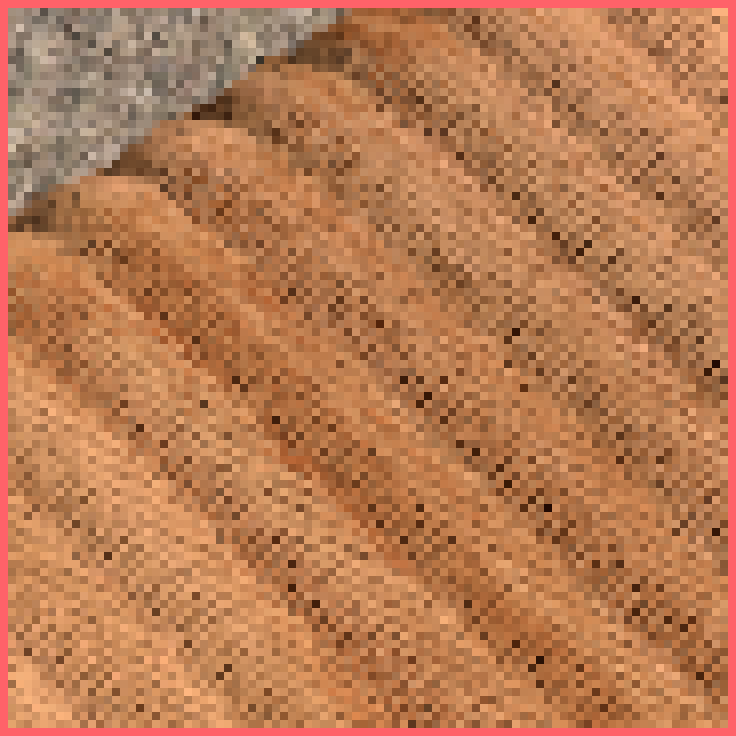}};
      \node[img] at (1cm,2\unitOther+2\gap) (insetRef)   {\includegraphics[height=1\unitOther]{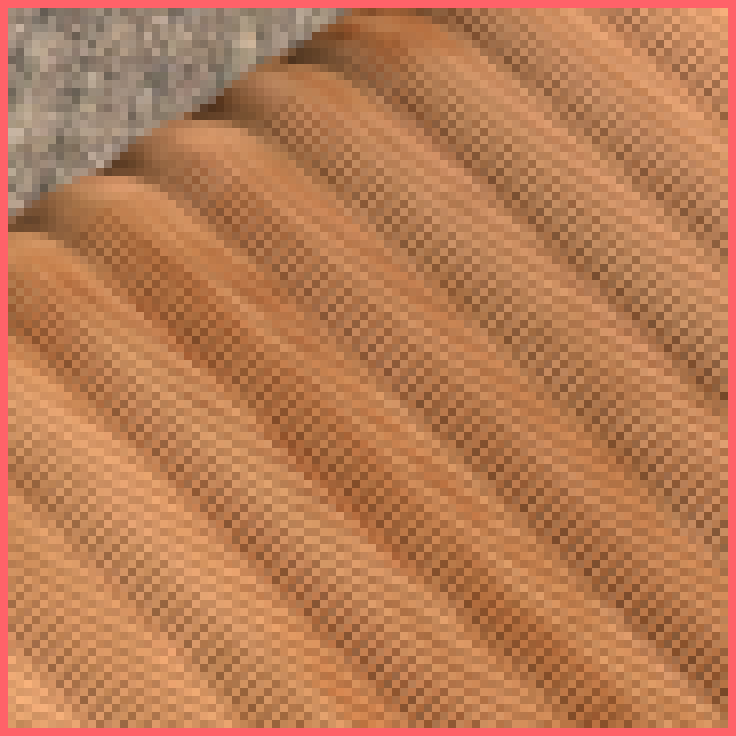}};
      \node[lblVertical] at (insetSobol.east) {Unscrambled};
      \node[lblVertical] at (insetOwen.east) {Scrambled};
      \node[lblVertical] at (insetRef.east) {Reference};
      \node[lbl, xshift=0.228522337\unit] at (all.south west) {Unscrambled};        % 0.228522337 = (0.5×133)÷291
      \node[lbl, xshift=0.743986254\unit] at (all.south west) {Scrambled};          % 0.743986254 = (133 + 0.5×167)÷291
      \node[lbl, xshift=1.317869416\unit] at (all.south west) {Reference};          % 0.228522337 = 0.5×133÷291
  \end{tikzpicture}\\
  \setlength{\gap}{2.25pt}%
  }%
  \caption{\label{fig:teaser}%
     With geometric details at the scale of a few pixels, unscrambled Sobol sampling points may give structured artifacts.
     This is a typical example of \emph{aliasing} artifacts \protect{\cite{Shannon1948}} where frequencies beyond the sampling rate manifest as low-frequency structures.
     Owen-scrambling the points causes the error to be higher-frequency,
     which is more visually pleasing.
     Our approach, \emph{ART-Owen Scrambling}, efficiently generates Owen-scrambled sample points using a grammar based on adaptive regular tiles (ART).
  }
\end{teaserfigure}
\begin{document}
% ==============================================================================
% ==============================================================================
% ==============================================================================
% ==============================================================================
% ==============================================================================
% ==============================================================================

\maketitle

% ==============================================================================
% ==============================================================================
% ==============================================================================
\section{Introduction\label{sec:introduction}}
% ==============================================================================
% ==============================================================================
% ==============================================================================

Sampling is an essential process that underlies many areas in computer graphics (CG) including rendering, halftoning and stippling, geometry processing, and machine learning.
The quality of the sampling patterns used can have a significant effect on the error and rates of convergence in the tasks they are applied to.
Sampling in CG is characterized by enormous sampling rates,
typically billions of samples to render a single image from a model.
Thus, not only is the quality of the points important, but the efficiency of generating them is important as well, as it affects the runtime of rendering-intensive applications such as movie production, computer games, and architectural modeling.
The problem is further complicated in rendering by a requirement to vary the samples between the sampled domains (time, camera lens, area lights, etc.) that constitute the light transport paths, as well as between neighboring pixels, in order to avoid moir\'e patterns, banding, and similar structured aliasing artifacts.
Similar constraints usually apply to other areas, e.g., halftoning or geometry processing.
This poses an extraordinary demand on the sample-generating process that it is required not only to deliver a well-distributed set of samples but also a different one in each call.
That is, a randomization method needs to be incorporated into the sample generation process while ensuring not to compromise the distribution quality or significantly degrade the speed performance.

Different techniques were proposed to answer to these challenging requirements in this open research area.
Popular early approaches~\cite{Glassner1995} include stratification, where individual samples are randomly placed over a grid of cells, and Poisson-disc sampling, where a minimal prescribed spacing is enforced between otherwise random samples. These techniques, and others, were mostly developed heuristically by the CG community, and they worked satisfactorily well, then.
Most of these solutions, however, fall short of scaling well to meet the demand of modern Monte-Carlo-based advanced rendering engines.
For example, stratification is cursed by the dimensionality of complex light paths, and look-up-based methods impose memory bottlenecks on GPUs.
These, and similar issues, lead to increasing adoption of the low-discrepancy (LD) sampling methods developed by the Monte Carlo community, and first introduced to computer graphics by Shirley~\shortcite{Shirley91Discrepancy}.
A mathematical recipe is used to compute the samples in these distributions, making them efficient in terms of both memory and speed. In addition, they scale well with dimensions and attain provably excellent performance in numerical integration, making them an attractive candidate for use in rendering and similar applications.

Of special interest are Sobol sequences \shortcite{sobol1967}.
Thanks to their bit-arithmetic nature, they offer high computational performance, sometimes outperforming high-quality pseudo-random number generators, while also exhibiting excellent numerical integration performance.
The interested reader is referred to Keller \shortcite{Keller13Nutshell} for a survey of how these sequences may be tailored to rendering problems.

Randomized Sobol sequences offer further benefits, including reduction of the structured error that can be seen at low sampling rates with Sobol points, as well as making it possible to apply randomized quasi-Monte Carlo (RQMC) techniques.
Intuitive randomization methods, such as Cranely-Patterson toroidal shifting \shortcite{Cranley76Randomization}, compromise the structure of these distributions, degrading their performance.
The best-known approach for randomizing Sobol sequences is known as Owen Scrambling, after its inventor~\shortcite{Owen95Randomly}.
Beyond the benefits of randomization, Owen scrambling offers further benefits, including asymptotically higher rates of convergence than regular Monte Carlo for smooth functions.
We will describe the key ideas of Owen scrambling in Section~\ref{sec:owen scrambling}.
Implementing the concept, however, is not simple.
The problem, in abstract terms, involves distributing random bits over a tree.
It may sound easy, and in fact is, but not so without losing the enormous throughput offered by the vanilla Sobol sampler.

Implementing Owen scrambling is a well-known challenging problem, as noted by Owen himself \shortcite{Owen98Scrambling}.
Only a few solutions were proposed over the past years, offering different trade-offs between quality, speed, and versatility.
Combined with a lack of user control over the generated samples, Christensen et al. \shortcite{Christensen18Progressive}, followed by Pharr~\shortcite{Pharr19Efficient}, made an attempt to skip Owen-scrambling altogether by synthesizing a Sobol-like sequence.
Ahmed and Wonka \shortcite{Ahmed2021Optimizing}, however, subsequently revealed that Owen scrambling spans all the space of Sobol-like sequences, bringing Owen scrambling back as the one-and-only way to go with Sobol sequences.
Thus, a satisfactory implementation of Owen scrambling is highly desirable.

Numerical error is not the only metric that matters when evaluating sampling techniques.
For example, sampling patterns with blue noise power spectra generally appear to have lower perceived error than those that do not, thanks to characteristics of the human visual system~\cite{Ulichney87Digital,Mitchell1987Antitliased}.
It is also useful to be able to invert the mapping, which allows applying a single global sampling pattern across the image plane and then being able to enumerate the samples that overlap a single pixel~\cite{Gruenschlos12Enumerating}.
Doing so is crucial for both adaptive sampling and parallel implementation.

In this paper, we present a novel approach to implementing Owen scrambling efficiently that combines the random bit generation with tree traversal in one process.
Unlike all known methods, our approach provides a generous amount of user control over the scrambling process, and our model is quite flexible, offering different trade-offs between quality, speed, memory, and control, all in one framework.
Thus, rather than offering a single working implementation, we provide a flexible framework that can be reconfigured and extended with different algorithmic choices.
We start by reviewing the most related work in Section~\ref{sec:related work}, followed by a brief review of the technical elements required to understand the paper in Section~\ref{sec:background}. We describe our core model in Section~\ref{sec:our method}, along with a discussion of the possible variations in Section~\ref{sec:grammar}. We then evaluate the model by highlighting various practical aspects in Section~\ref{sec:evaluation}, and make concluding remarks in Section~\ref{sec:conclusion}.

% ==============================================================================
% ==============================================================================
% ==============================================================================
\section{Related Work\label{sec:related work}}
% ==============================================================================
% ==============================================================================
% ==============================================================================

Owen scrambling was originally presented in the context of Monte Carlo integration as a way to impose randomness onto semi-regular quasi-Monte samples, so as to enable variance estimation as with the vanilla Monte Carlo method \cite{Owen95Randomly}.
It therefore relates to a bulk of literature in Monte Carlo and numerical integration areas of research.
Most of that literature, however, is devoted to provable discrepancy and error bounds, whereas the focus in CG is more on the perceivable visual quality and spectral profiles.
We, therefore, confine this section only to the literature related to CG, or cited there.

Many alternatives were proposed to emulate Owen scrambling. For example, XOR-scrambling~\cite{Kollig02Efficient} uses a single scrambling bit per level of the tree. Tezuka~\shortcite{Tezuka1994Generalization} proposed a slightly richer matrix-based scrambling.
His work actually preceded Owen's, but may still be seen as a special case of it.
These alternatives, however, are far less powerful than the full-tree Owen scrambling, and can only realize a smaller subset of possible Owen scramblings, trading quality for speed. 

An early efficient implementation of Owen scrambling was presented by Friedel and Keller~\shortcite{Friedel02Fast}.
Randomization bits are generated and consumed on the go by sorting the sample points,
leading to optimal space complexity and $\mathcal{O}(N \log(N))$ performance, mainly bottlenecked by the sorting pre-process.
Helmer et al.~\shortcite{Helmer2021Stochastice} recently presented an orders-of-magnitude faster implementation by ``hacking'' the sample-generation algebraic recipe and extracting a very simple indexing rule to replace the sorting process.
Their implementation is actually tangibly faster than the typical common implementation \cite{Pharr23PBRT} of the vanilla Sobol sequence.
Both techniques realize a true, uncompromised Owen scrambling---as far as the used random number generator is random.
On the downside, these techniques compute all the samples at once, violating the parallelization requirements in rendering, for example, where the leaves down the scrambling tree need to access consistent scrambling information in ancestor nodes.
We consider Helmer's implementation the current state of the art in delivering high-quality Owen-scrambled sets, and we use it for benchmarking in Fig.~\ref{fig:periodograms}.

Owen~\shortcite{Owen03Variance} himself proposed a hash-based implementation, currently adopted in PBRT~\cite{Pharr23PBRT},  that computes a hash function after each digit is generated to determine whether to flip the subsequent bit.
More recently, Burley~\shortcite{Burley2020Scrambling} presented a different hash-based algorithm.
It exploits an observation that the long-multiplication process, as taught in primary school, produces a nested tree, but in the opposite direction of bit significance.
The trick, then, is to reverse the bits, multiply by a scrambling code, and reverse the bits again.
Through this simple process, Burley managed to obtain a very efficient implementation of Owen scrambling.
The main shortcoming of the method is the limited control.
The method employs a selection of mixed-bits integers, but it is not quite clear, so far, how these exactly map to the actual scrambling bits.
Based on our empirical testing, we noted some spectral distortion in the resulting sets, as may be seen in Fig~\ref{fig:periodograms}(e).

Gruenschlo\ss\ et al. identified the importance of being able to invert low-discrepancy constructions in order to enumerate the low discrepancy sample points that overlap a selected pixel~\shortcite{Gruenschlos12Enumerating}.
They presented algorithms that do so for both Halton and Sobol points, though do not support Owen scrambling of these points.
Thus, for example, typical practice in current rendering frameworks is not to apply scrambling for the first two dimensions of low-discrepancy points used for image plane sampling, even if the remaining dimensions are scrambled~\cite{Pharr23PBRT}.

The mentioned implementation methods offer different trade-offs between speed, memory footprint, quality, and coding complexity. In this paper, we propose a single parametrized model that avails flexibility to adapt to different budgets and targets just by tuning the parameters. In addition, our model uniquely offers a fast inversion of the scrambling where needed.

% ==============================================================================
% ==============================================================================
% ==============================================================================
\section{Essential Background\label{sec:background}}
% ==============================================================================
% ==============================================================================
% ==============================================================================

Before presenting our method, in this section, we make a brief review of a few underlying concepts.
Readers familiar with the titles may skip the respective subsections.

% ==============================================================================
\subsection{Nets and Sequences\label{sec:nets and sequences}}
% ==============================================================================

Owen scrambling is closely associated with low-discrepancy nets and sequences \cite{Niederreiter92Random}.
A full understanding of nets and sequences is not really needed to understand the paper, but it is worthwhile reviewing them to put our method into context.
These are best described in terms of stratification mentioned in the introduction.
To choose 16 2D sample points, for example, we may intuitively divide the sampled domain into $4\times 4$ cells, also known as \emph{strata}, and pick a sample point inside each stratum.
Nets, however, are designed to simultaneously satisfy all possible stratifications: $16\times 1$, $8\times 2$, $4\times 4$, $2\times 8$, and $1\times 16$, which implies much-improved uniformity and copes with a wider range of sampled signals that vary differently along the two axes.
Such a layout is called a $(0, 4, 2)$-net in base $2$, which means having $2^0 = 1$ sample point in each of the possible $2^{-4}$-large strata in 2 dimensions.
The model is known as $(t, m, s)$-nets in base $b$, and extends analogously in each of these parameters.
The idea is extended further into $(t, s)$-sequences in base $b$ that, for any $m$, maintain a $(t, m, s)$-net for the first and all subsequent $b^m$ blocks in sequence.

Such a complex non-obvious sample distribution model primarily emerged thanks to the existence of comprehensible algebraic recipes to construct it.
The interested reader is referred to Dick and Pillichshammer's book \shortcite{Dick10Digital}.
The model developed in this paper primarily targets Sobol sequences, which synthesize high-dimensional sequences of samples from carefully designed one-dimensional $(0, 1)$-sequences in base 2 along constituent dimensions.
The first two dimensions of Sobol sequences constitute a $(0, 2)$-sequence;
see Pharr et al.~\shortcite{Pharr23PBRT} for  an introduction to the topic.

% ==============================================================================
\subsection{Owen Scrambling\label{sec:owen scrambling}}
% ==============================================================================

\begin{figure}
% ------------------------------------------------------------------------------
\setlength{\unit}{1\columnwidth}
    {\centering\scriptsize
        \includegraphics[width=1\unit]{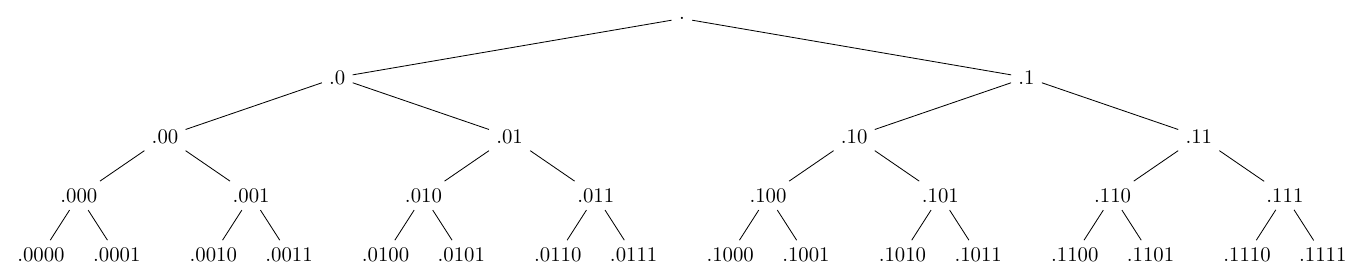}\\
        (a) Input: binary tree of the unit interval\\[2mm]
        \includegraphics[width=1\unit]{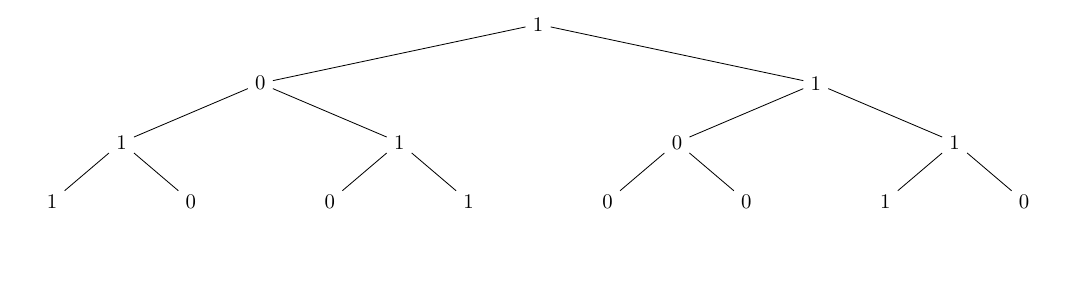}\\[-4mm]
        (b) Binary Scrambling tree\\[2mm]
        \includegraphics[width=1\unit]{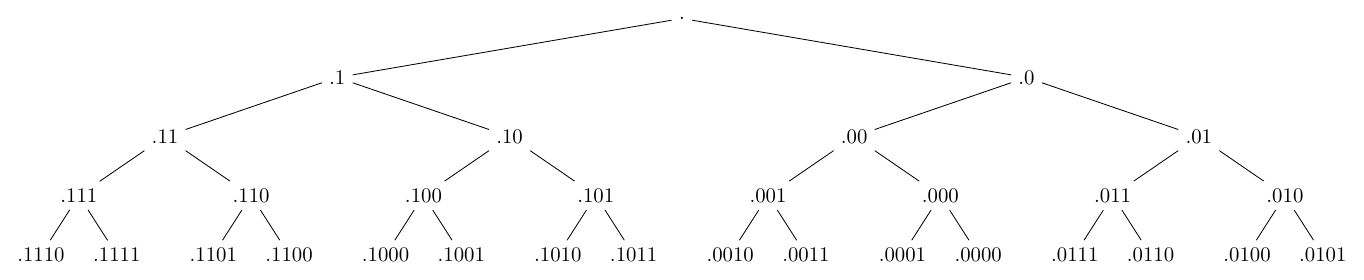}\\
         (c) Scrambled unit interval
    }
    \caption{\label{fig:os illustration}%
        Illustration of Owen scrambling of the unit interval $[0, 1)$.
        Each node in the scrambling tree instructs to swap the branches of the corresponding node of the tree to be scrambled, where indexing uses the original positions of the nodes, before scrambling.
        Note that all intervals stay contiguous, even though possibly scrambled.
    }
% ------------------------------------------------------------------------------
\end{figure}

\begin{figure*}
% ------------------------------------------------------------------------------
\newcommand{\grayout}[1]{{\color{gray!50}#1}}
\setlength{\unit}{(\textwidth - 5\gap)/6}
\newlength{\tick}
\setlength{\tick}{\unit/16}
{\scriptsize\centering
    \begin{tikzpicture}[img/.style = {inner sep=0, anchor=south west}, xlbl/.style={anchor=north}]
        \node[img] at (0\unit + 0\gap, 0) (X0) {\includegraphics[width=\unit]{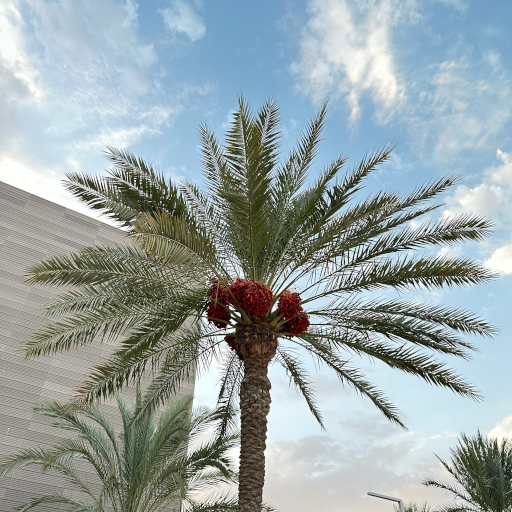}};
        \node[img] at (1\unit + 1\gap, 0) (X1) {\includegraphics[width=\unit]{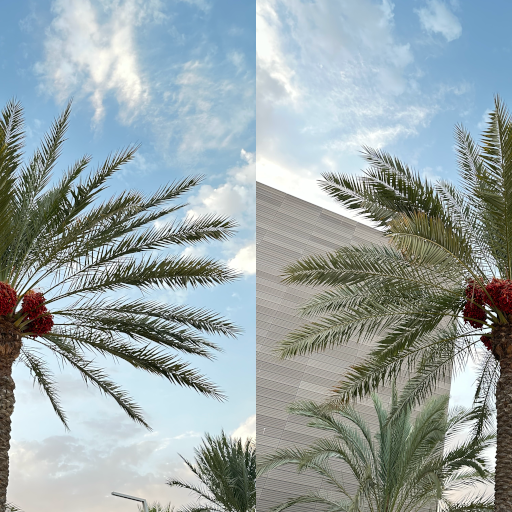}};
        \node[img] at (2\unit + 2\gap, 0) (X2) {\includegraphics[width=\unit]{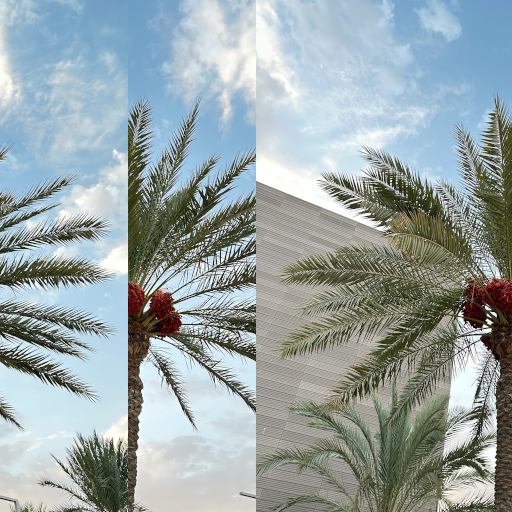}};
        \node[img] at (3\unit + 3\gap, 0) (X3) {\includegraphics[width=\unit]{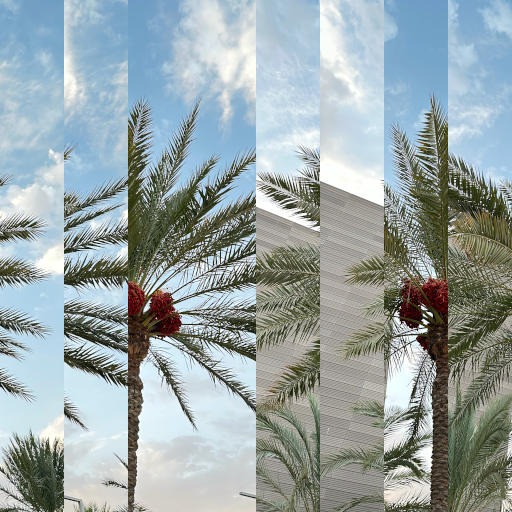}};
        \node[img] at (4\unit + 4\gap, 0) (X4) {\includegraphics[width=\unit]{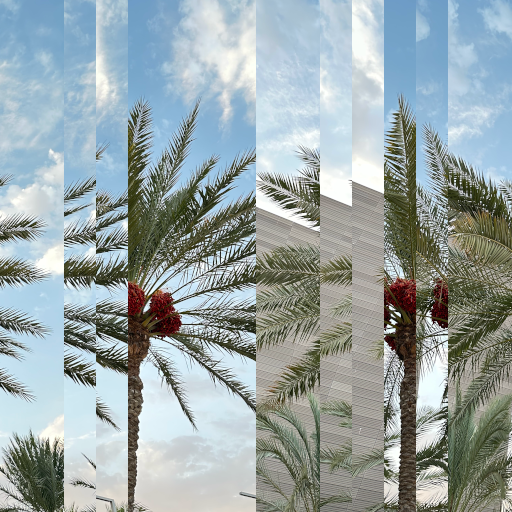}};
        \node[img] at (5\unit + 5\gap, 0) (XY) {\includegraphics[width=\unit]{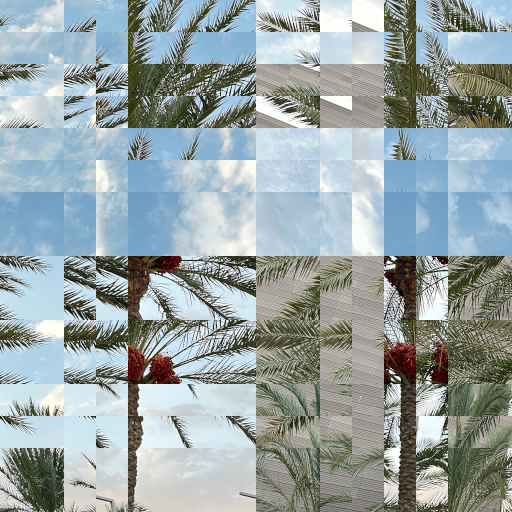}};
        \node[xlbl, xshift= 8\tick] at (X0.south west) {Input};
        \node[xlbl, xshift= 8\tick] at (X1.south west) {1};
        \node[xlbl, xshift= 4\tick] at (X2.south west) {1};
        \node[xlbl, xshift=12\tick] at (X2.south west) {0};
        \node[xlbl, xshift= 2\tick] at (X3.south west) {1};
        \node[xlbl, xshift= 6\tick] at (X3.south west) {0};
        \node[xlbl, xshift=10\tick] at (X3.south west) {1};
        \node[xlbl, xshift=14\tick] at (X3.south west) {1};
        \node[xlbl, xshift= 1\tick] at (X4.south west) {0};
        \node[xlbl, xshift= 3\tick] at (X4.south west) {1};
        \node[xlbl, xshift= 5\tick] at (X4.south west) {0};
        \node[xlbl, xshift= 7\tick] at (X4.south west) {0};
        \node[xlbl, xshift= 9\tick] at (X4.south west) {0};
        \node[xlbl, xshift=11\tick] at (X4.south west) {1};
        \node[xlbl, xshift=13\tick] at (X4.south west) {1};
        \node[xlbl, xshift=15\tick] at (X4.south west) {0};
        \node[xlbl, xshift= 8\tick] at (XY.south west) {Scrambling both axes};
    \end{tikzpicture}}%
    \caption{\label{fig:2d illustration}%
        \edited{Illustration of Owen scrambling applied to an image spanning the unit square,
        using the same ``1,01,1101,10010010'' scrambling tree in Fig~\ref{fig:os illustration} for the $x$-axis, applied step-wise to each depth.
        Please note that the convention is to index the scrambling bits after the unscrambled nodes of the domain as if the scrambling is applied bottom-up to the tree, which might be confusing at the beginning.
        The scrambling of the $y$-axis uses a ``0,10,1010,01110010'' tree.
        Note that all intervals stay contiguous, even though possibly scrambled.}
    }
% ------------------------------------------------------------------------------
\end{figure*}

Owen scrambling is an algorithm that derives new nets and sequences from given ones.
It is best explained in terms of a tree, where it can be described concisely as:
% ------------------------------------------------------------------------------
\begin{quote}
% ------------------------------------------------------------------------------
    Shuffle the branches while keeping them together.
% ------------------------------------------------------------------------------
\end{quote}
% ------------------------------------------------------------------------------
Considering the binary case of Sobol, for example, Owen scrambling is applied as post-processing to the coordinates of the computed sample points in the unit domain, treating each axis as a binary tree, as illustrated in Fig.~\ref{fig:os illustration}, and recursively shuffling the halves.
Fig.~\ref{fig:2d illustration} shows a more visual illustration of the abstract process in Fig.~\ref{fig:os illustration}.
% ------------------------------------------------------------------------------
% ------------------------------------------------------------------------------
% ------------------------------------------------------------------------------
% ------------------------------------------------------------------------------
Even though the scrambling is applied independently to each axis, the ``keep together'' instruction ensures that two- or higher-dimensional strata stay contiguous, hence preserving the net or sequence properties.

Owen scrambling gives rise to a binary scrambling tree, with a single bit of information in each node instructing whether to (0) leave or (1) swap the halves.
The whole tree carries $2^m-1$ bits of information for an $m$-bit resolution of coordinates, giving rise to $2^{2^m-1}$ different shufflings along each axis.
This completely solves the variation problem discussed in the introduction.
The challenge, then, is in how to efficiently supply the enormous amount of scrambling bits while ensuring an acceptable level of randomness.

% ==============================================================================
\subsection{Adaptive Regular Tiles\label{sec:ART}}
% ==============================================================================

Our model is inspired by Adaptive Regular Tiles (ART) \cite{Ahmed17ART}, hence the title of the paper.
A full understanding of that work is not needed to understand our paper, so we only give a high-level abstraction of the relevant part.

The goal of ART is to be able to discriminate tiles and neighborhoods in a recursive regular-lattice tiling.
Earlier treatments of similar problems included identifying by the geometry of the neighborhoods in complex-shaped tiles \cite{Ostromoukhov07Polyominoes,Wachtel14Fast}, and color-coding the edges of regular tiles, along with a matching rule \cite{Kopf06Recursive}, but Ahmed et al. approached the problem differently by treating the dimensions separately.
That is, a one-dimensional matching rule is extended as a Cartesian product to higher dimensions, which makes the concept work with the similarly structured Owen scrambling.
For the one-dimensional problem, they maintain a set of identification symbols, e.g., $\{A, B, C, D\}$, and define a \emph{context-free grammar}, e.g.,
% ------------------------------------------------------------------------------
\begin{equation}
% ------------------------------------------------------------------------------
    \psi:\begin{array}{c}
        A\mapsto AD\\
        B\mapsto BC\\
        C\mapsto AB\\
        D\mapsto BA
    \end{array}\,,\label{eq:art-prodcution}
% ------------------------------------------------------------------------------
\end{equation}
% ------------------------------------------------------------------------------
to map each symbol to a pair of symbols from the same set. The goal is to define a set of production rules for laying child tiles on top of larger ones.
Starting from a single tile with any id, we note that the recursive application of Eq.~\eqref{eq:art-prodcution} produces a binary tree of symbols, which is then matched to a binary tree of tiles defining a recursive tiling.
This assigns an id to each tile at every level.
The original goal of Ahmed et al. was to control the neighborhood around each distinct tile, and towards that end, they developed the grammar from the repetition-avoiding low-complexity Thue-Morse word \cite{Lothaire02Algebraic}, as we will review in Section~\ref{sec:th grammar}.

% ==============================================================================
% ==============================================================================
% ==============================================================================
\section{Our Method\label{sec:our method}}
% ==============================================================================
% ==============================================================================
% ==============================================================================

While the ART model described in Section~\ref{sec:ART} above was originally meant to distribute sample points over tiles, we note that the concept is generic, and can be used to distribute any information over the nodes of a tree, as abstracted in Algorithm~\ref{alg:art model}.
% ------------------------------------------------------------------------------
\begin{algorithm} [tb]
% ------------------------------------------------------------------------------
    \SetKwInOut{KwIn}{Input}
    \SetKwInOut{KwOut}{Output}
    \caption{
        ART methodology for distributing information over the nodes of a tree.
    }
    \label{alg:art model}
    \KwIn{(1) A target tree data structure of a fixed branching rate $q$ and an arbitrary depth.\\(2) A prescribed data storage budget $N$.}
    \KwOut{An assignment of samples of data to nodes of the tree.}
    Define an alphabet $\Sigma = \{S_i\}_{i=1}^{N}$ of $N$ symbols\;
    Define a grammar, i.e., a production rule, that maps each symbol to a $q$-tuple of symbols from the same set\label{alg:step grammar}\;
    Store a data element in each symbol\label{alg:step storage}\;
    Starting from an arbitrary symbol, apply the production rule recursively to produce a tree of symbols that matches the target tree\;
    For each node in the target tree assign the data element stored in the symbol of the corresponding node in the matched tree.
% ------------------------------------------------------------------------------
\end{algorithm}
% ------------------------------------------------------------------------------
For practical implementation, the alphabet is typically defined implicitly as the set $\{0..N-1\}$ of the first $N$ integers, and the grammar is encoded as an $N\times q$ array populated with integers in the designated range.

ART methodology is simple, intuitive, and efficient, and naturally matches the problem of distributing information over a tree.
It is not obvious, though.
Indeed, the only application of it we are aware of, beyond the original paper, is the one by Ahmed and Wonka~\shortcite{Ahmed20Screen}, who used it to scramble a quad-tree of pixels, a problem akin to Owen scrambling.
Our idea is very similar, and is inspired by theirs, but introduces an important modification to solve a shortcoming in the original model; discussed next.

The power of ART methodology comes from the reasonable assumption that reusing the same information over different levels in the hierarchy is not a problem.
There is still a problem, though, that the same symbol might repeat at the same level.
The context-free nature of the grammar, then, implies that all the descendant subtrees would behave identically, since they have the same hierarchy of children, each carrying the same information.
This was not deemed a problem in \cite{Ahmed20Screen} possibly thanks to their relatively generous alphabet size of 4K symbols.
The problem, however, becomes quite serious with the very small alphabets we would consider for an efficient implementation of Owen scrambling.
Indeed, repetition is inevitable once the breadth of the tree tops the number of symbols.
In addition, carrying a single information bit per entry is relatively inefficient, and does not help a lot in building a GPU-friendly solution.
Thus, despite its elegance, the plain ART methodology in Algorithm~\ref{alg:art model} falls short of meeting the excessive information bandwidth required to implement Owen scrambling.

% ==============================================================================
\subsection{\texorpdfstring{ART\raisebox{1pt}{++}}{ART++}\label{sec:art++}}
% ==============================================================================

To overcome the limitation of the original ART model, we introduce a small modification that makes a big difference.
We superimpose a level of context awareness onto the information distribution tree.
Instead of placing a single bit of information in each symbol to encode the swapping of its immediate pair of children, we equip each symbol with a whole vector of bits that affects all descendants, one bit for each level.
Readers familiar with net scrambling may be able to identify this with XOR scrambling \cite{Kollig02Efficient}: each symbol stores a scrambling code that applies to the corresponding interval the same way XOR-scrambling is used over the whole domain.

The essence of our modification is that, even though two nodes at the same level may carry the same symbol, and the same scrambling vector as well, they may still behave differently thanks to the information inherited from their different ancestry.
Assuming a random assignment of scrambling vectors, this model only fails significantly if the grammar assigns an ``identical twin'' pair of children to a symbol, which may easily be avoided in the grammar construction step~\ref{alg:step grammar} of Algorithm~\ref{alg:art model}.
The interaction of the random scrambling vectors embodies the randomization element, and with a careful choice of the grammar it can generate a rich set of quasi-random scrambling bits, as demonstrated empirically in Fig.~\ref{fig:periodograms}.

This idea, along with the abstract ART methodology in Algorithm~\ref{alg:art model}, constitute the core of our approach to solve the Owen scrambling problem.
Algorithm~\ref{alg:scrambling} lists the actual run-time code.
% ------------------------------------------------------------------------------
\begin{algorithm} [tb]
% ------------------------------------------------------------------------------
    \let\oldnl\nl% Store \nl in \oldnl
    \newcommand{\nonl}{\renewcommand{\nl}{\let\nl\oldnl}}% Remove line number for one line
    \caption{\label{alg:scrambling}%
        C-code implementation of ART-Owen scrambling.
    }
    \nonl{\includegraphics[width=0.91\columnwidth]{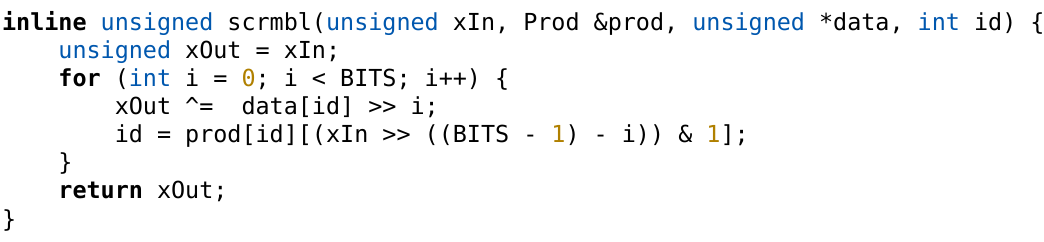}}
% ------------------------------------------------------------------------------
\end{algorithm}
% ------------------------------------------------------------------------------
While the reader may find the concept quite simple, as we do, it is by no means trivial.
In the rest of this paper, we discuss various aspects of the method and highlight some of the available handles to unlock its full potential.

% ==============================================================================
% ==============================================================================
% ==============================================================================
\section{Grammar\label{sec:grammar}}
% ==============================================================================
% ==============================================================================
% ==============================================================================

We start our discussion with a few choices for the grammar, which is an intrinsic part of the model.
As we mentioned, the structure of our model is deceptively simple.
The two levels of referencing: traversing the information tree, and evaluating the scrambling bits, are especially confusing, and might mislead one into making the wrong conclusions about the size of the design space.
Even though the model uses only $m\cdot N$ bits for applying an $m$-bit deep scrambling, the size of the design space is actually much larger, thanks to the factorial growth of the grammar choices.
Specifically, we have $N^{2N}$ possible grammars, which is comparable to the huge size of Owen scrambling trees.
To give an example, a 256-symbol grammar size generates \mbox{$256^{512} = 2^{4096} = 2^{2^{12}}$}, which tops a 12-bits Owen-scrambling tree.
On top of that, then, comes the actual scrambling data fed to symbols.
The resultant size, however, is not a Cartesian product, since not all the grammars have the same capacity to distribute information.
To give an example, we take the extreme case of a grammar, or actually `the' grammar, that maps each symbol to itself on both branches.
This grammar is equivalent to a single-symbol grammar, irrespective of the size.

\begin{figure*}
% ------------------------------------------------------------------------------
    \setlength{\unit}{0.3\textwidth}
    {\centering\scriptsize
    \begin{tabular*}{1\textwidth}{@{}c@{\extracolsep{\fill}}c@{\extracolsep{\fill}}c@{\extracolsep{\fill}}c@{\extracolsep{\fill}}c@{}}
        \begin{tikzpicture}
            \node[inner sep=0, anchor=south west] at (0, 0.505\unit) {%
                \includegraphics[height=0.495\unit]{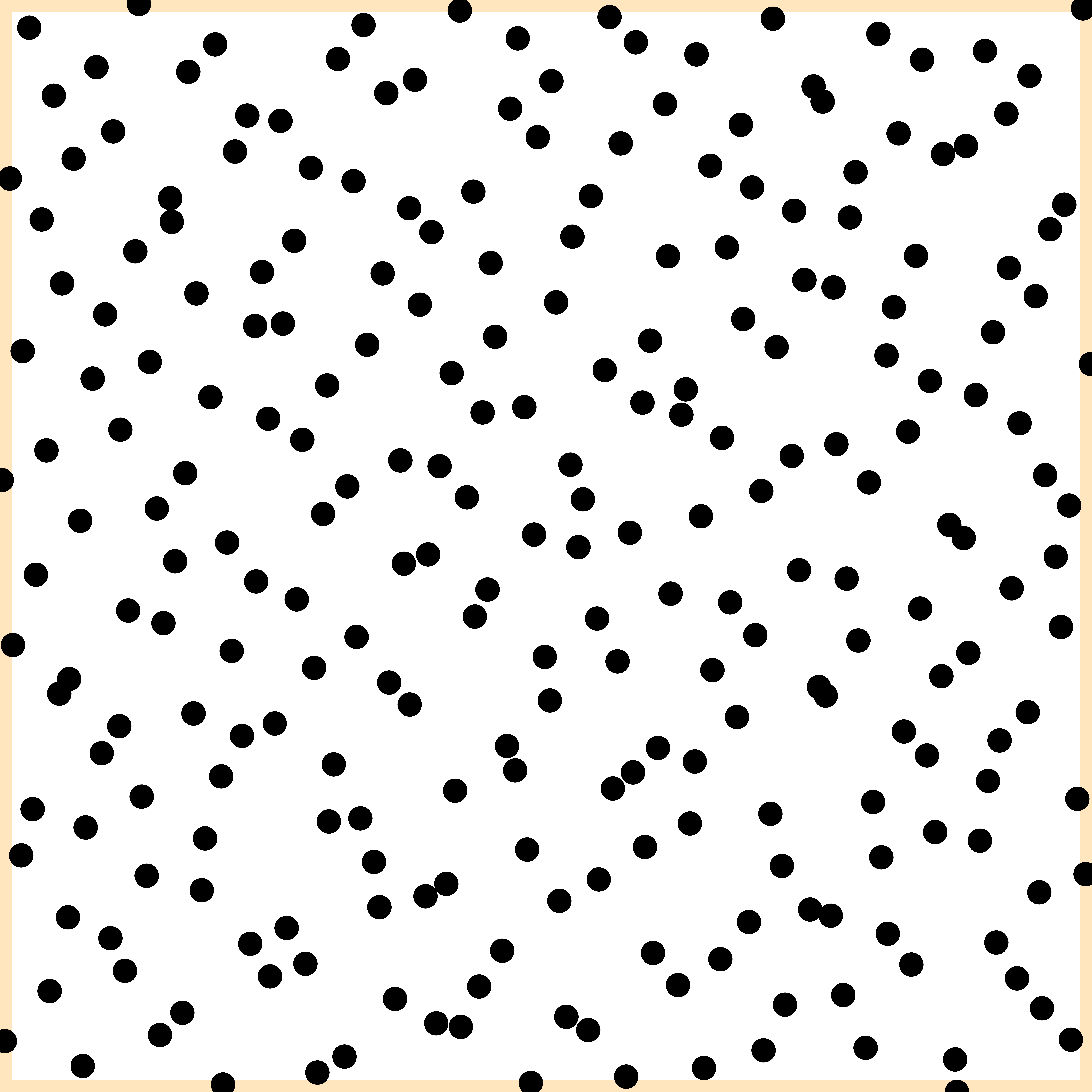}
            };
            \node[inner sep=0, anchor=south west] at (0, 0) {%
                \includegraphics[height=0.495\unit]{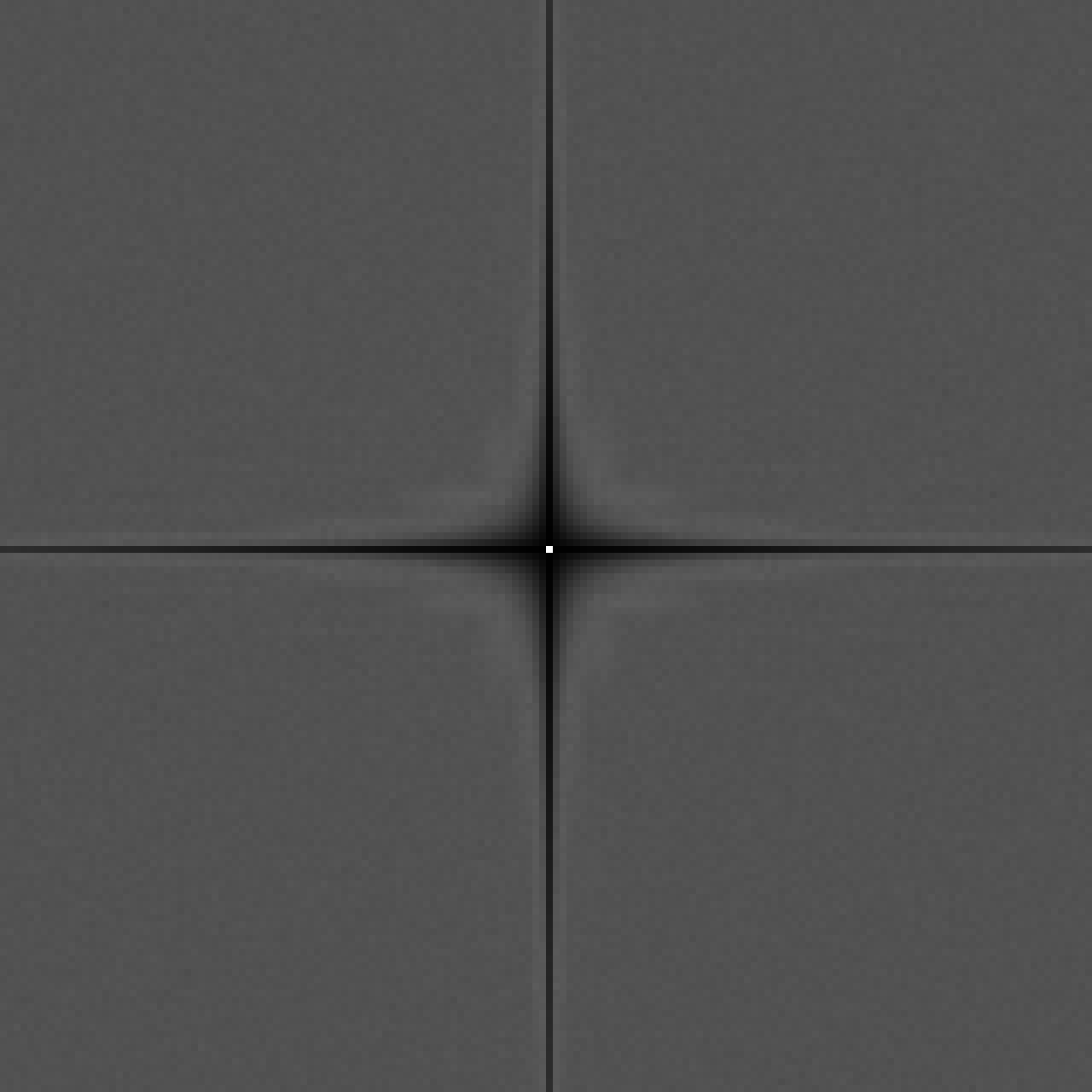}
            };
            \node[inner sep=0, anchor=south west] at (0.51\unit, 0) {%
                \includegraphics[height=1\unit]{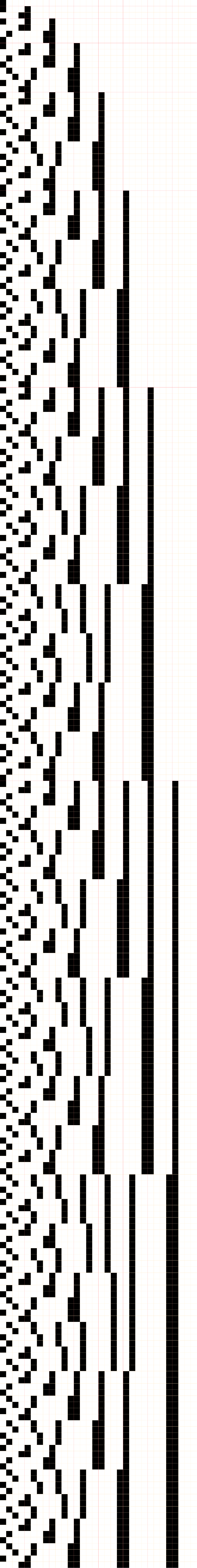}
            };
        \end{tikzpicture}&%
        \begin{tikzpicture}
            \node[inner sep=0, anchor=south west] at (0, 0.505\unit) {%
                \includegraphics[height=0.495\unit]{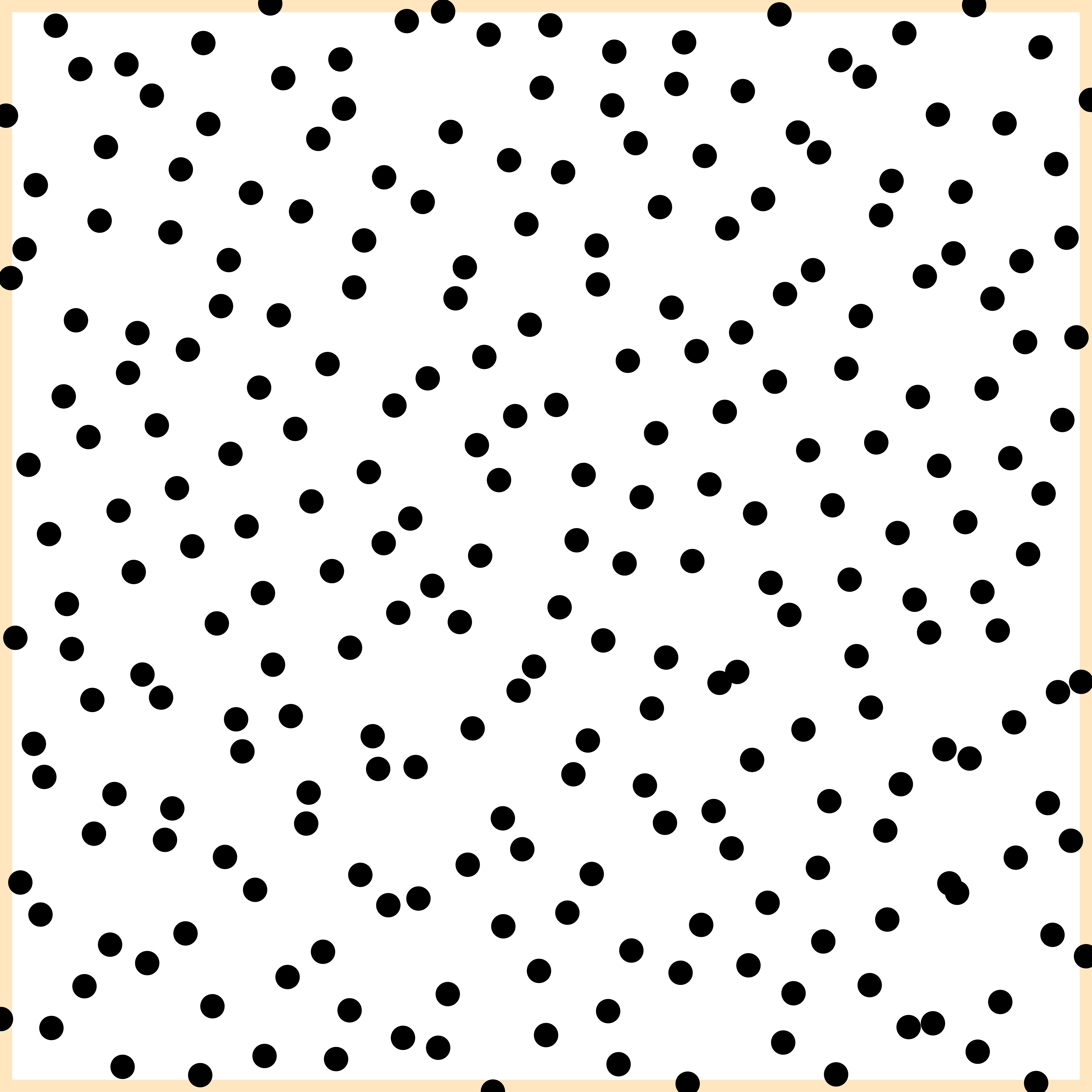}
            };
            \node[inner sep=0, anchor=south west] at (0, 0) {%
                \includegraphics[height=0.495\unit]{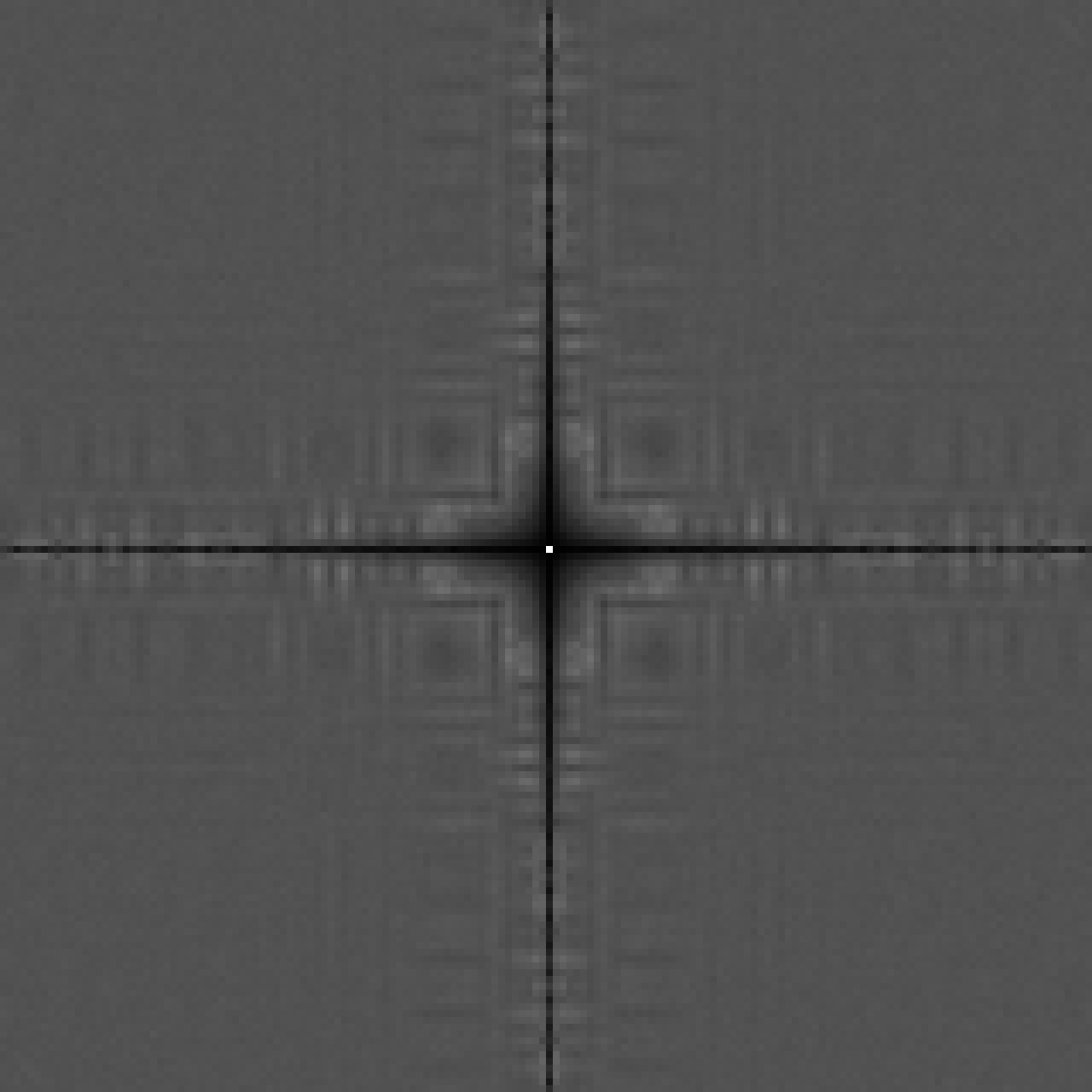}
            };
            \node[inner sep=0, anchor=south west] at (0.51\unit, 0) {%
                \includegraphics[height=1\unit]{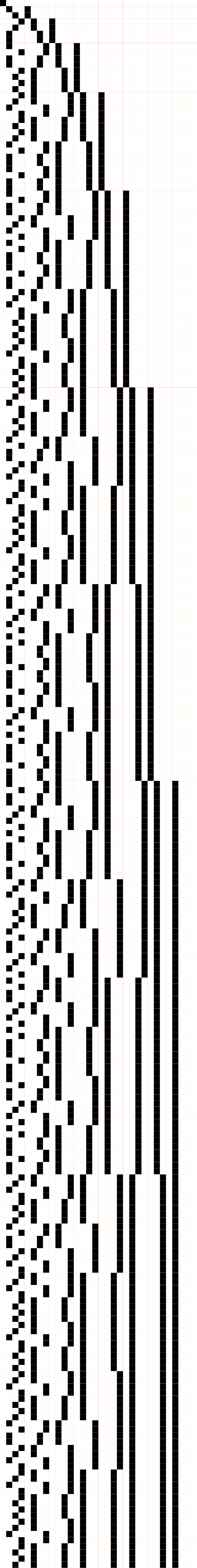}
            };
        \end{tikzpicture}&%
        \begin{tikzpicture}
            \node[inner sep=0, anchor=south west] at (0, 0.505\unit) {%
                \includegraphics[height=0.495\unit]{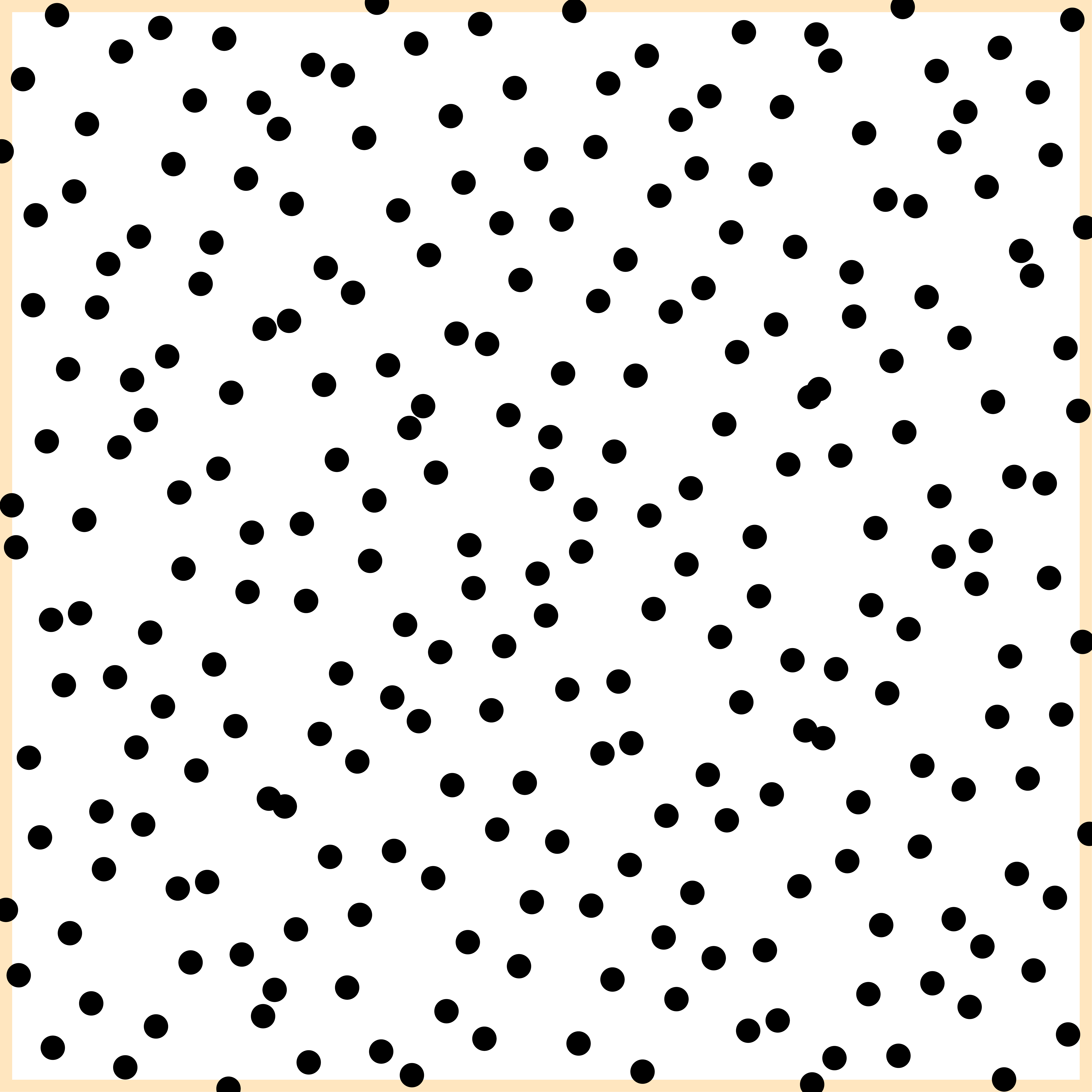}
            };
            \node[inner sep=0, anchor=south west] at (0, 0) {%
                \includegraphics[height=0.495\unit]{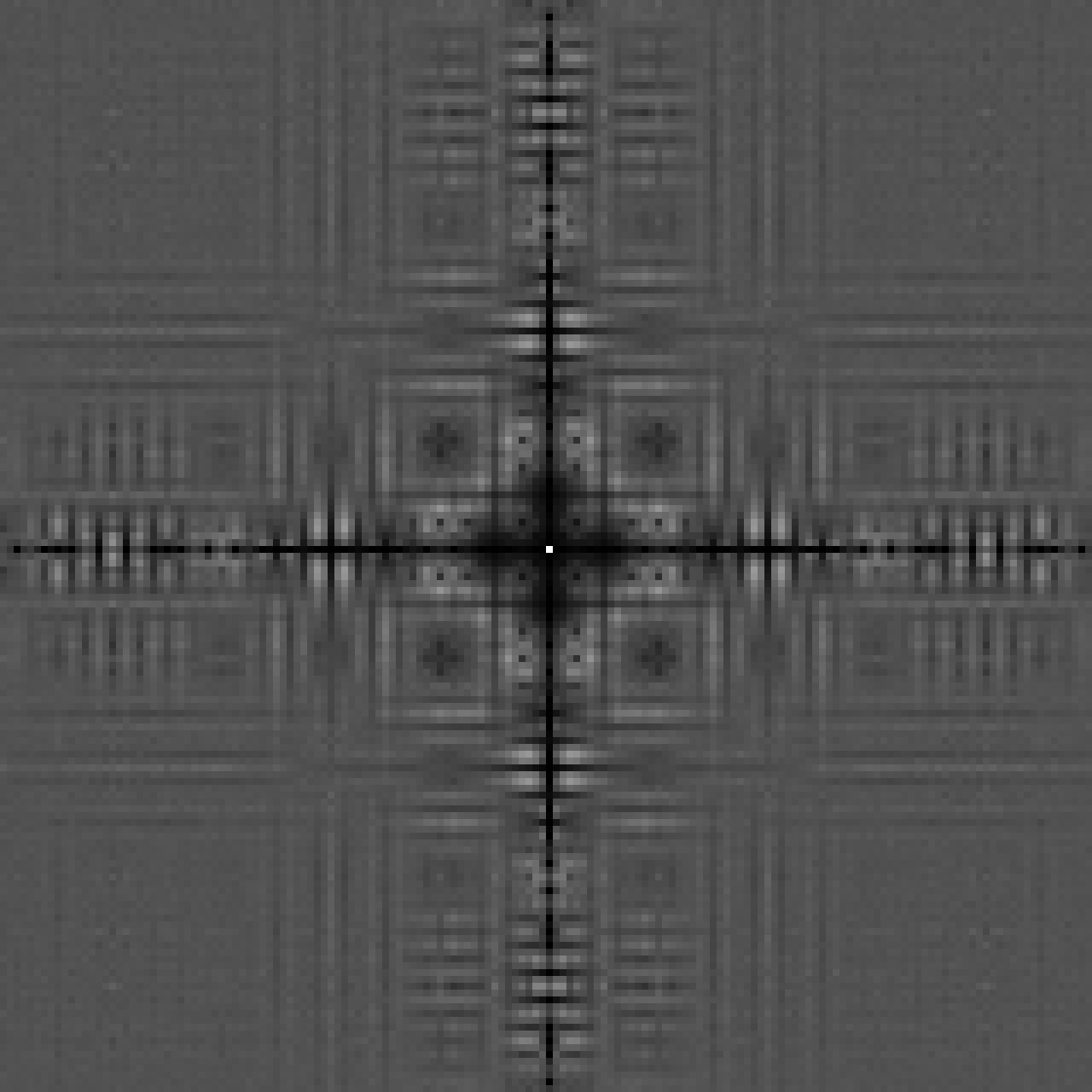}
            };
            \node[inner sep=0, anchor=south west] at (0.51\unit, 0) {%
                \includegraphics[height=1\unit]{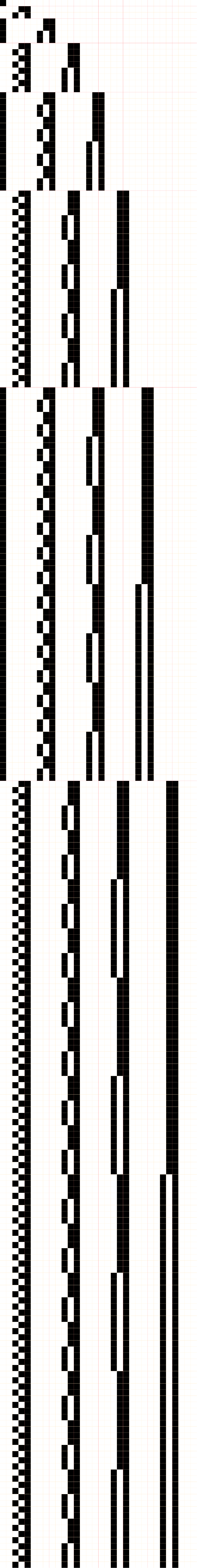}
            };
        \end{tikzpicture}&%
        \begin{tikzpicture}
            \node[inner sep=0, anchor=south west] at (0, 0.505\unit) {%
                \includegraphics[height=0.495\unit]{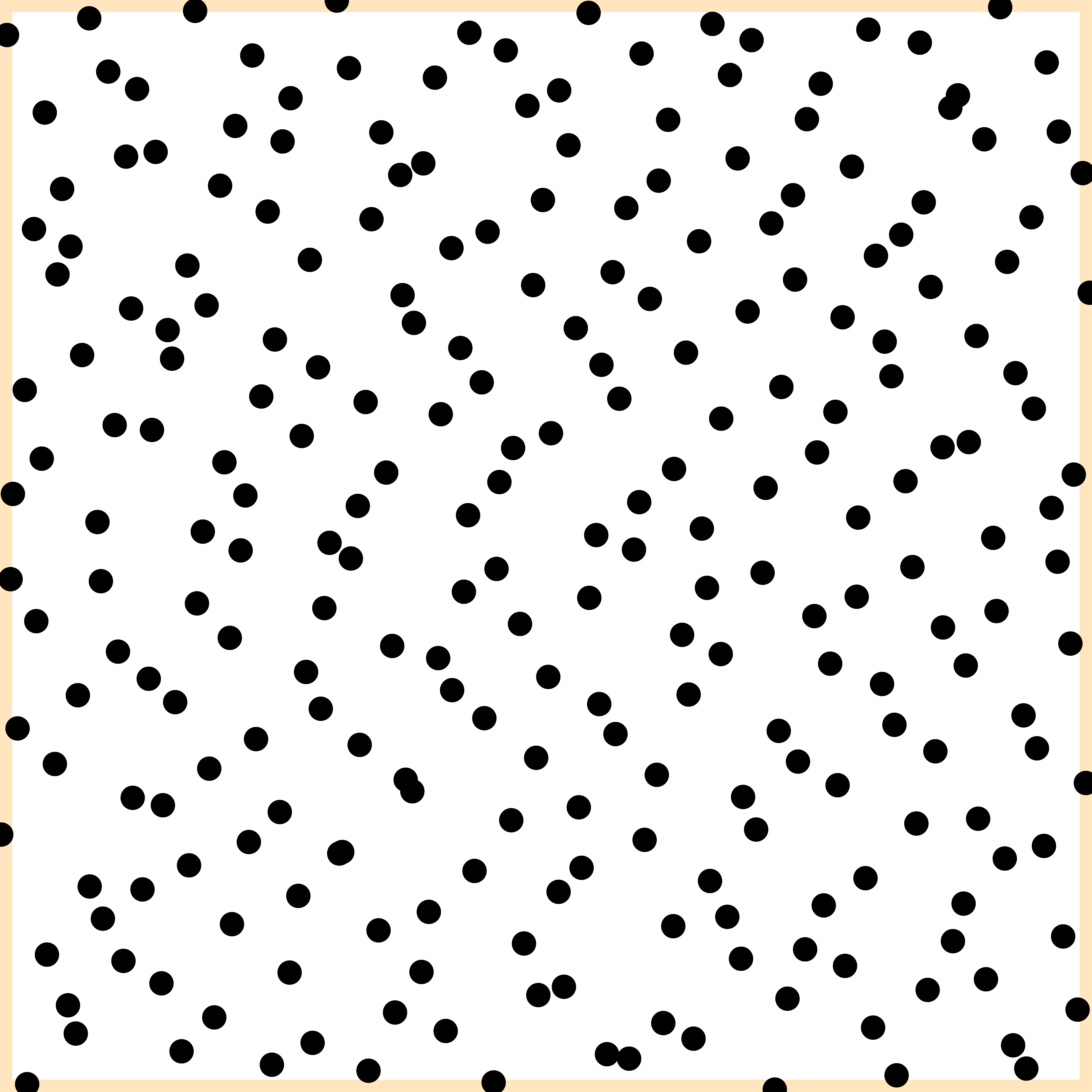}
            };
            \node[inner sep=0, anchor=south west] at (0, 0) {%
                \includegraphics[height=0.495\unit]{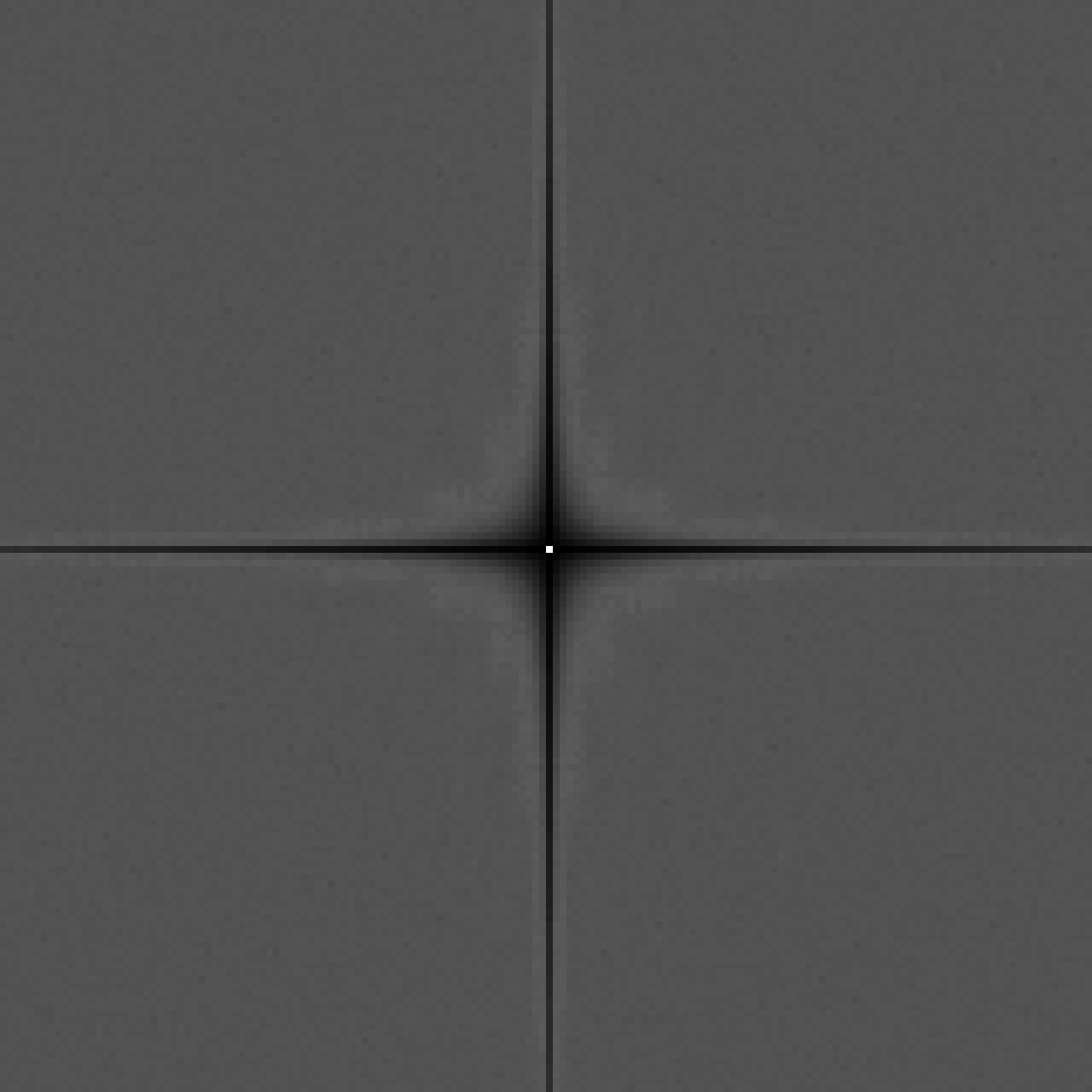}
            };
            \node[inner sep=0, anchor=south west] at (0.51\unit, 0) {%
                \includegraphics[height=1\unit]{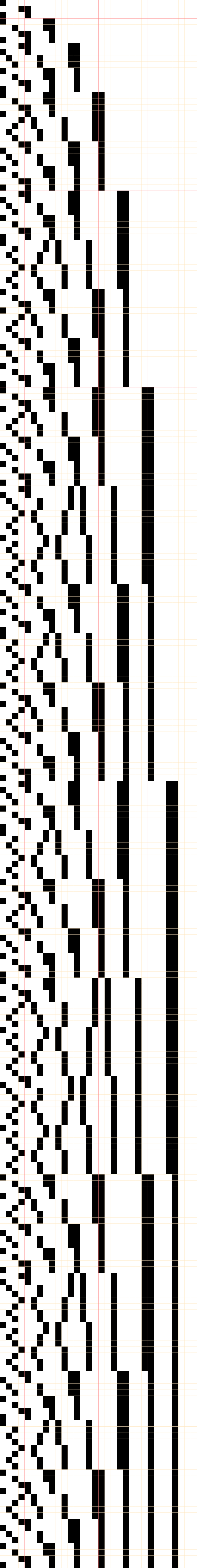}
            };
        \end{tikzpicture}&%
        \begin{tikzpicture}
            \node[inner sep=0, anchor=south west] at (0, 0.505\unit) {%
                \includegraphics[height=0.495\unit]{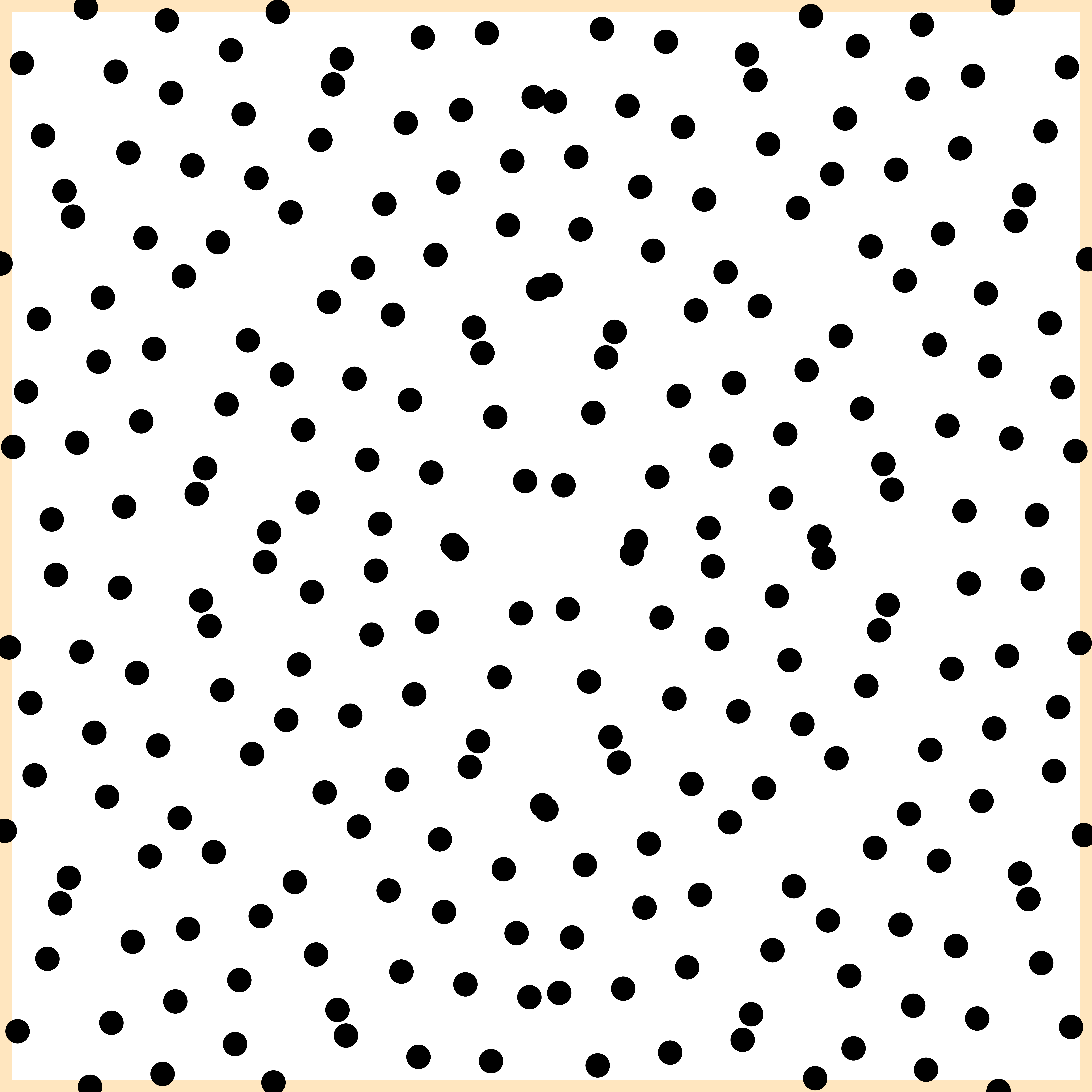}
            };
            \node[inner sep=0, anchor=south west] at (0, 0) {%
                \includegraphics[height=0.495\unit]{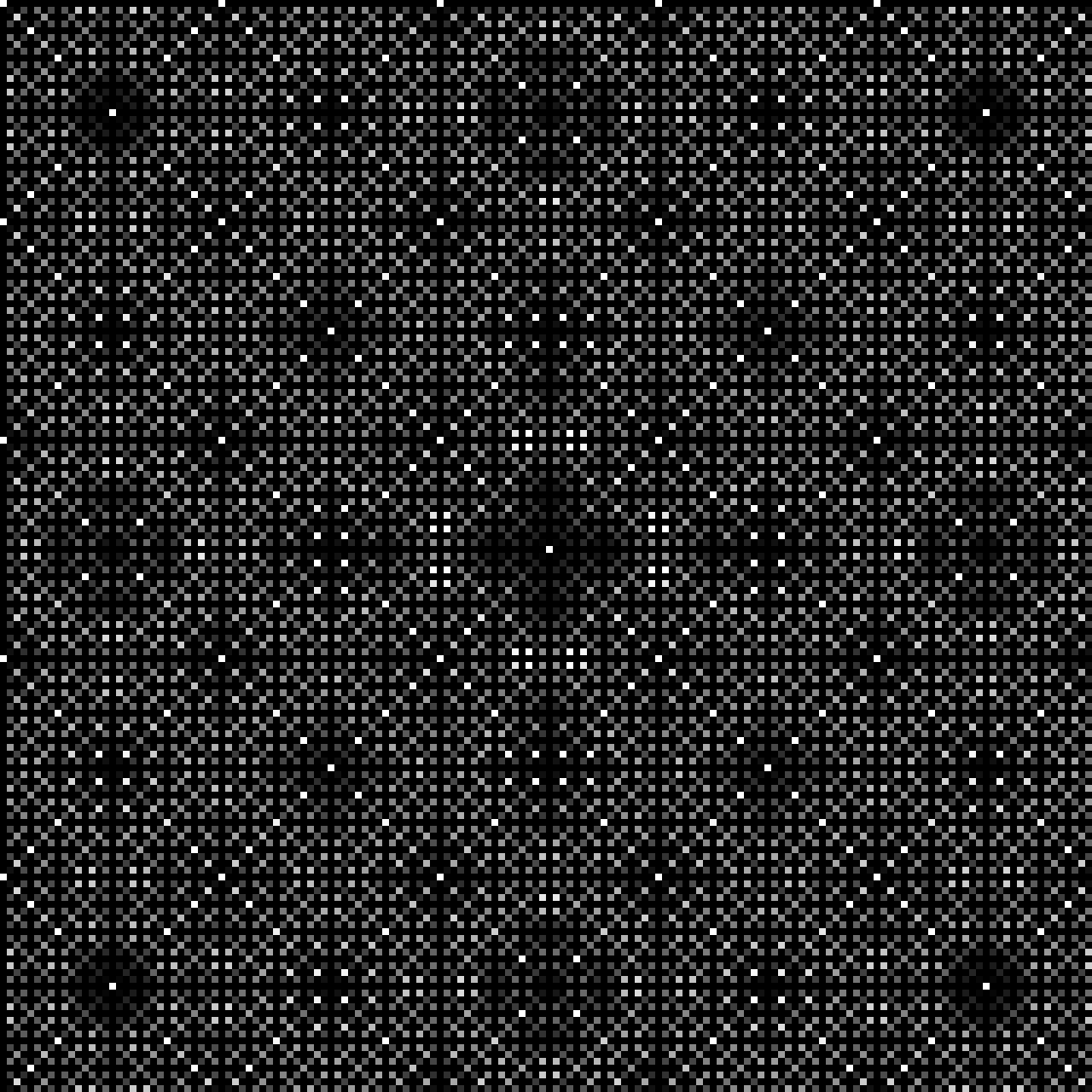}
            };
            \node[inner sep=0, anchor=south west] at (0.51\unit, 0) {%
                \includegraphics[height=1\unit]{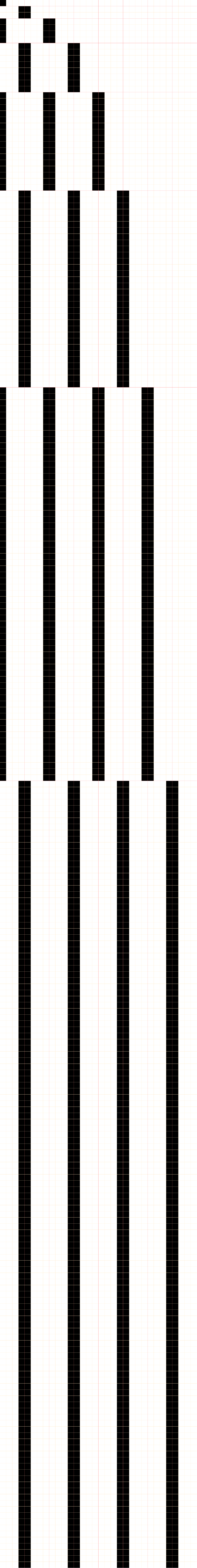}
            };
        \end{tikzpicture}\\[1mm]
        (a) TM:         ((0, 3), (1, 2), (0, 1), (1, 0)) &
        (b) Ordered:    ((1, 2), (3, 2), (1, 1), (1, 3)) & 
        (c) Random:     ((3, 2), (2, 2), (0, 0), (0, 0)) & % rand1
        (d) Random:     ((3, 0), (3, 2), (1, 2), (0, 2)) & % rand2
        (e) Random:     ((3, 3), (3, 1), (3, 0), (0, 0)) % rand4
    \end{tabular*}}
    \caption{\label{fig:grammar}%
        Example ART-Owen scramblings of the first 256 points of 2D Sobol sequence, showing (top-left) a sample point distribution, (bottom-left) the frequency power spectrum obtained by averaging periodograms over 1K realizations, and (right) the mapping of the information bits to the actual scrambling bits;
        comparing:
        (a) Thue-Morse grammar, (b) ordered grammar, and (c, d, e) three random grammars.
        Only four symbols are used, and the actual grammar is shown below each set.
        Each setting of the randomization data represents a realization.
        The mapping plot is essentially just a linear system of GF2 equations relating the final scrambling bits to the tabulated data bits.
        Each column of the grid corresponds to an assigned data bit of a symbol in the grammar, grouped by significance then depth.
        Each row corresponds to an output bit in the scrambling tree.
        The horizontal section lines mark the boundaries of tree levels.
        Finally, a dot in an intersection means that the scrambling bit in the row is affected by the data bit in the column.
    }
% ------------------------------------------------------------------------------
\end{figure*}

The redundant size of the design space is not necessarily very good news, though.
Indeed, noting that the size of the grammar space may exceed the size of the target scrambling trees for a smaller memory size simply means that there are duplicates, which is not a big issue in itself, but raises concerns about the presence of gaps as well; that is, the existence of some scrambling trees that are never realizable by the method, or by a specific realization of it.
This motivates the importance of studying the grammars in our model.
In the following we discuss three possible choices of grammars, illustrated in Fig.~\ref{fig:grammar}.
% ------------------------------------------------------------------------------
% ------------------------------------------------------------------------------

% ==============================================================================
\subsection{Thue-Morse Grammar\label{sec:th grammar}}
% ==============================================================================

Since our model is inspired by Ahmed et al. \shortcite{Ahmed17ART}, their chosen grammar comes as a natural choice for us to consider first.
The grammar is obtained by extending the binary Thue-Morse (TM) grammar
% ------------------------------------------------------------------------------
\begin{equation}
% ------------------------------------------------------------------------------
    \mu:\begin{array}{c}
        0\mapsto 01\\
        1\mapsto 10\\
    \end{array}\label{eq:t-prodcution}
% ------------------------------------------------------------------------------
\end{equation}
% ------------------------------------------------------------------------------
to a larger set of symbols.
Starting from 0, repeated application of Eq.~\eqref{eq:t-prodcution} leads to the Thue-Morse word
% ------------------------------------------------------------------------------
\begin{equation}
% ------------------------------------------------------------------------------
    T = 01101001100101101001011001101001\ldots \label{eq:t}
% ------------------------------------------------------------------------------
\end{equation}
% ------------------------------------------------------------------------------
as a steady point.
This is one of the most studied words in the Combinatorics of Words field of study \cite{Lothaire02Algebraic}.
It is characterized by its repetition-avoidance properties, and that is what made it attractive for use in sample distribution, and also promises good performance in our model.
As discussed in detail in \cite{Ahmed17ART}, an extended grammar may be derived from $T$ as follows.
The symbols are identified by the distinct sub-strings of a chosen length in $T$, the number of which decides the alphabet size.
The production rules are deducted by applying Eq.~\eqref{eq:t-prodcution} to the identification strings, and reading the child identification strings.
Since $T$ is a fixed point, all produced strings exist in $T$, hence this process maps to children from the same set.
We will avoid the distracting details here, and confine ourselves to an abstract idea that this non-obvious process enforces the resulting grammar to maintain a more-or-less similar sequence of symbols at all levels of the tree.
The essence of this is that, if the tree works well at a given level, it works equally well at all levels.

We tested the TM grammar in our model, and it produced excellent results, as far as we can judge from the frequency spectrum.
For more reassurance, we developed a visualization to see how the scrambling data is distributed over the scrambling tree, as may be seen in the vertical grids in Fig.~\ref{fig:grammar}.
While it is not easy to extract conclusions from this visualization, we may still note that the TM grammar produces a good coverage of the space, which suggest a reasonable shuffling of the data bits to produce the final scrambling bits.
To highlight a contrast, for example, note that the first column in Fig.~\ref{fig:grammar}(b) is almost empty, indicating that this bit is only used once.
Cross-checking with the grammar readily reveals that 0 is not produced by any rule, hence only found in the root node.
This \emph{under-utilization} gives a good example of the problems that may arise in an arbitrary grammar.

From the preceding discussion we may see that a tested-and-working grammar is invaluable; pending the development of a more objective solutions to choose from the intractable design space.
We therefore nominate the TM grammar as our default.
This also helps in maintaining a reference for bench-marking.
We have experimented with this grammar for different alphabet sizes, including 2, which reduces to Eq.~\eqref{eq:t-prodcution}, and we have not seen any failure case.
The plots in Fig.~\ref{fig:periodograms} use this grammar.
All that being said, the TM grammar is still not optimal in all aspects, as we will reveal next.
Please note that scrambling data is generated randomly, irrespective of the grammar.

% ==============================================================================
\subsection{Ordered Grammar\label{sec:ordered grammar}}
% ==============================================================================

The bit-mapping visualizations in Fig.~\ref{fig:grammar} actually represent a linear system that may be solved for the data bits, reversing the model.
This makes it possible to select data bits so as to reproduce a specific scrambling tree.
Trying to test this idea with our nominated TM grammar, we were disappointed to realize that the system fails early, consistently producing an-all zeros equation in the eights row, no matter how large the alphabet is.
We tried many tricks in the grammar extraction to avoid this destination, but they all failed.
This possibly comes from the fact that the grammar is inherited from a binary one.
This by no means implies that the TM grammar is not a good one.
It still distributes the scrambling data fairly over the whole tree, at all levels.
Just that it does not prioritize the leading bits enough as needed to solve our current problem.

\begin{figure}
% ------------------------------------------------------------------------------
    {\centering\scriptsize
        \includegraphics[width=1\columnwidth]{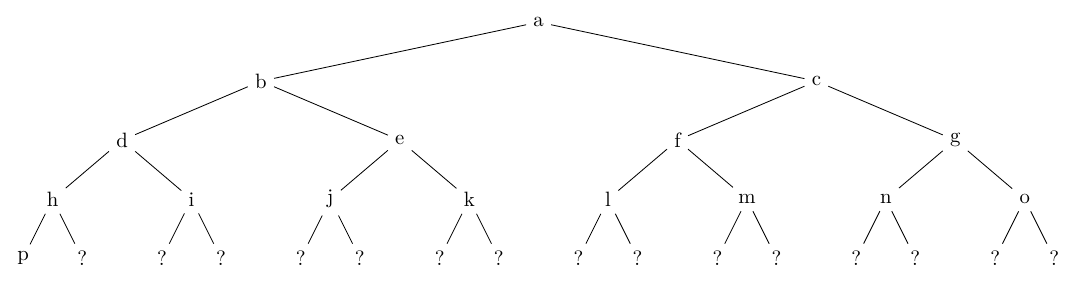}\\
    }
    \caption{\label{fig:ordered grammar}%
        An illustration of a tree generated by a 16-symbol ordered grammar.
        The essence is to allocate a distinct symbol, hence ensure an independent bit setting, for the topmost nodes of the scrambling tree; that is, the most significant bits of the scrambled data.
        The grammar is read back from the tree:
        $(a, b, \ldots, g, h, \ldots, p) \mapsto (bc,de,\ldots,no,p?,\ldots,??)\,,$
        and the missing entries (question marks) are populated arbitrarily.
    }
% ------------------------------------------------------------------------------
\end{figure}

To solve this shortcoming with the TM grammar, we conceived a brute-force grammar model that is guaranteed not to fail in this specific problem of reproducing a given scrambling tree, obtained by placing the symbols orderly such that they fill the top of the tree, hence we call it ordered grammar. See Fig.~\ref{fig:ordered grammar}.
% ------------------------------------------------------------------------------
% ------------------------------------------------------------------------------
The production rules are then read back from the tree, and the missing rules are filled arbitrarily.
Given a sufficiently-large alphabet, an ordered grammar, by construction, can reproduce any given Owen scrambling of any depth.
Not necessarily efficiently, though, and we are not claiming optimality.
Actually, the first sample ordered grammar we drew randomly for illustration in Fig.~\ref{fig:grammar}(b) readily exhibited an under-utilization, as discussed in the preceding section, that the root symbol is not reproduced by any rule.
This happened by chance, and we embraced it to highlight this potential inefficiency.
Please note that this is not necessarily an inherent deficiency with this grammar model; it warrants more research to find out, which we leave for future followup.
Our main goal for now is to prove this important aspect of our model that it spans all the universe of Owen scrambling, answering to the gaps question we raised at the beginning of the section.

% ==============================================================================
\subsection{Random Grammar\label{sec:random grammar}}
% ==============================================================================

Finally, we discuss the intuitive choice to use a random grammar.
A random production rule essentially spans the universe of grammars, which may sound good for our goal of random shuffling.
There is a subtle point that we need to be cautious about here, however, buried in the two levels of referencing.
Indeed, composing two randomized operations would not necessarily produce more or equally random results: they can counter each other.
Notably, we may already see noticeable regularity in Fig.~\ref{fig:grammar}(c, e) produced by supposedly random grammars.
Thus, there are certain rules and cases that we would like to avoid in our model.
A clear one is the twin rule discussed earlier.
Another one is fragmenting the alphabet into smaller disjoint sets.
This case becomes quite serious, and also quite likely, in a small alphabet.
We therefore do not recommend using a random grammar for a small alphabet.
At least it should be manually inspected.
We note that our favored TM grammar provably avoids these two problems.
In summary, a random grammar is not a first choice in memory-conservative applications; e.g., GPU-based ones.

Despite all the mentioned warnings, a random grammar still remains an attractive choice for scenarios that can afford a relatively large alphabet size.
For example, CPU-based applications.
Of special interest is the 256 alphabet size. The symbols fit exactly in one byte, making the assignment as simple as a single memory read.

% ==============================================================================
% ==============================================================================
% ==============================================================================
\section{Evaluation\label{sec:evaluation}}
% ==============================================================================
% ==============================================================================
% ==============================================================================

In the following, we discuss various aspects of our method.
See Table~\ref{table:comparison} for a summary of the performance and differences among the various techniques for generating Sobol points that we have evaluated.
Our model trades a marginal loss of speed for other benefits; the key ones are scalability, invertibility, and the opportunity for optimization.

% raw data:
%  Generate ray samples - SobolSampler                1024 launches     21.27 ms /   1.3% (avg  0.021, min  0.015, max   0.033)
% Total rendering time:   1591.73 ms
%  Generate ray samples - ARTOwenSampler              1024 launches     34.92 ms /   2.2% (avg  0.034, min  0.024, max   0.059)
%Total rendering time:   1590.74 ms
% 1.64x
%
% xor (permuted digits)
%  Generate ray samples - SobolSampler                1024 launches     21.95 ms /   1.4% (avg  0.021, min  0.015, max   0.034)
%Total rendering time:   1583.54 ms
% 1.03x
% 
% fastowen (~burley)
%  Generate ray samples - SobolSampler                1024 launches     22.37 ms /   1.4% (avg  0.022, min  0.016, max   0.034)
% 1.05x
% 
%owen 03 hashed
%  Generate ray samples - SobolSampler                1024 launches     34.50 ms /   2.2% (avg  0.034, min  0.022, max   0.060)
%Total rendering time:   1592.62 ms  
% 1.62x
%

% ------------------------------------------------------------------------------
\begin{table}[tb]
% ------------------------------------------------------------------------------
    \caption{\label{table:comparison}Summary of the characteristics and performance of various methods for generating Sobol sample points. Performance is measured by rendering the scene in Fig.~\ref{fig:teaser} on an NVIDIA 4090 RTX GPU.
    Although high-quality scrambling algorithms increase the sample generation time by a factor of $1.6$, sample generation is still just 2.2\% of total rendering time.
    }
    {%
    \centering
    \small
    \setlength\tabcolsep{4.5pt}
    \begin{tabular}{l|rccc}
    Sobol Scrambling&
    Time&%
    Invertible?&%
    Quality&%
    Optimizable?\\
    \hline
    None & $1\times$ & Yes & n/a & No\\
    XOR & $1.03\times$ & Yes & Poor & No\\
    Burley~\shortcite{Burley2020Scrambling} & $1.05\times$ & No & Good & No\\
    Hashed~\cite{Owen03Variance} & $1.62\times$ & No & Excellent & No\\
    ART Owen & $1.64\times$ & Yes & Excellent & Yes
    \end{tabular}
    }
% ------------------------------------------------------------------------------
\end{table}
% ------------------------------------------------------------------------------

% ==============================================================================
\subsection{Rendering\label{sec:rendering}}
% ==============================================================================

\begin{figure*}
% ------------------------------------------------------------------------------
  \setlength{\unit}{0.19\textwidth}
  {\scriptsize\centering
    \begin{tabular*}{1\textwidth}   {@{}c@{\extracolsep{\fill}}c@{\extracolsep{\fill}}c@{\extracolsep{\fill}}c@{\extracolsep{\fill}}c@{\extracolsep{\fill}}c@{\extracolsep{\fill}}c@{}}
        \includegraphics[height=1\unit]{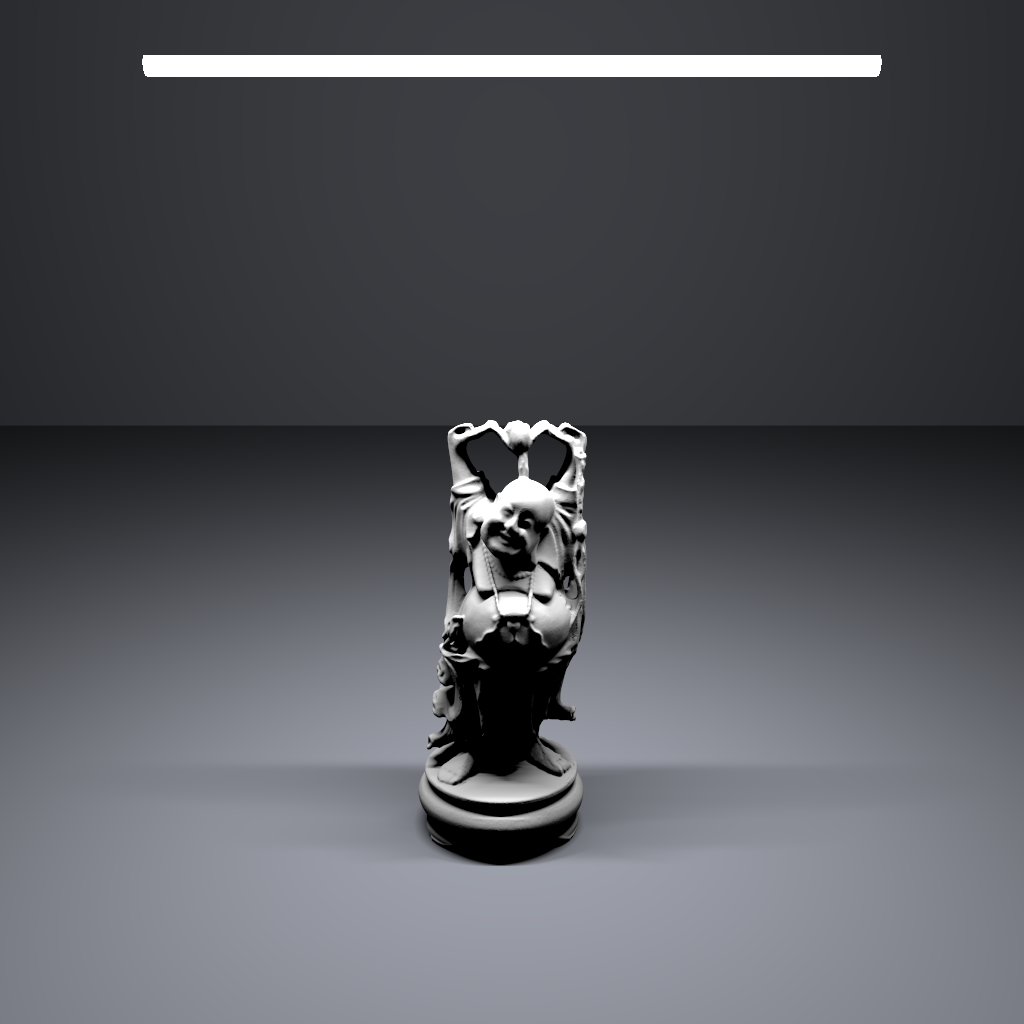}&%
        \includegraphics[height=1\unit]{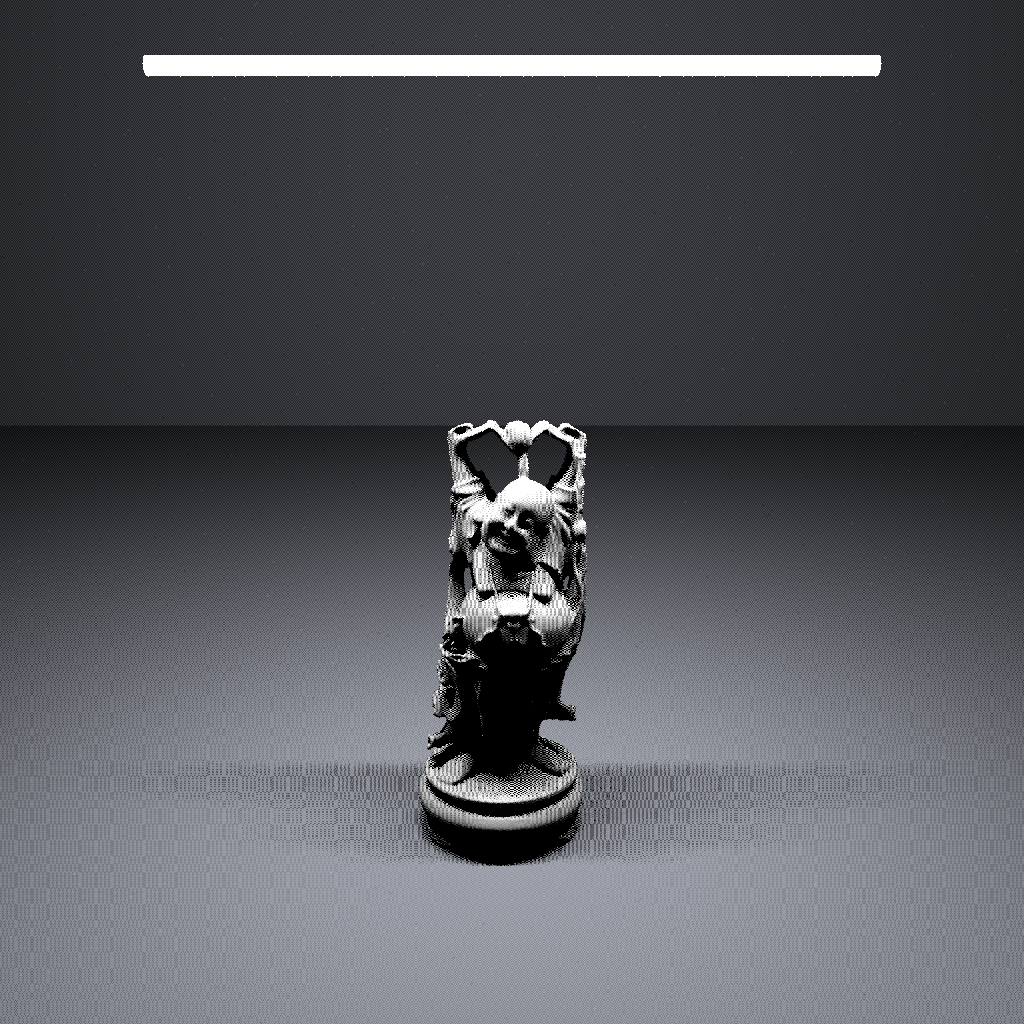}&%
        \includegraphics[height=1\unit]{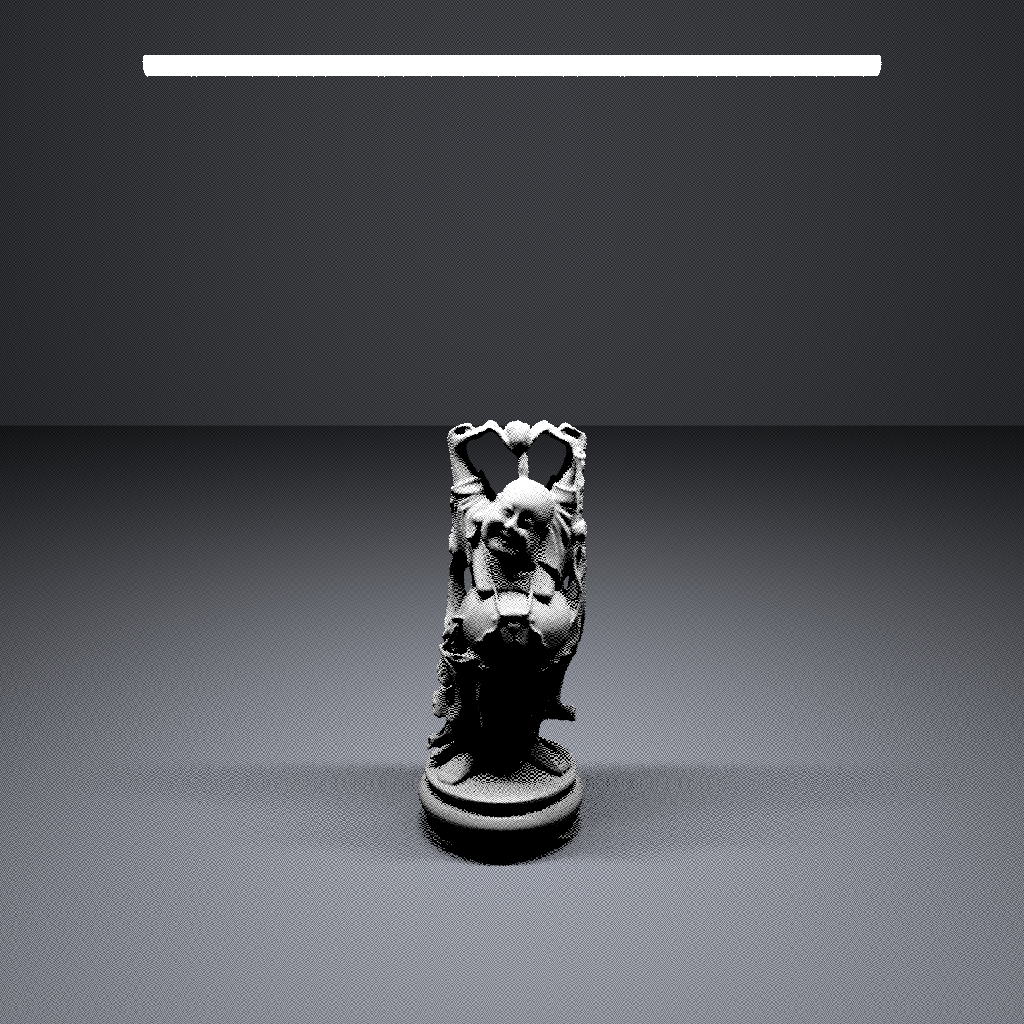}&&%
        \includegraphics[height=1\unit]{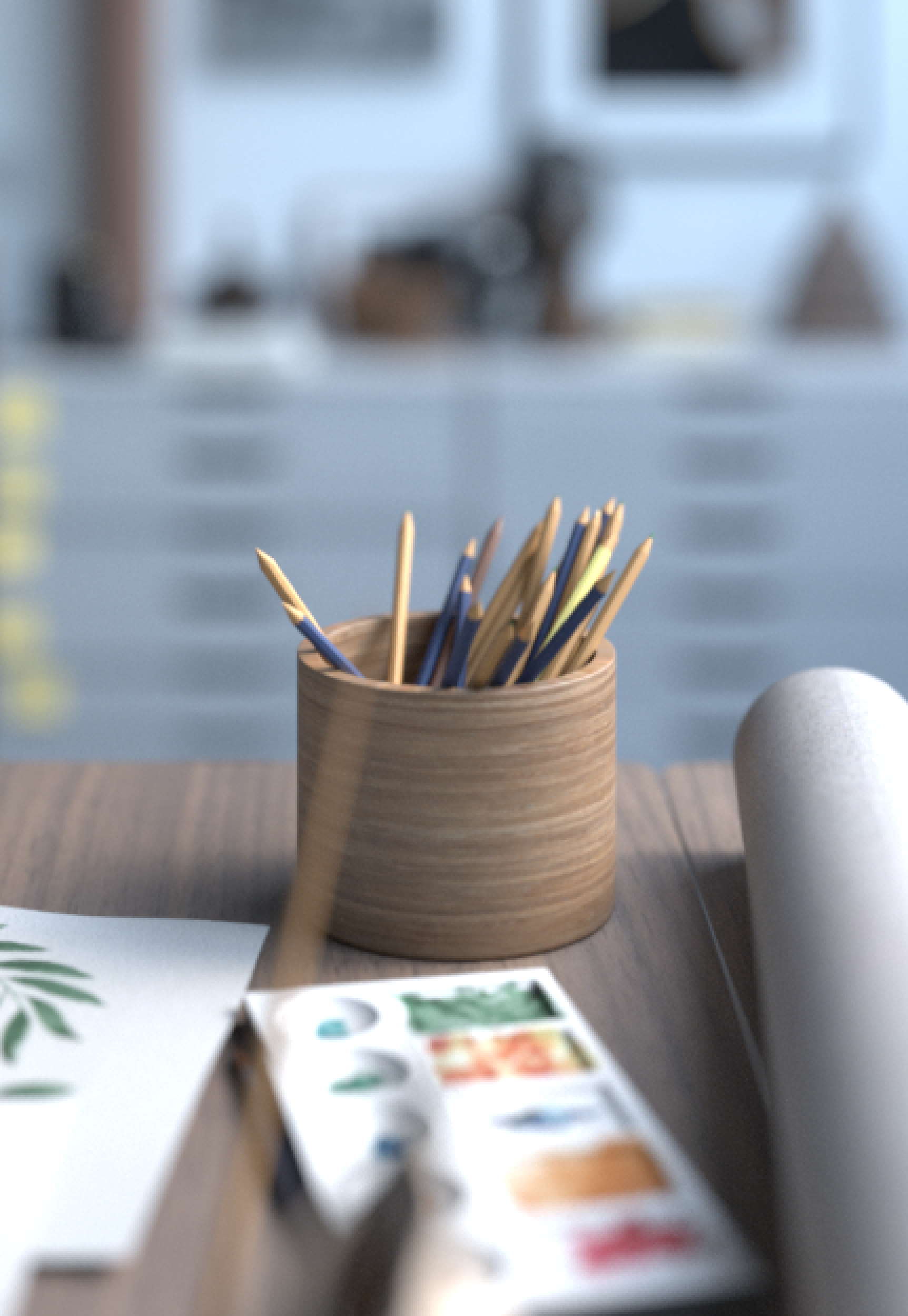}&%
        \includegraphics[height=1\unit]{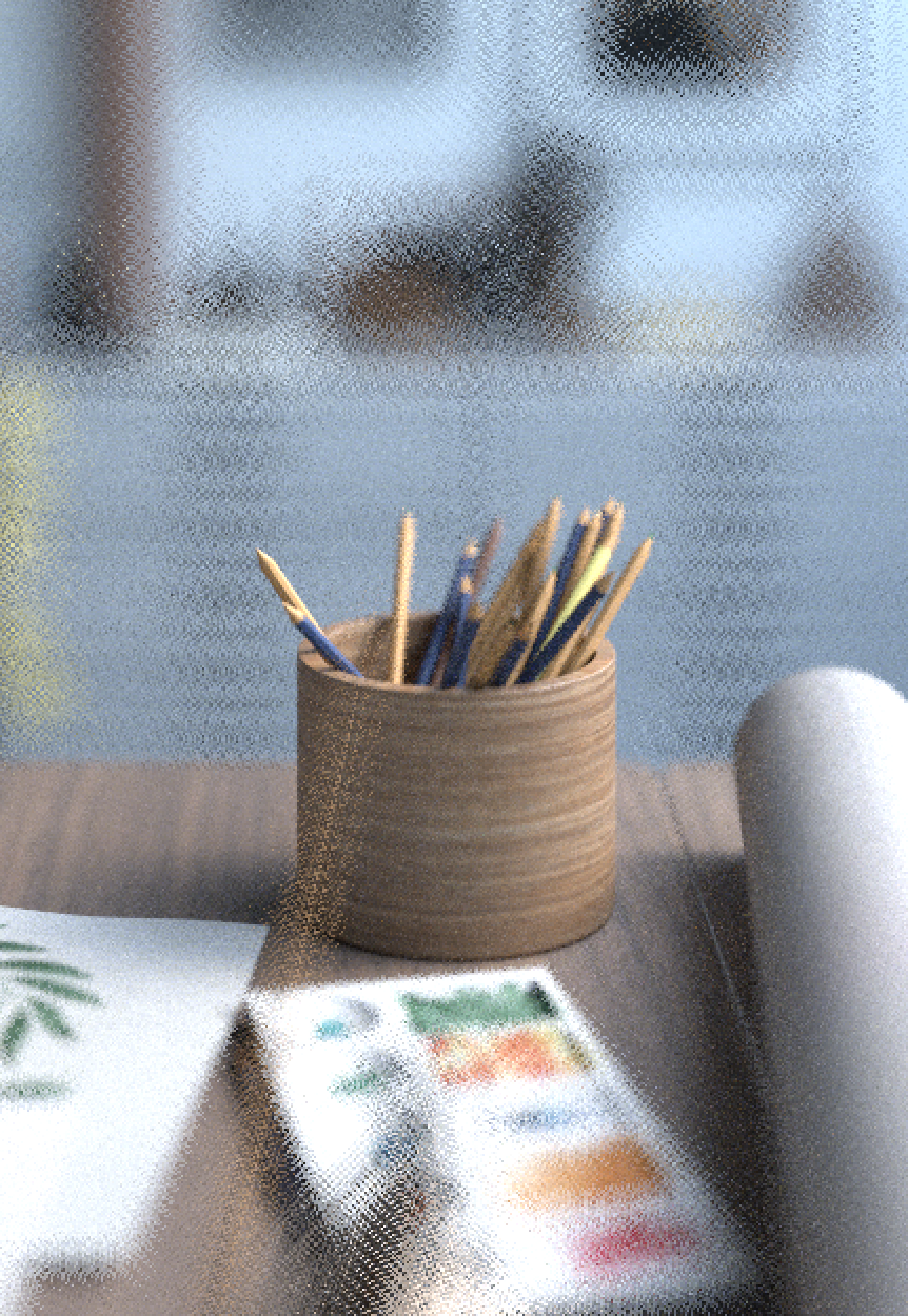}&%
        \includegraphics[height=1\unit]{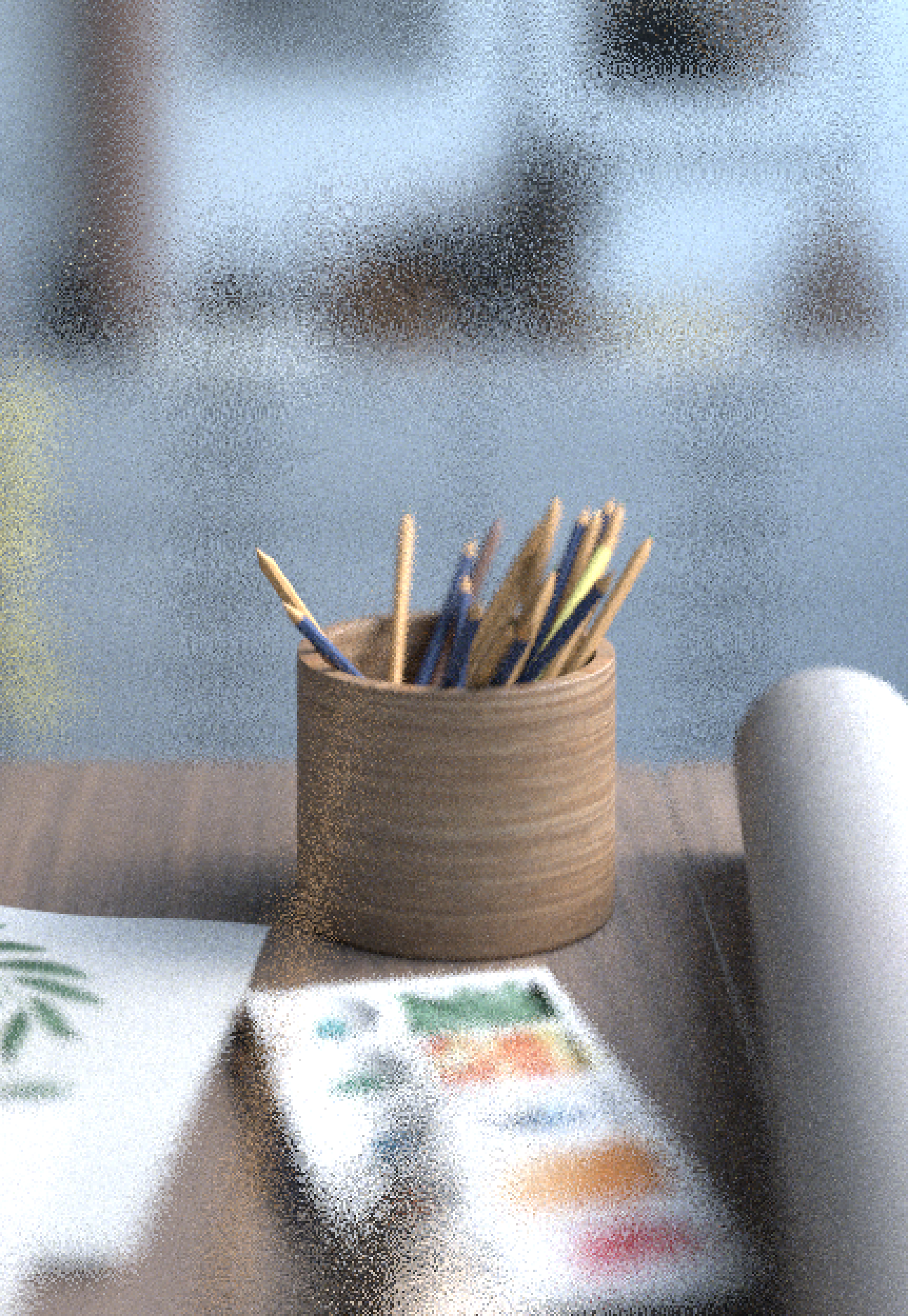}\\
        (a) Reference & (b) Unscrambled & (c) ART-Owen scrambled &&
        (d) Reference & (e) Unscrambled & (f) ART-Owen scrambled
    \end{tabular*}}
  \caption{\label{fig:rendering}%
     Renderings to demonstrate the potential improvements of integrating our ART-Owen scrambler with a Sobol sampler.
     Showing (left) area-light and (right) depth-of-field sampling effects at two and four samples per pixel, respectively.
     The unscrambled Sobol points cause structured artifacts that are greatly reduced with Owen scrambling.
  }
% ------------------------------------------------------------------------------
\end{figure*}

The primary application of Sobol sequences and Owen scrambling is rendering.
We integrated our scrambling code as a new sampler in PBRT \cite{Pharr23PBRT}.
Fig.~\ref{fig:teaser} demonstrates the benefit of Owen scrambling for anti-aliasing in the image plane, while
Fig.~\ref{fig:rendering} demonstrates similar perceptual improvements in sampling area lights and camera lens.
Please refer to the full-resolution images of these and other scenes in the supplementary materials.
%The left image in Fig.~\ref{fig:rendering} shows the structured error that is typical of Sobol sample points before convergence.
%Here, structure in the higher dimensions of the Sobol sequence that are used to sample the light source leads to error in the shadows.
%On the right, the scene is rendered using our Owen scrambled Sobol sequence, which successfully breaks up the visible error.
% ------------------------------------------------------------------------------
% ------------------------------------------------------------------------------

Even though our model is very efficient, it can by no means be faster than the unscrambled sampler;
see the sampling ``Time'' measurements in Table~\ref{table:comparison}.
Our measurements show that our scrambling causes Sobol sample generation to be $1.64\times$ times slower than unscrambled Sobol sample generation, measured on an NVIDIA RTX 4090 GPU.
However, sample generation is only a small fraction of the overall rendering time, especially for complex scenes.
For example, for the chair model in Fig.~\ref{fig:teaser}, it is less than 2\% of total runtime.
Thus, ART-Owen scrambling increases overall rendering time by less than 1\%.
For very simple scenes where sample generation consumes a larger fraction of runtime or for non-rendering applications, it may be more efficient to increase the sampling rate than to scramble.
We note that, given that scrambling requires just the few lines of code in Algorithm~\ref{alg:scrambling}, our scrambler may easily be selected on a scene-by-scene or even dimension-by-dimension basis.

% ==============================================================================
\subsection{Inversion\label{sec:inversion}}
% ==============================================================================

While it is straightforward to generate independent samples in each pixel, superior results are generally achieved using a \emph{global sampler} that generates points across the entire image plane.
In this way, not only is there a good distribution of points within each pixel, but samples in adjacent pixels are also well-distributed with respect to each other.
In order to be used with parallelism, global samplers must be \emph{invertible}, which allows enumerating the samples within a selected pixel.
Although the invertibility of unscrambled Halton and Sobol sequences was demonstrated by Gruenschlo\ss\ et al.~\shortcite{Gruenschlos12Enumerating}, to our knowledge, inversion of Owen scrambled sequences has not been demonstrated previously.
Therefore, for example, even though PBRT supports a number of scrambling algorithms in its Sobol sampler, it only uses unscrambled points for image plane sampling.

\begin{figure}
% ------------------------------------------------------------------------------
  {%
    \scriptsize\centering
    \setlength{\unit}{(\columnwidth - 2\gap)/2}
    \begin{tikzpicture}[
                img/.style = {inner sep=0, anchor=south west},
                lbl/.style = {align=center,yshift=-2mm},
            ]
            \node[img] at (0           ,0) (mask) {\includegraphics[width=1\unit]{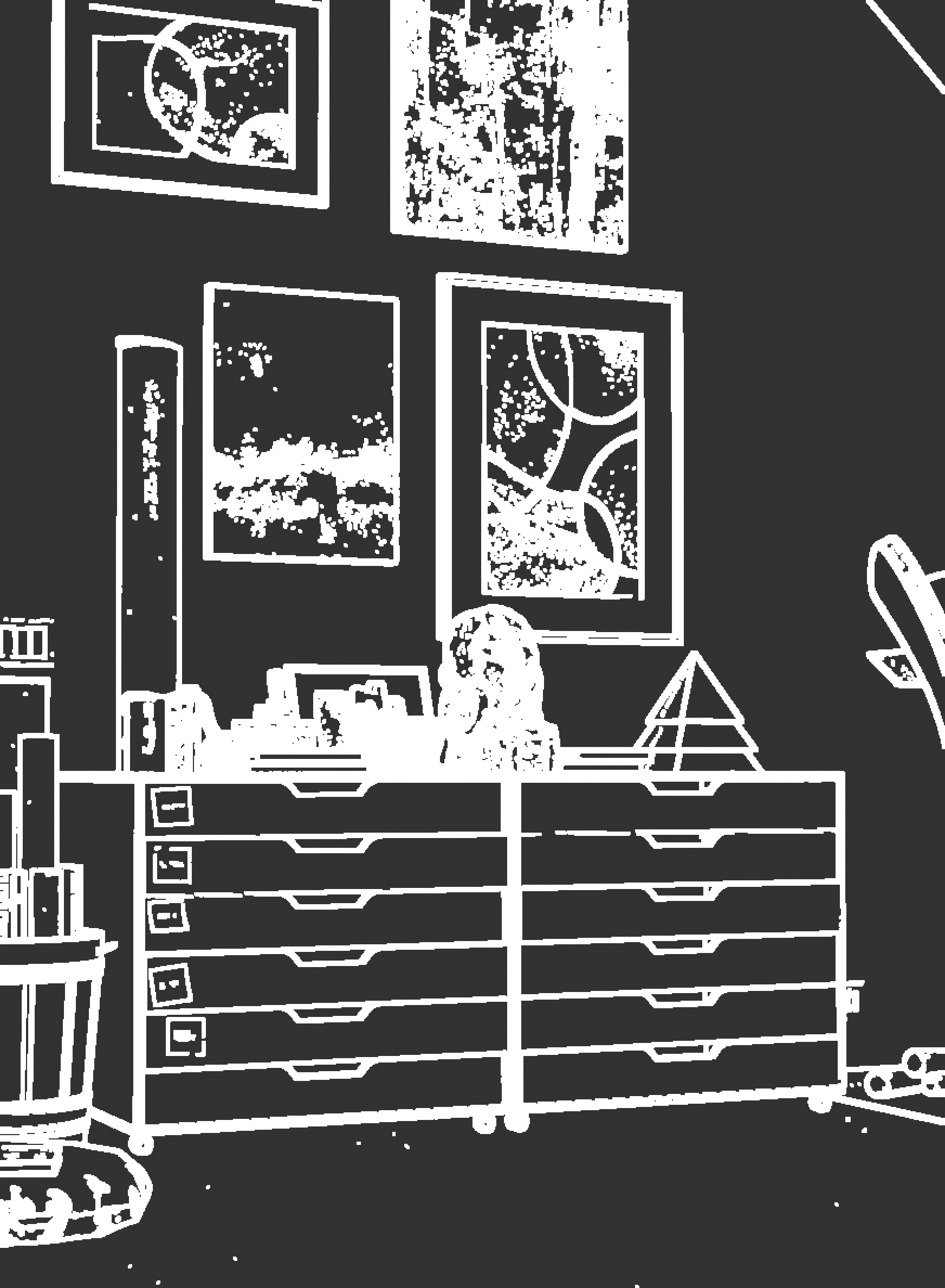}};
            \node[img] at (1\unit+2\gap,0) (all)  {\includegraphics[width=1\unit]{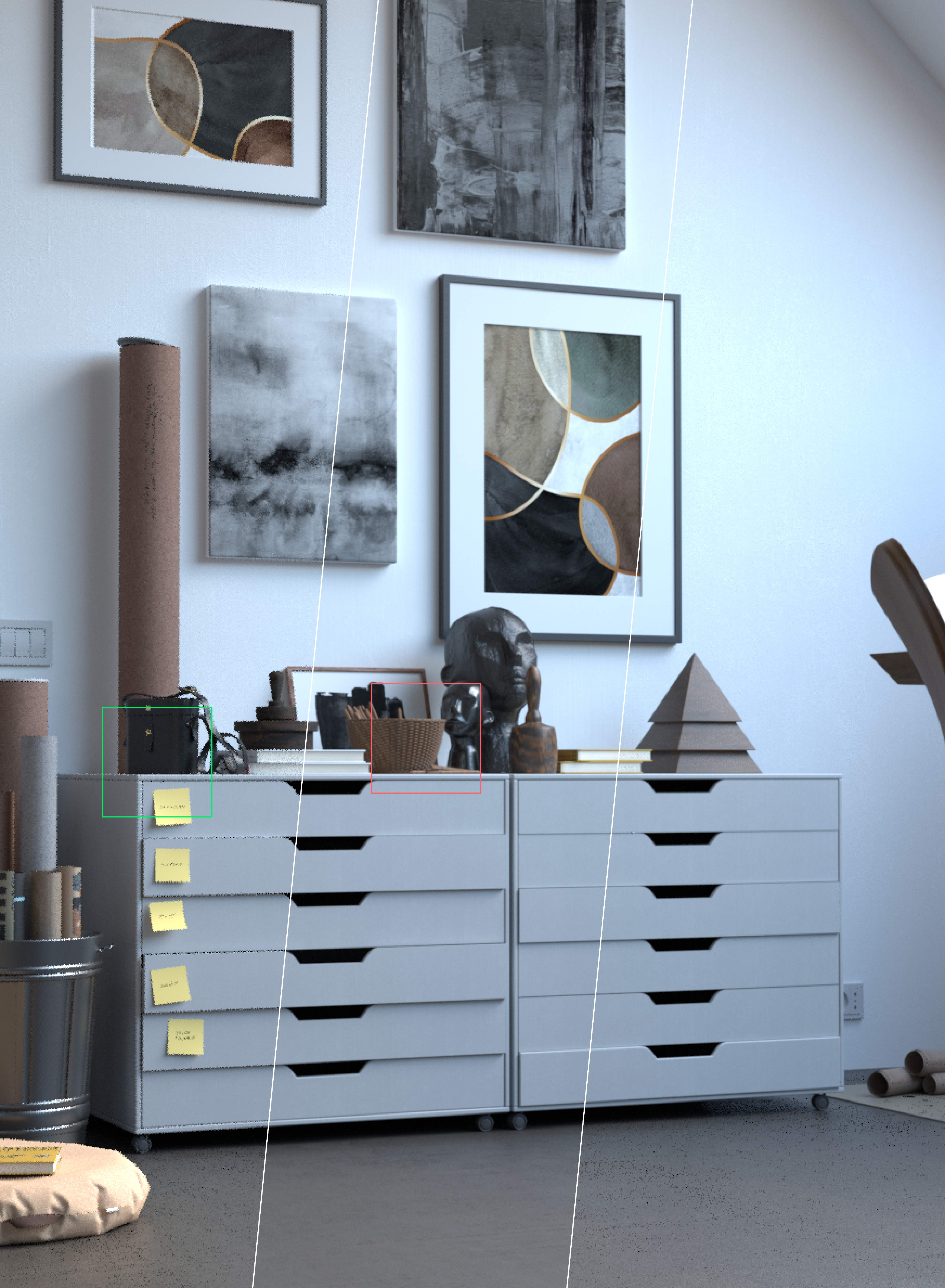}};
            \node[lbl] at (mask.south) {Mask};
            \node[lbl, xshift=0.133600917\unit] at ( all.south west) {2 spp};        % (0+0.5×233)÷872
            \node[lbl, xshift=0.433486239\unit] at ( all.south west) {64 spp};       % (233+0.5×290)÷872
            \node[lbl, xshift=0.766055046\unit] at ( all.south west) {Adative spp};  % (523+0.5×290)÷872
    \end{tikzpicture}\\[2mm]%
    \setlength{\unit}{(\columnwidth - 6\gap)/6}%
    \begin{tabular*}{1\columnwidth}{@{}c@{\extracolsep{\fill}}c@{\extracolsep{\fill}}c@{\extracolsep{\fill}}c@{\extracolsep{\fill}}c@{\extracolsep{\fill}}c@{\extracolsep{\fill}}c@{}}
        \includegraphics[width=1\unit]{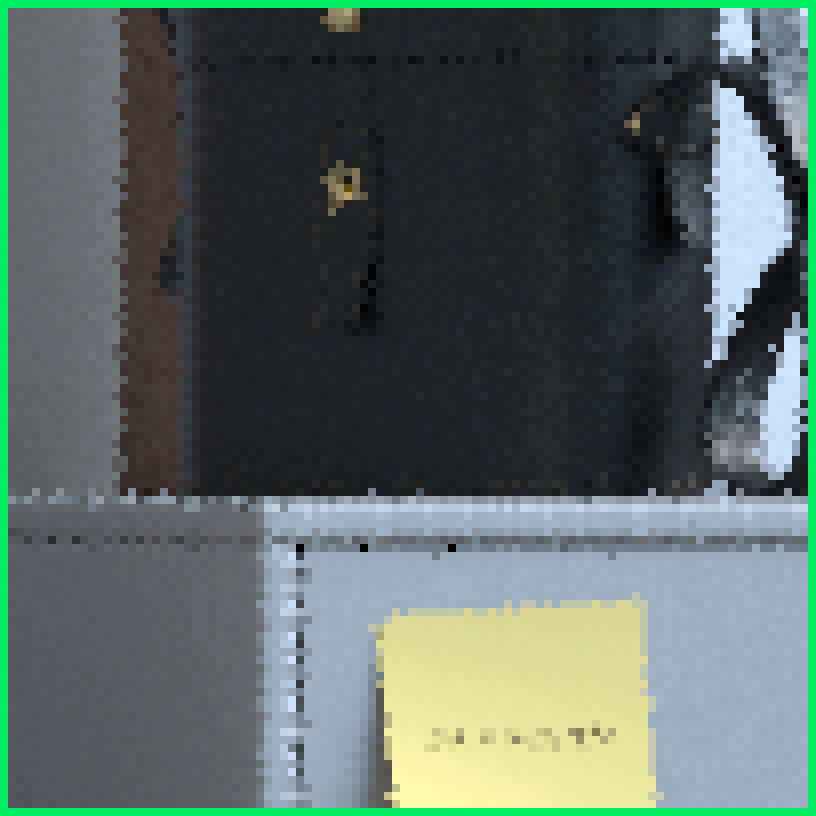}&%
        \includegraphics[width=1\unit]{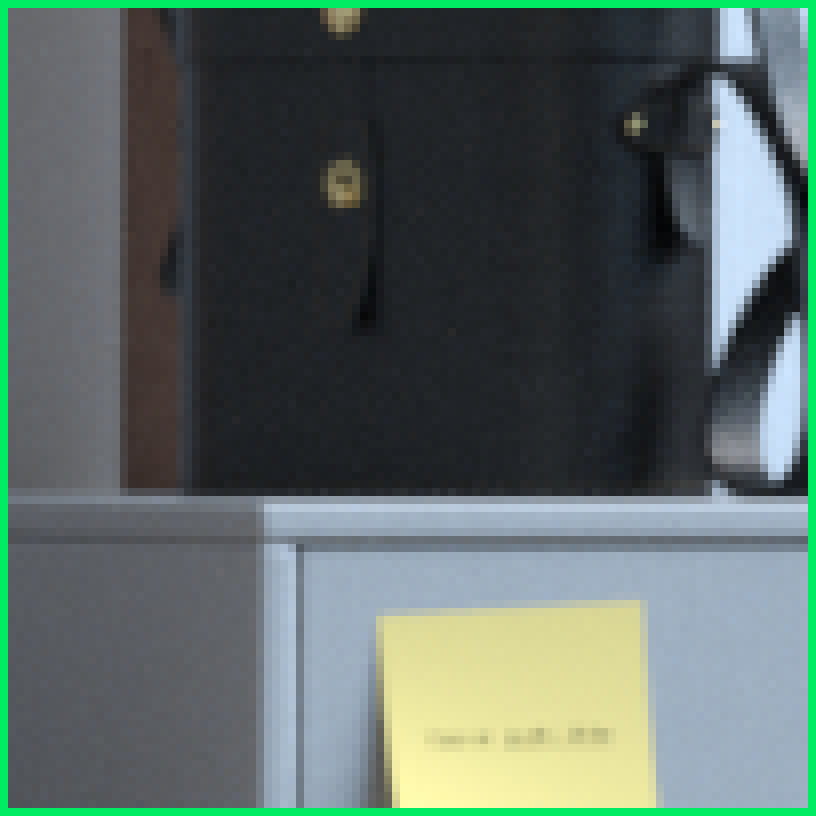}&%
        \includegraphics[width=1\unit]{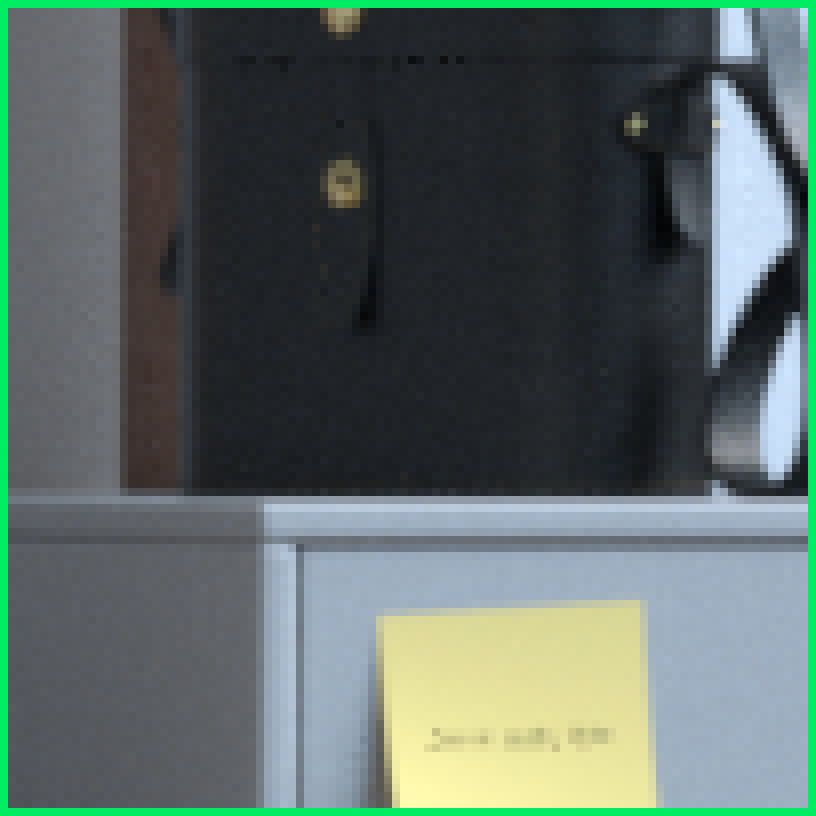}&&%
        \includegraphics[width=1\unit]{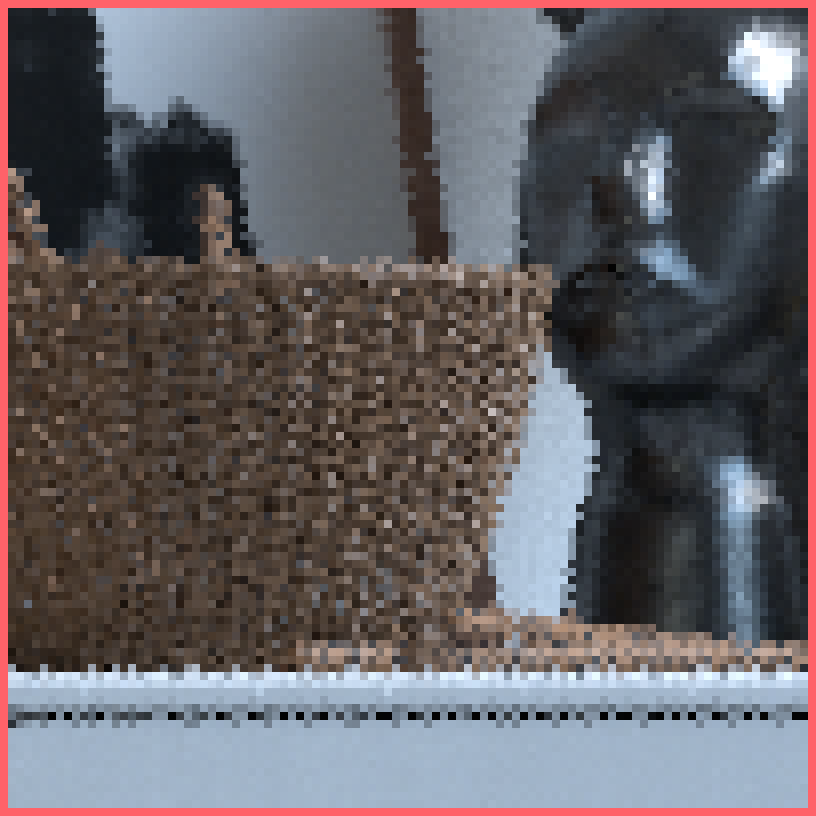}&%
        \includegraphics[width=1\unit]{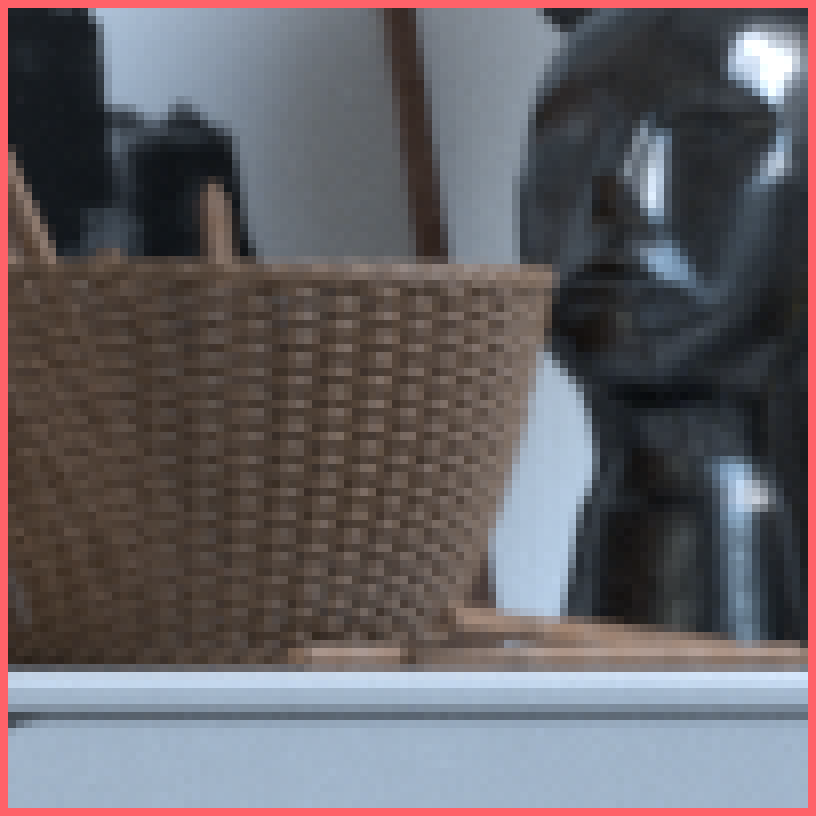}&%
        \includegraphics[width=1\unit]{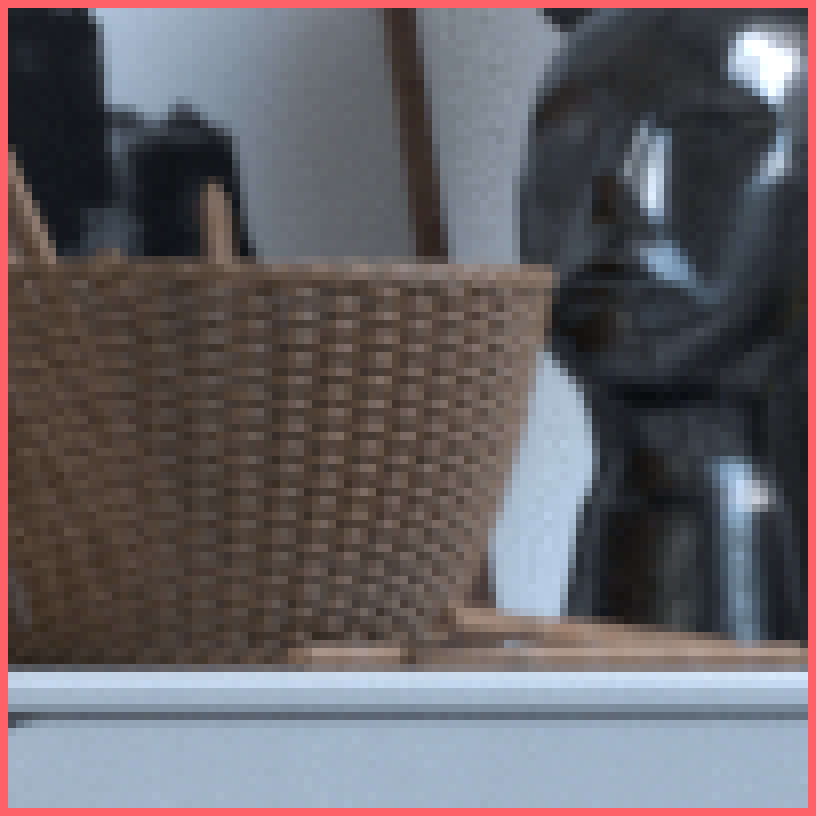}\\
        2 spp & 64 spp & Adaptive spp && 2 spp & 64 spp & Adaptive spp
    \end{tabular*}%
  }
  \caption{\label{fig:adaptive-sampling}%
    An invertible scrambling allows adaptive sampling. 
    The ``mask'' on the left is a visualization of the per-pixel sampling rates used, where black corresponds to 2~spp (samples per pixel) and white 64~spp.
    The scene on the right is rendered at, from left to right, a fixed rate of 2~spp, a fixed rate of 64~spp, and adaptive sampling based on the shown mask, at an average of 15.7~spp.
    Crops demonstrate the effectiveness of adaptive sampling.
  }
% ------------------------------------------------------------------------------
\end{figure}

Our model is not only invertible, but inversion is efficient; the algorithm to recover the original sample location given the scrambled location and grammar and data tables is shown in Algorithm~\ref{alg:unscrambling}.
Note that the unscrambling process is almost identical to the scrambling one.
% ------------------------------------------------------------------------------
\begin{algorithm} [tb]
% ------------------------------------------------------------------------------
    \let\oldnl\nl% Store \nl in \oldnl
    \newcommand{\nonl}{\renewcommand{\nl}{\let\nl\oldnl}}% Remove line number for one line
    \caption{\label{alg:unscrambling}%
        C code implementation of unscrambling.
        It is almost identical to the scrambling Algorithm~\ref{alg:scrambling}.
    }
    \nonl{\includegraphics[width=0.91\columnwidth]{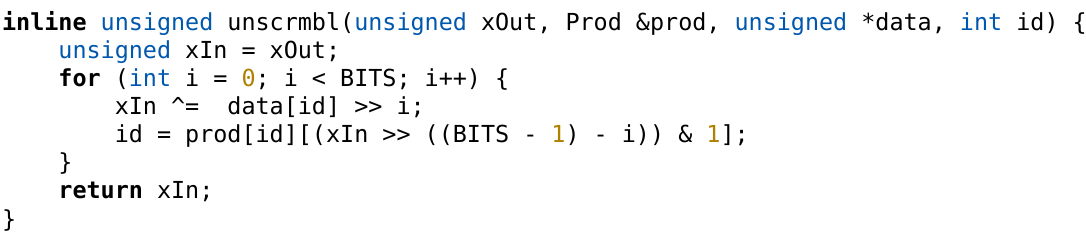}}
% ------------------------------------------------------------------------------
\end{algorithm}
% ------------------------------------------------------------------------------
\edited{The scrambling is undone to the pixel level with Algorithm~\ref{alg:unscrambling}, then inversion is completed as in Gr{\"u}nschlo{\ss} et al. \shortcite{Gruenschlos12Enumerating}.}
Fig.~\ref{fig:teaser} shows the benefit of this capability with an example where fine geometric detail and structure in the first two dimensions of the Sobol sequence lead to errors in the image.
It is evident that error from the fine detail in the chair's seat is pushed to higher frequencies with Owen scrambling, giving improved blue noise characteristics and an error that is visually closer to the reference.\footnote{For both Fig.~\ref{fig:teaser} and~\ref{fig:adaptive-sampling},
we applied \emph{splitting}, tracing many secondary rays after each primary hit. In this way, these comparisons highlight the image-plane sampling benefits of scrambling with consistently high-quality estimates of indirect lighting.}

Another useful application of invertibility is adaptive sampling: it is straightforward to produce as many samples as are required in each pixel, potentially doing so incrementally until convergence criteria are met.
Fig.~\ref{fig:adaptive-sampling} shows a proof of concept: at each pixel, we take either 2 or 64 image samples, with the higher sampling rate selected based on geometric edges and discontinuities and the color contrast of the albedo compared to neighboring pixels.
An average of 15.7 samples per pixel are taken with the adaptive sampler, giving a result that is nearly the same as 64 samples at every pixel, yet with $4\times$ fewer pixel samples.
% ------------------------------------------------------------------------------
% ------------------------------------------------------------------------------

% ==============================================================================
\subsection{Convergence\label{sec:convergence}}
\begin{figure}[tb]
% ---------------------------------------
  {\scriptsize\centering
  \includegraphics[width=\columnwidth]{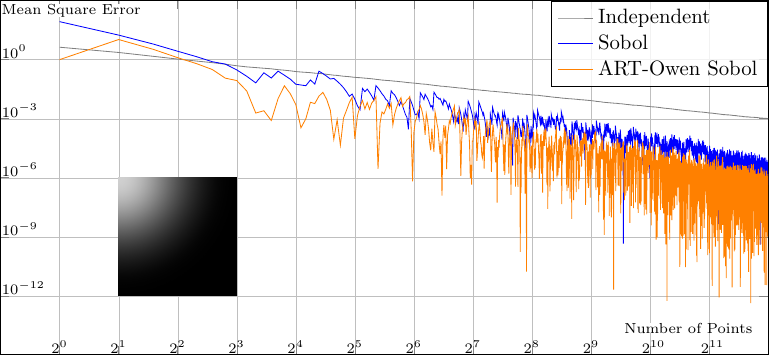}
   }
  \caption{\label{fig:convergence}%
Mean squared error when integrating the Gaussian function shown in the inset with three different point sets.
Low discrepancy Sobol points have much lower error than independent uniform points that converge at $O(1/\sqrt{n})$.
Owen scrambling offers an asymptotically higher rate of convergence, $O(1/n^{3/2})$, than unscrambled Sobol points, $O(1/n)$ \edited{\protect{\cite{Burley2020Scrambling}}}.
  }
% ---------------------------------------
\end{figure}

% ==============================================================================
One of the advantages of Owen scrambling is superior asymptotic rates of convergence with smooth functions.
Fig.~\ref{fig:convergence} shows a synthetic example of integrating a smooth function, following the examples of Christensen et al.~\cite{Christensen18Progressive}.
Independent uniform samples have the highest error, with error decreasing by $O(1/\sqrt{n})$, as is standard with Monte Carlo integration.
For this smooth function, unscrambled Sobol points converge at the higher rate of $O(1/n)$, but ART Owen scrambled points have a remarkable $O(1/n^{3/2})$ rate of convergence, with dramatically lower error at power-of-two sample counts.
We expect similar benefits in rendering given smooth integrands such as unoccluded light sources.
% ---------------------------------------
% ---------------------------------------

% ==============================================================================
\subsection{Complexity\label{sec:complexity}}
% ==============================================================================

The time complexity of our method is $\mathcal{O}(m)$, the bit depth, which may
be considered constant for most applications, but we meant to expose the bit
depth as a degree of freedom to control the process.
Our method clearly stands out when it comes to the coding complexity of the runtime part.
The variation is large, however, in the grammar-design part.
For example, a random grammar is trivial to implement, while the TM grammar is considerably more complex.
Finally, the space complexity is user-prescribed.

% ==============================================================================
\subsection{Spectral Analysis\label{sec:quality}}
\begin{figure*}
% ------------------------------------------------------------------------------
  \setlength{\unit}{(\textwidth - 8\gap)/9}
  {\tiny
  \begin{tabular*}{1\textwidth}{@{}c@{\extracolsep{\fill}}c@{\extracolsep{\fill}}c@{\extracolsep{\fill}}c@{\extracolsep{\fill}}c@{\extracolsep{\fill}}c@{\extracolsep{\fill}}c@{\extracolsep{\fill}}c@{\extracolsep{\fill}}c@{}}
  \centering
	  \includegraphics[width=1\unit]{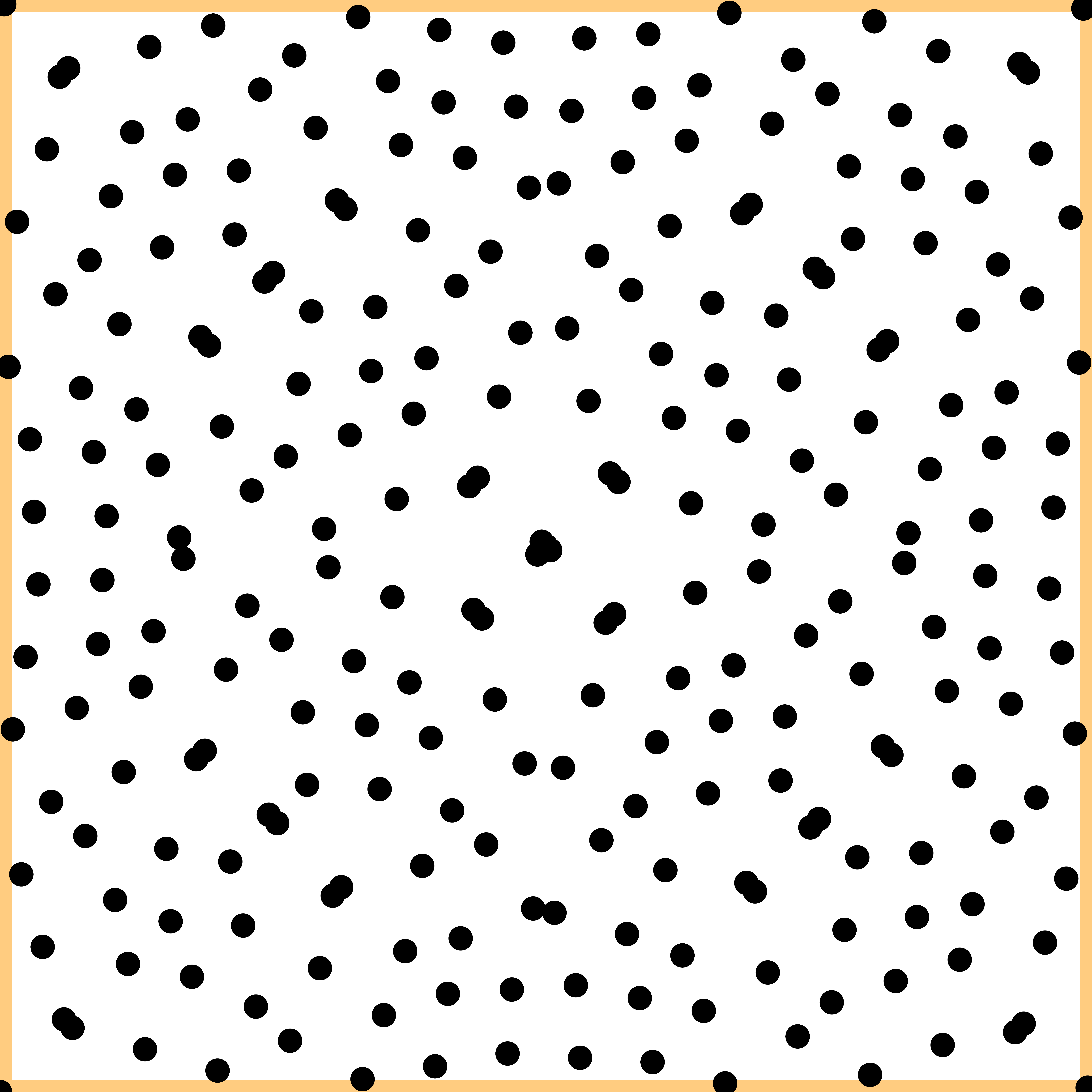}&%
	  \includegraphics[width=1\unit]{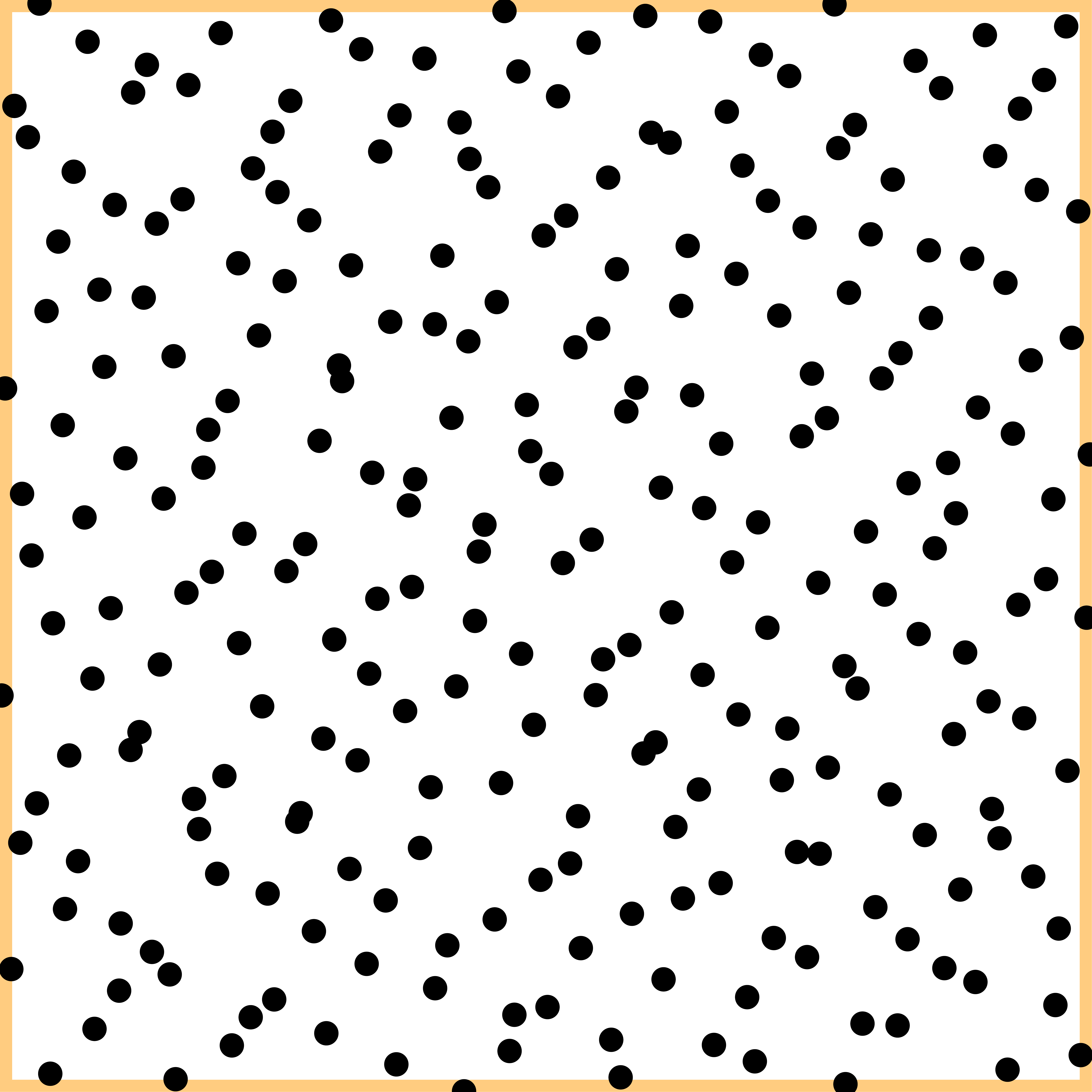}&%
	  \includegraphics[width=1\unit]{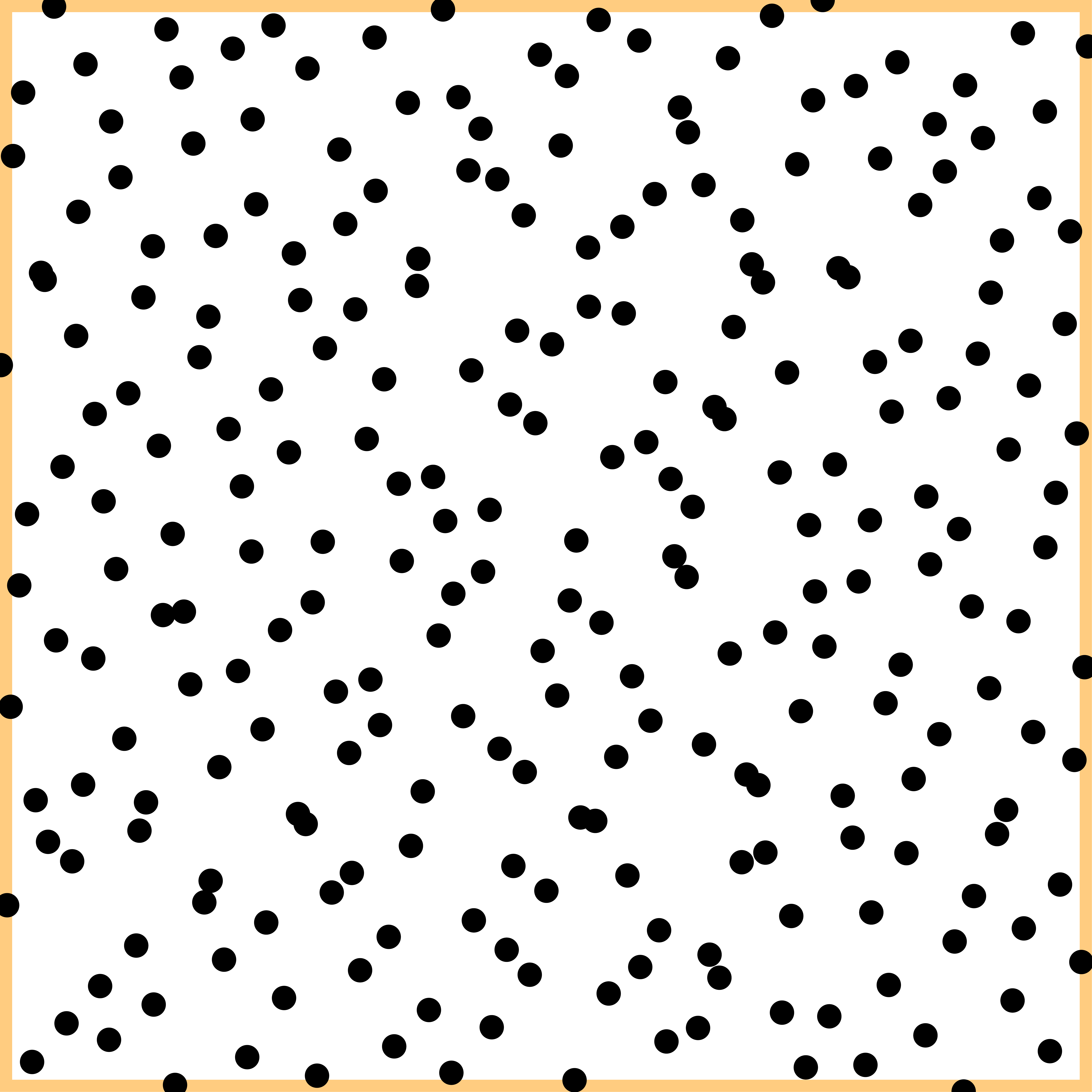}&%
	  \includegraphics[width=1\unit]{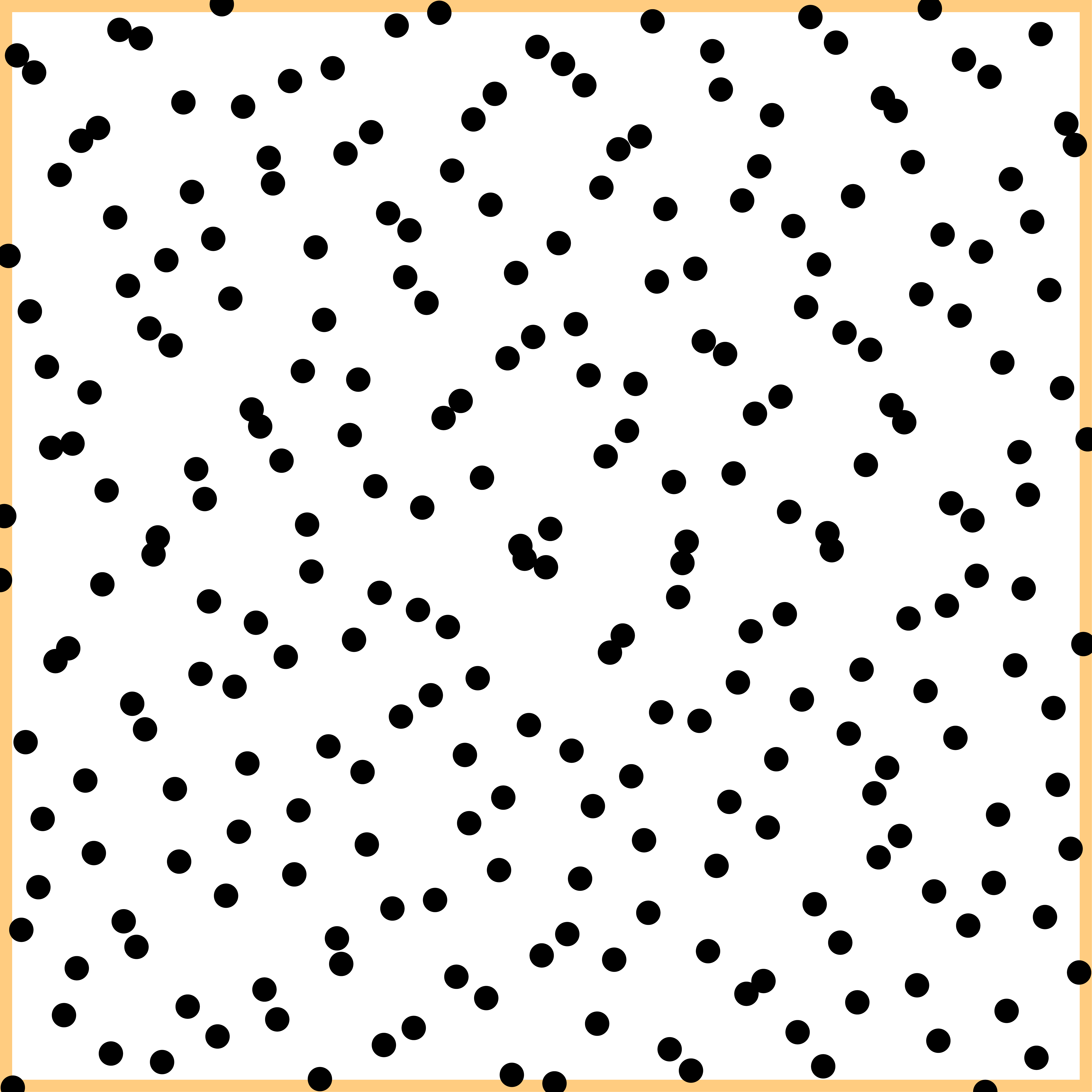}&%
	  \includegraphics[width=1\unit]{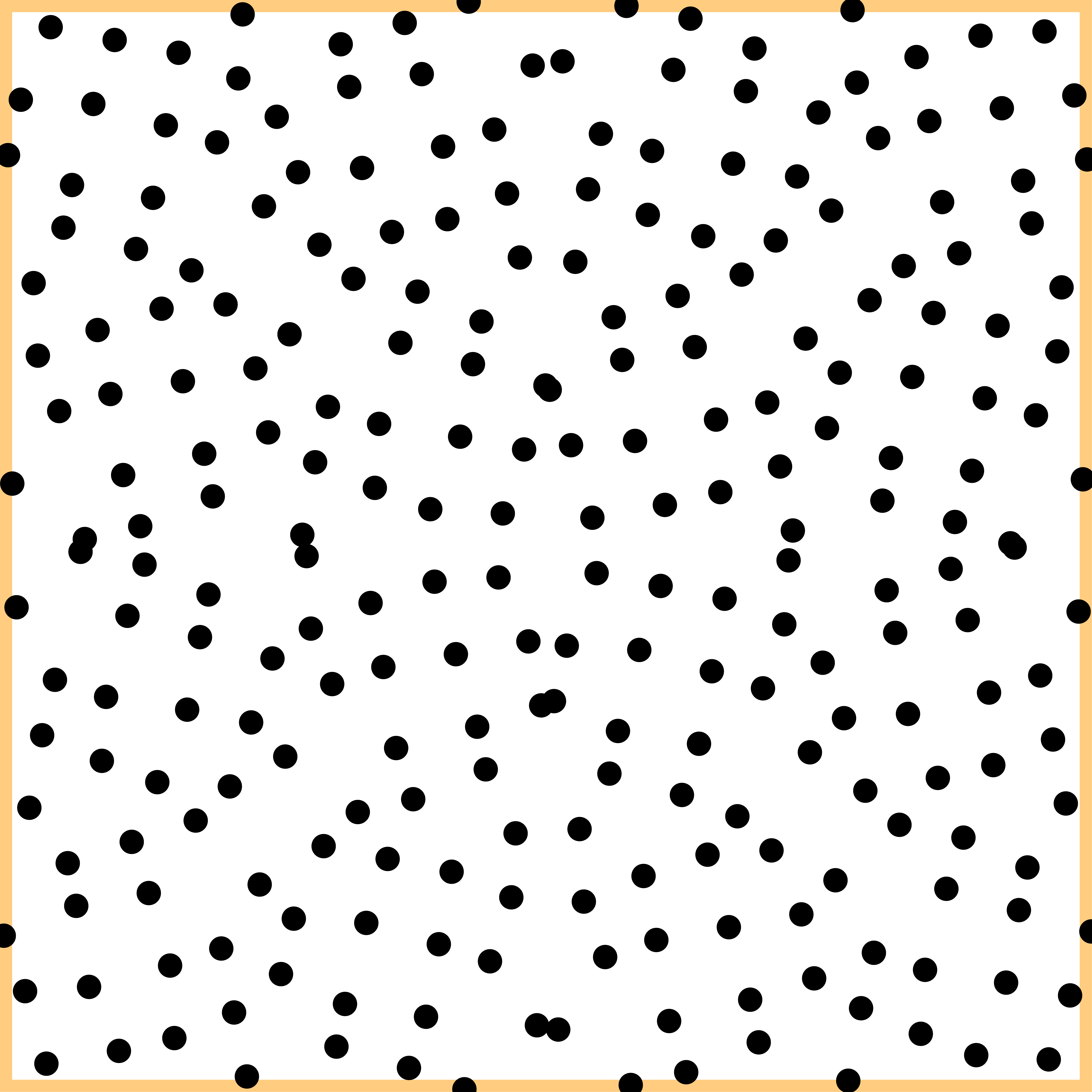}&%
	  \includegraphics[width=1\unit]{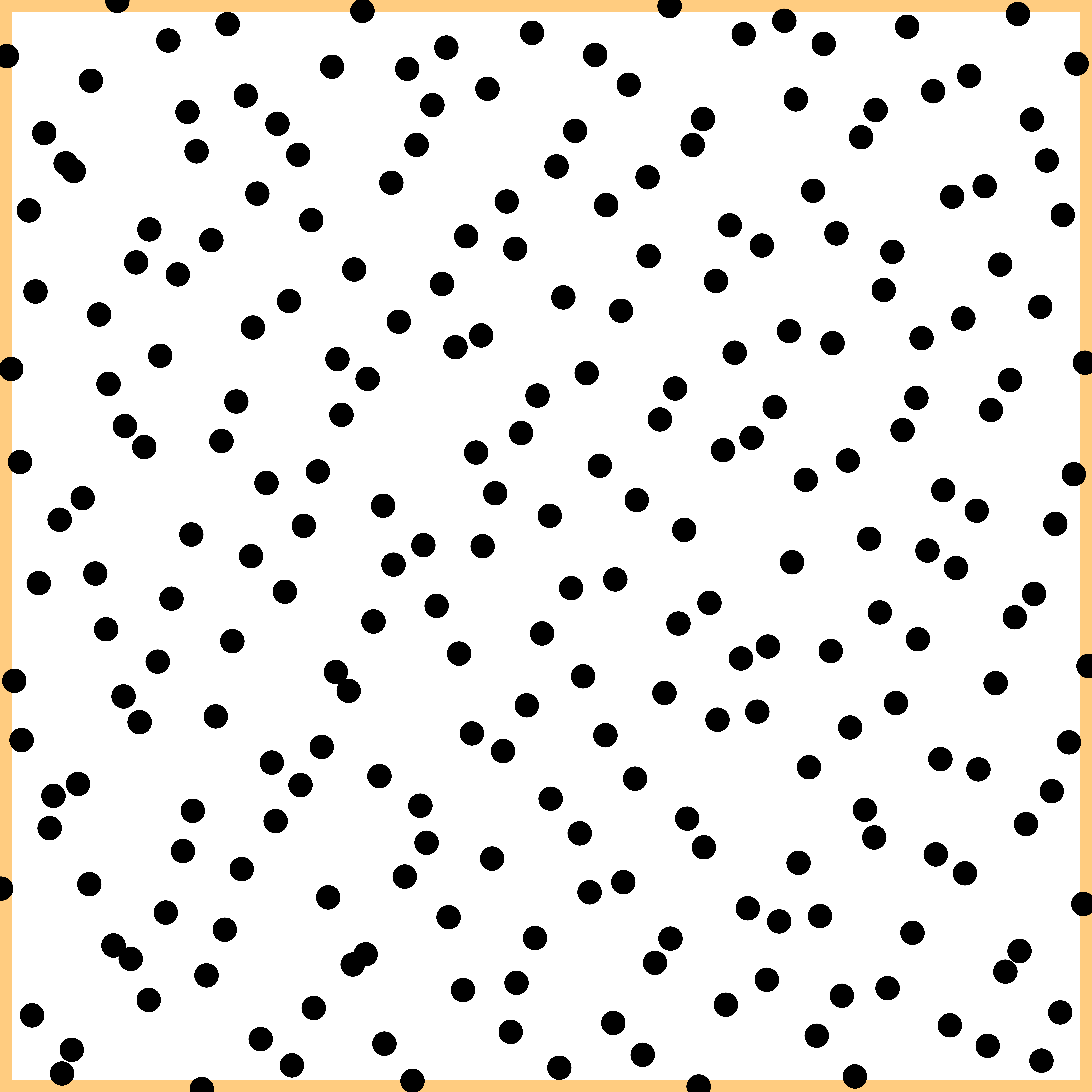}&%
	  \includegraphics[width=1\unit]{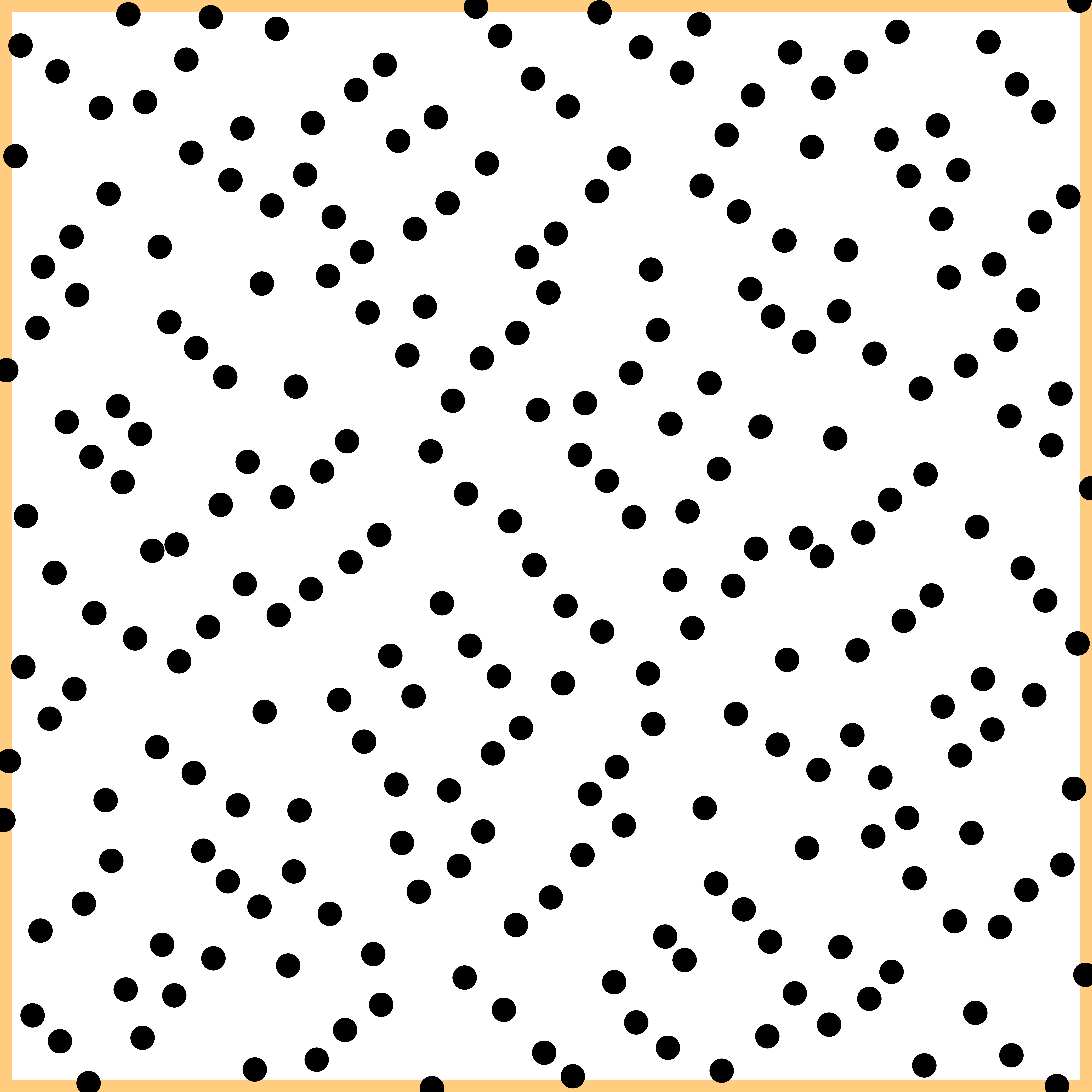}&%
	  \includegraphics[width=1\unit]{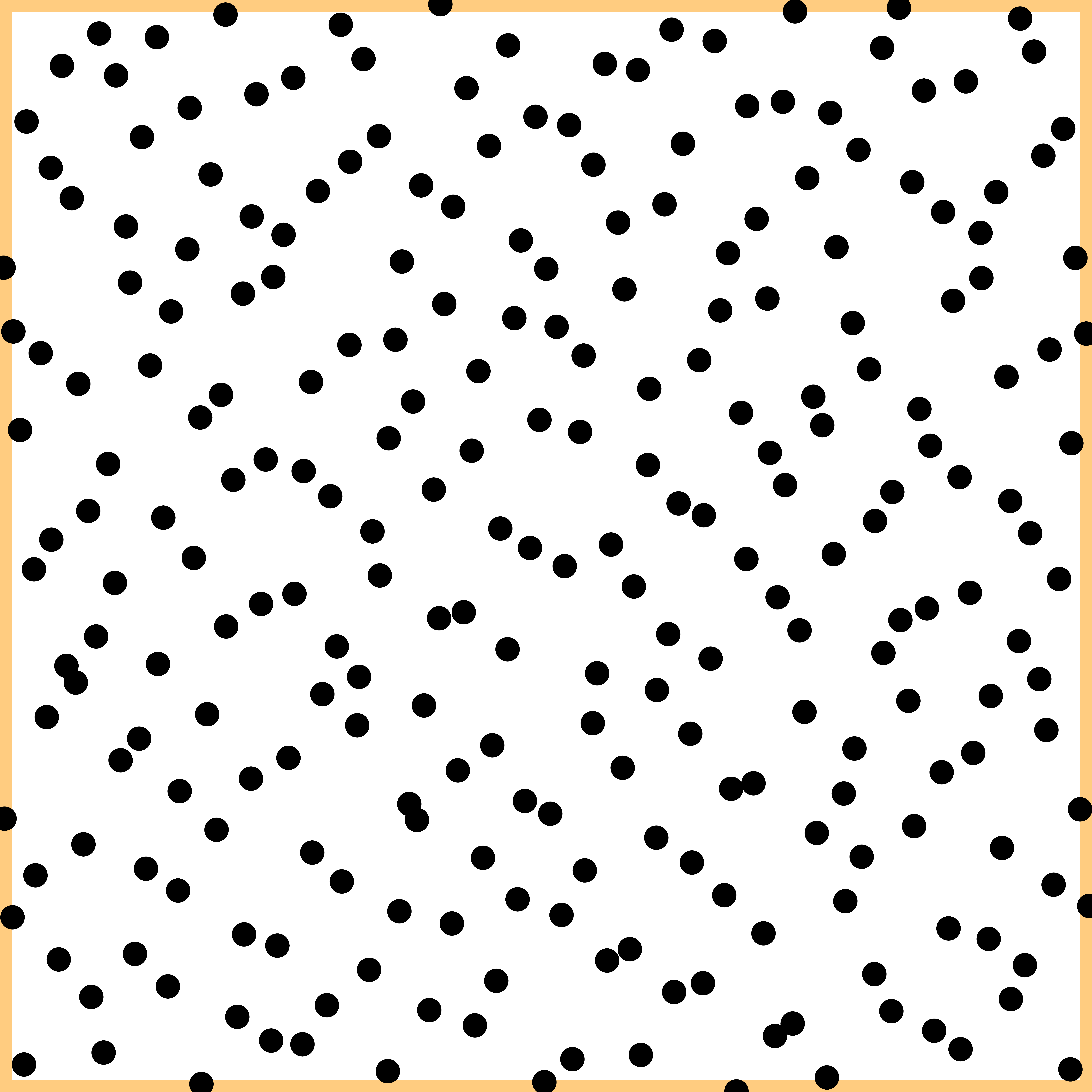}&%
	  \includegraphics[width=1\unit]{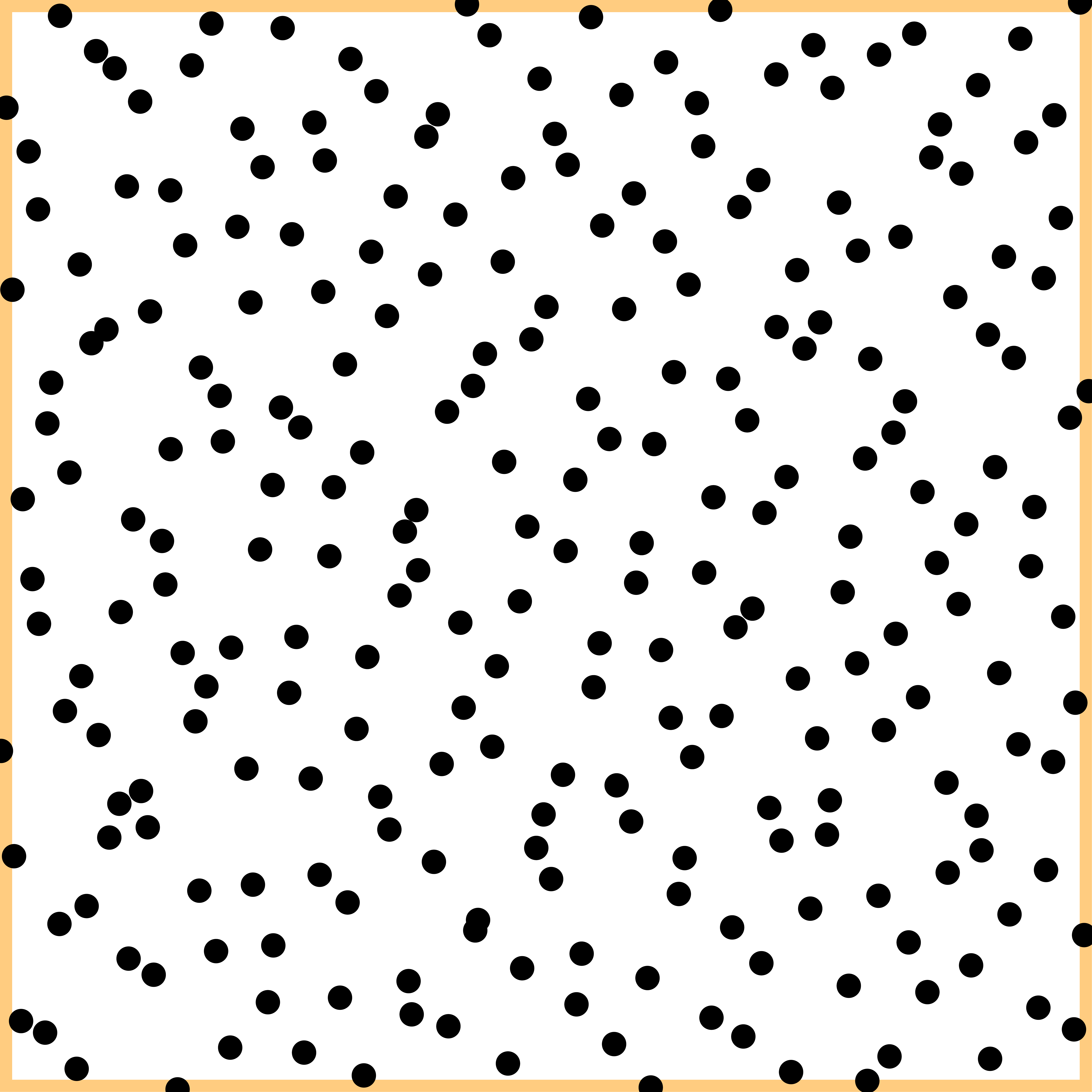}\\
	  \includegraphics[width=1\unit]{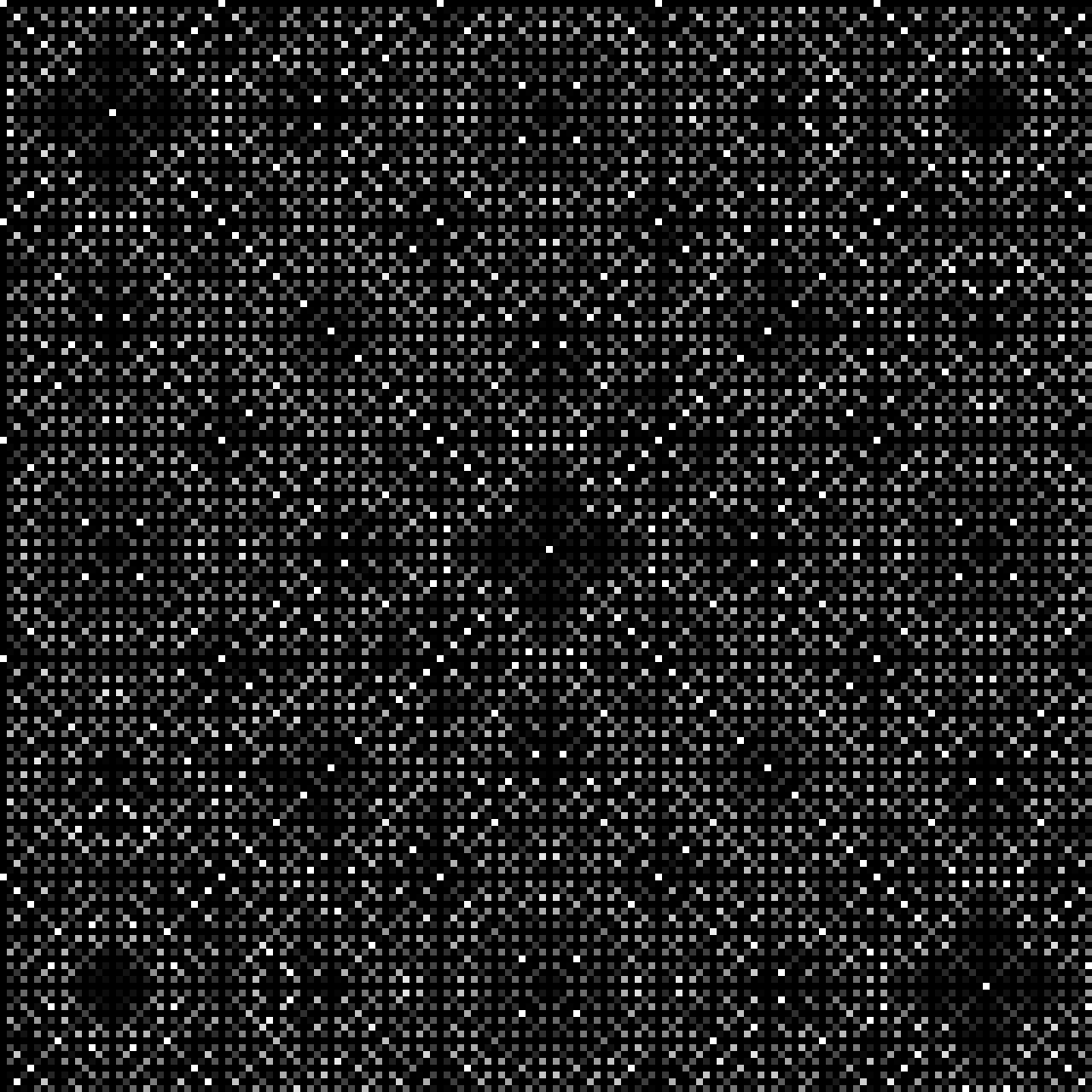}&%
	  \includegraphics[width=1\unit]{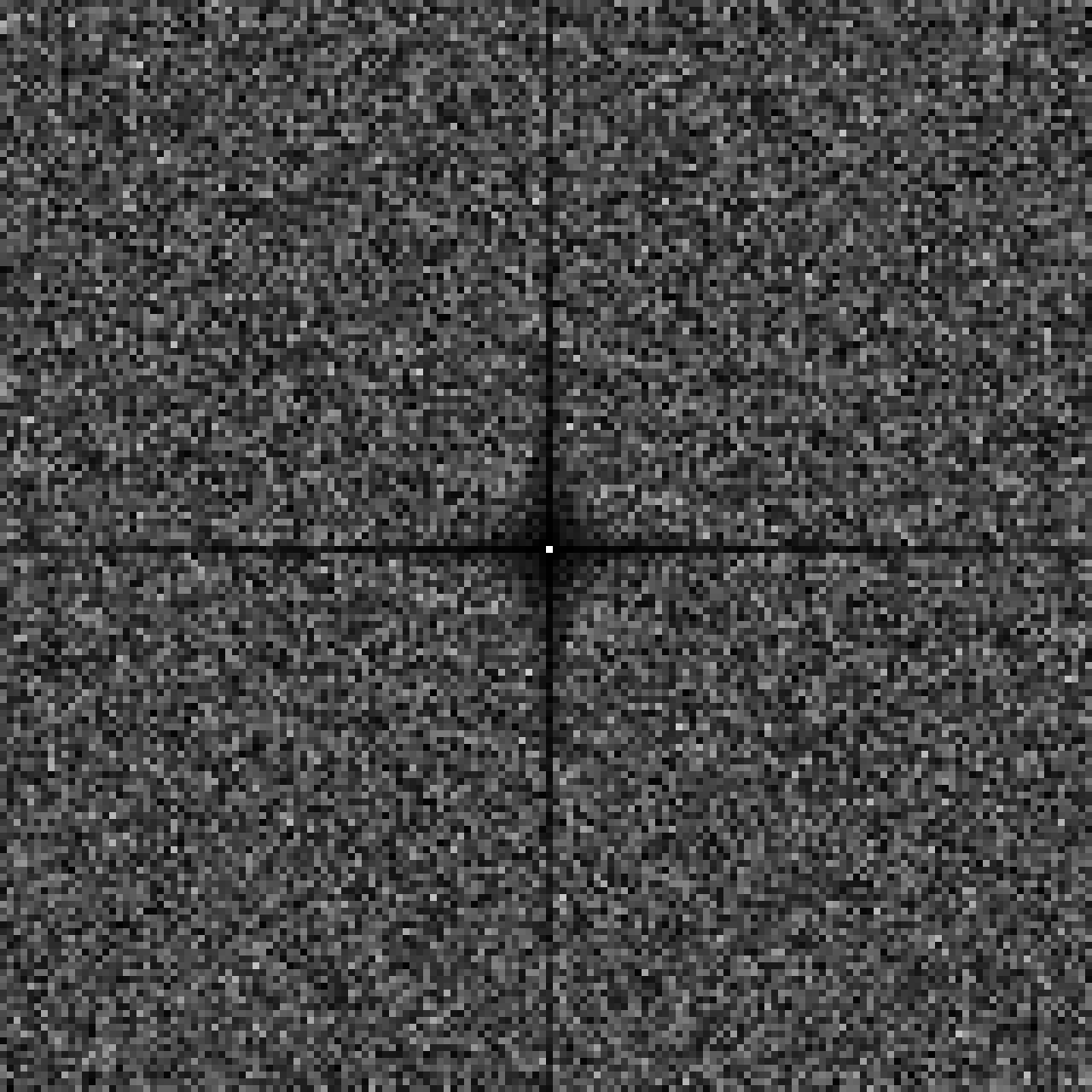}&%
	  \includegraphics[width=1\unit]{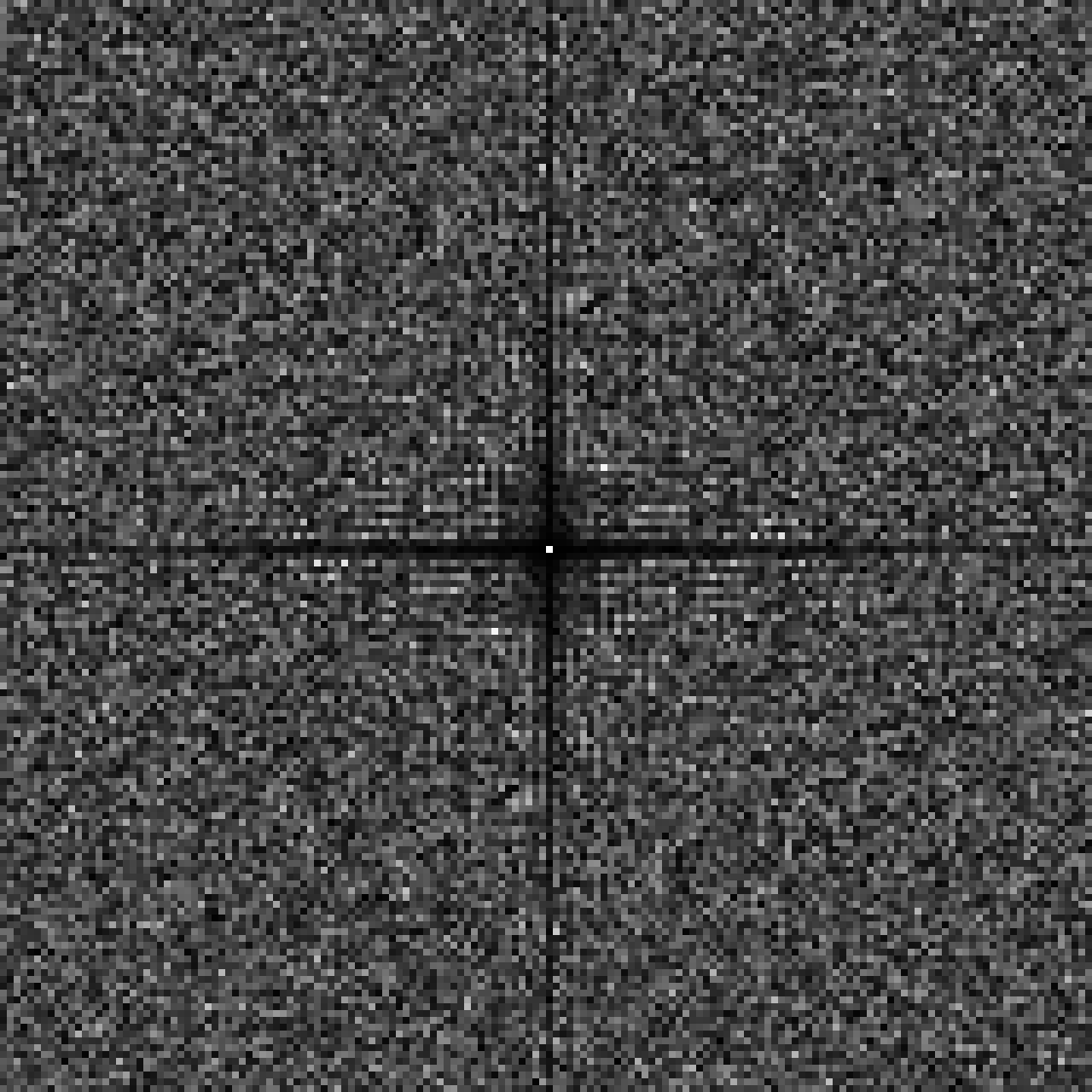}&%
	  \includegraphics[width=1\unit]{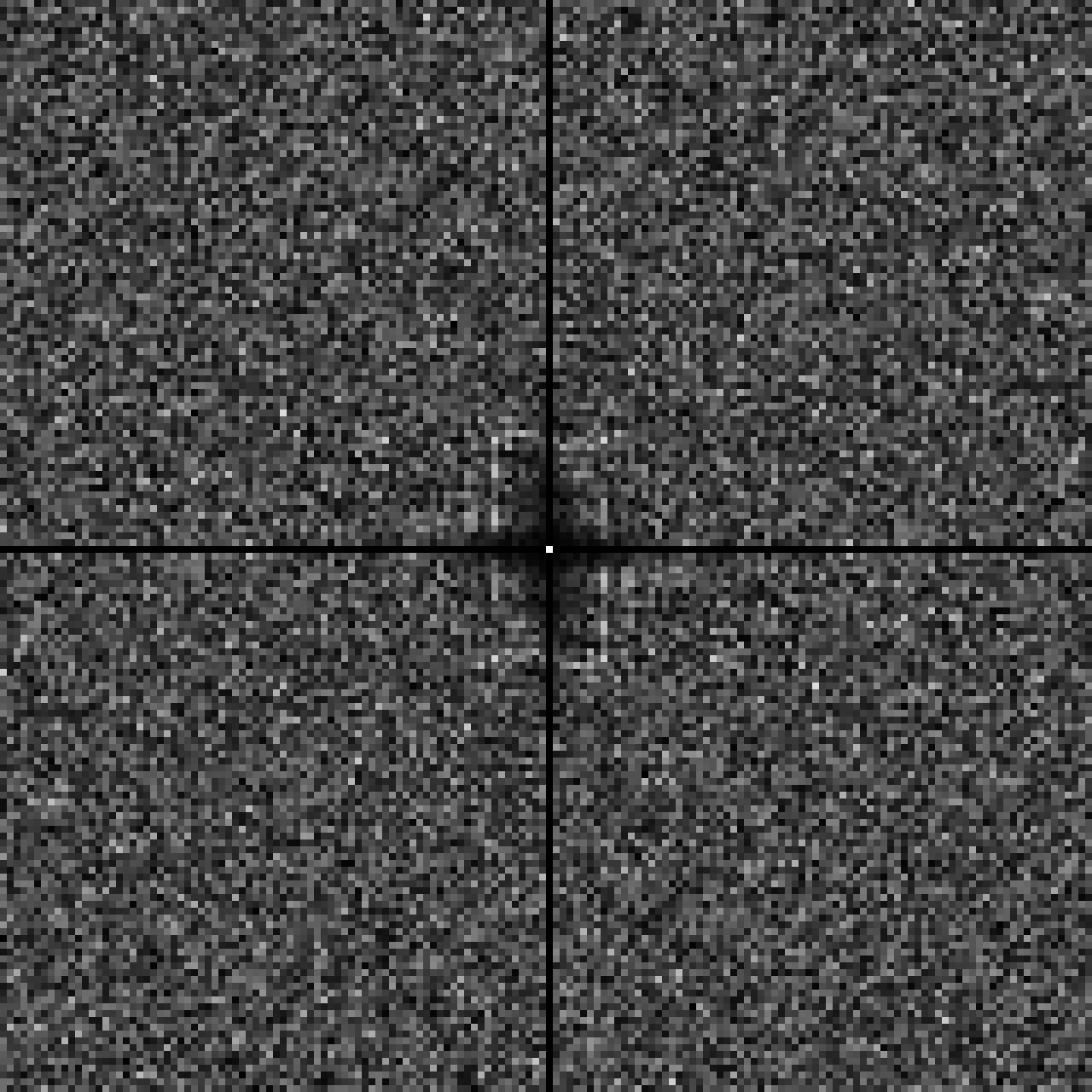}&%
	  \includegraphics[width=1\unit]{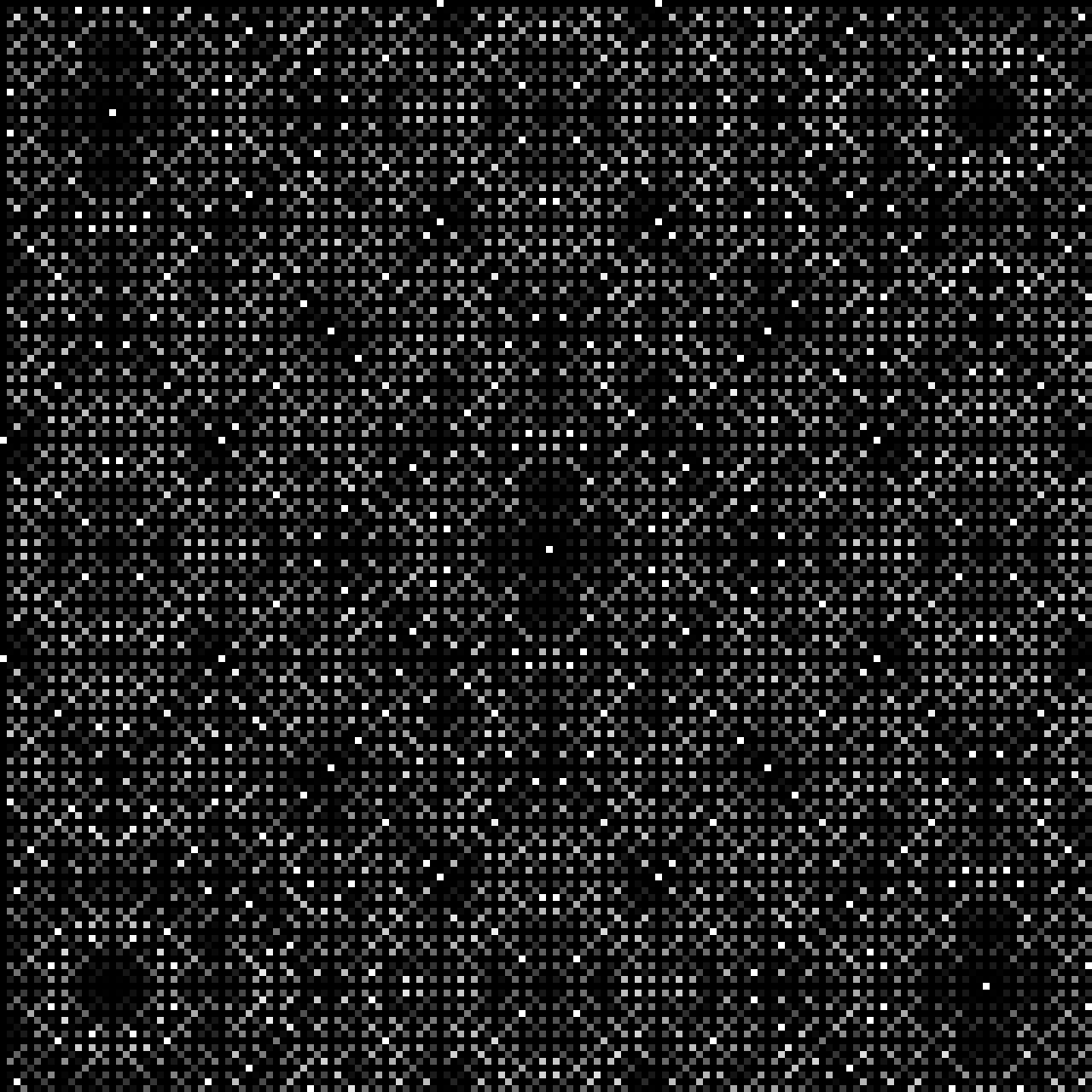}&%
	  \includegraphics[width=1\unit]{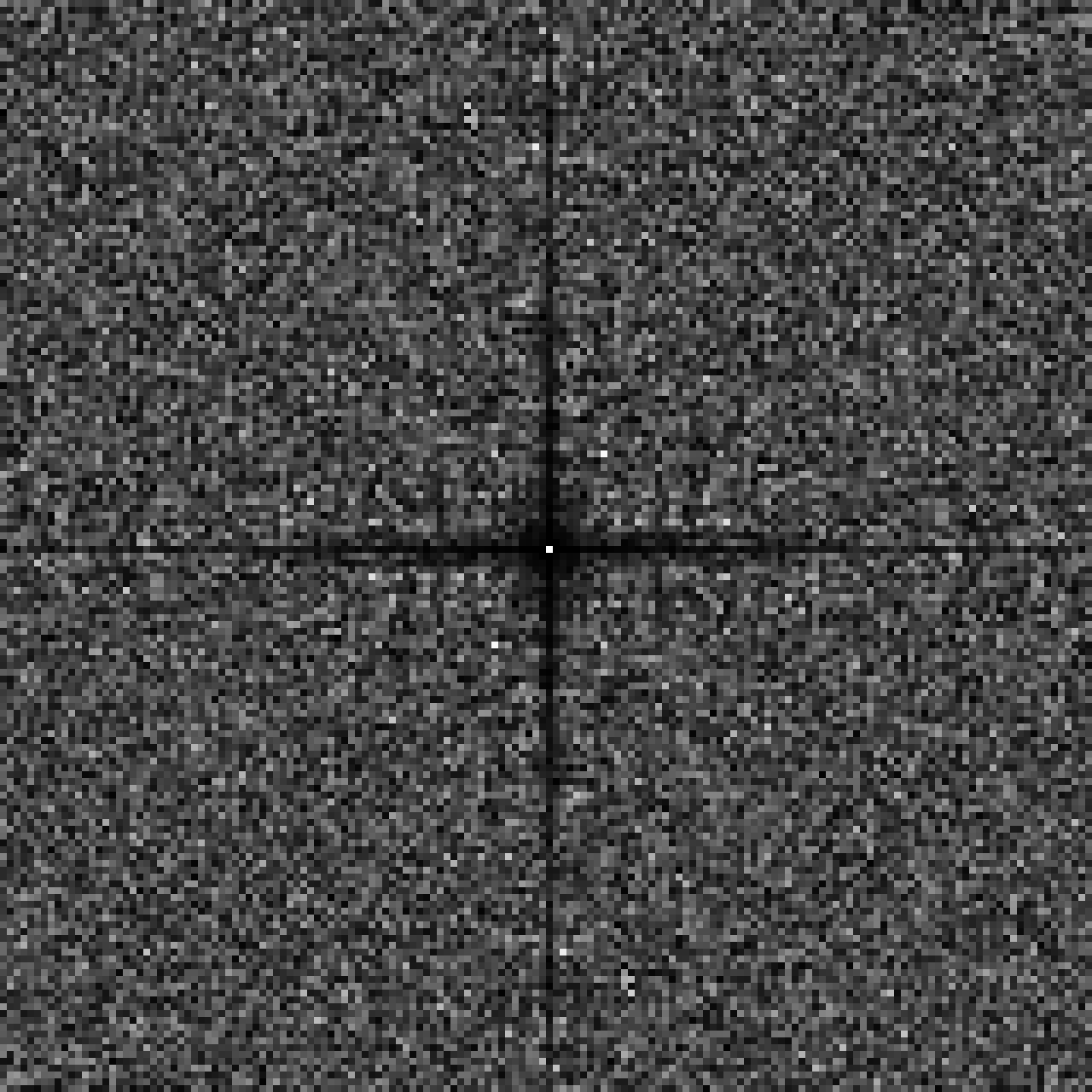}&%
	  \includegraphics[width=1\unit]{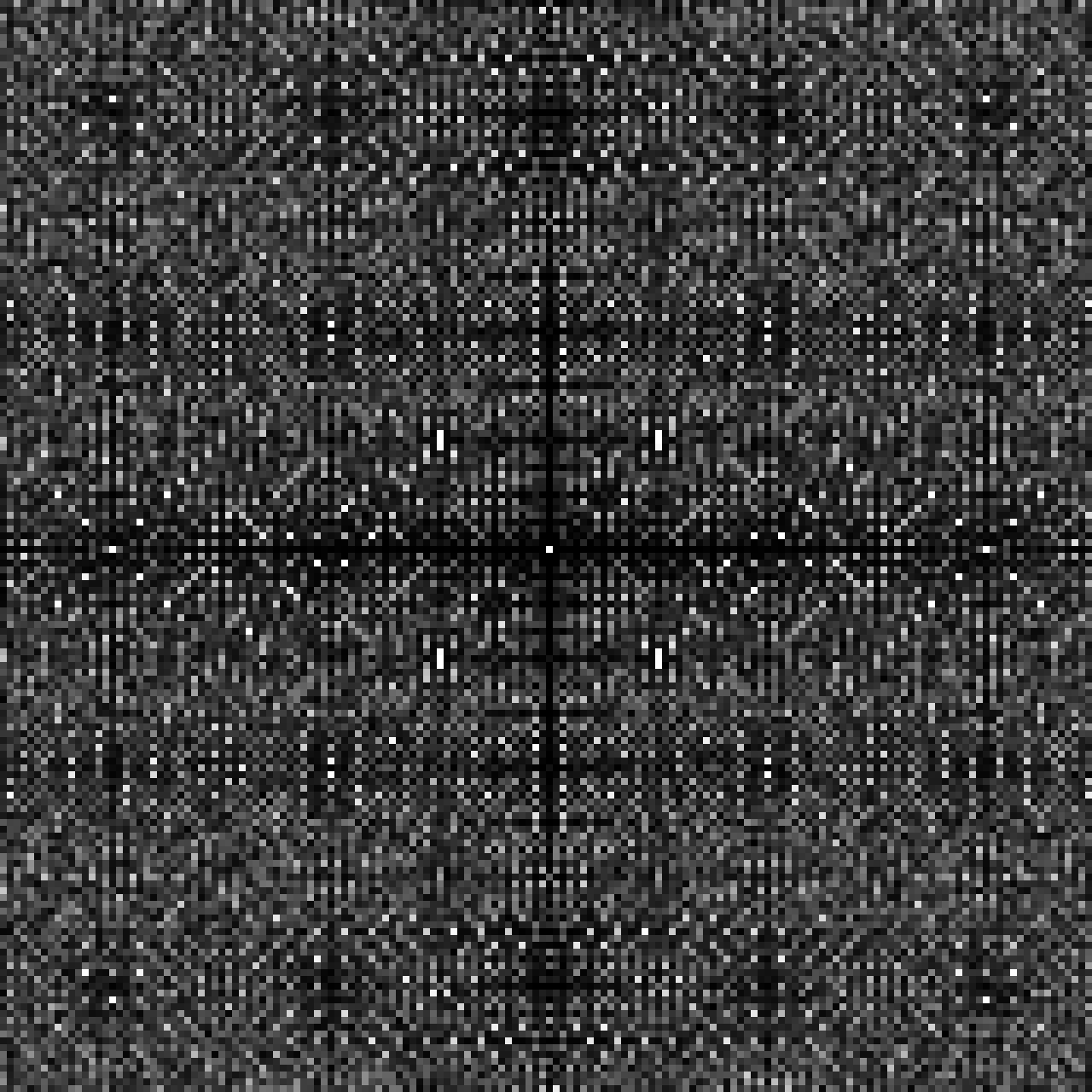}&%
	  \includegraphics[width=1\unit]{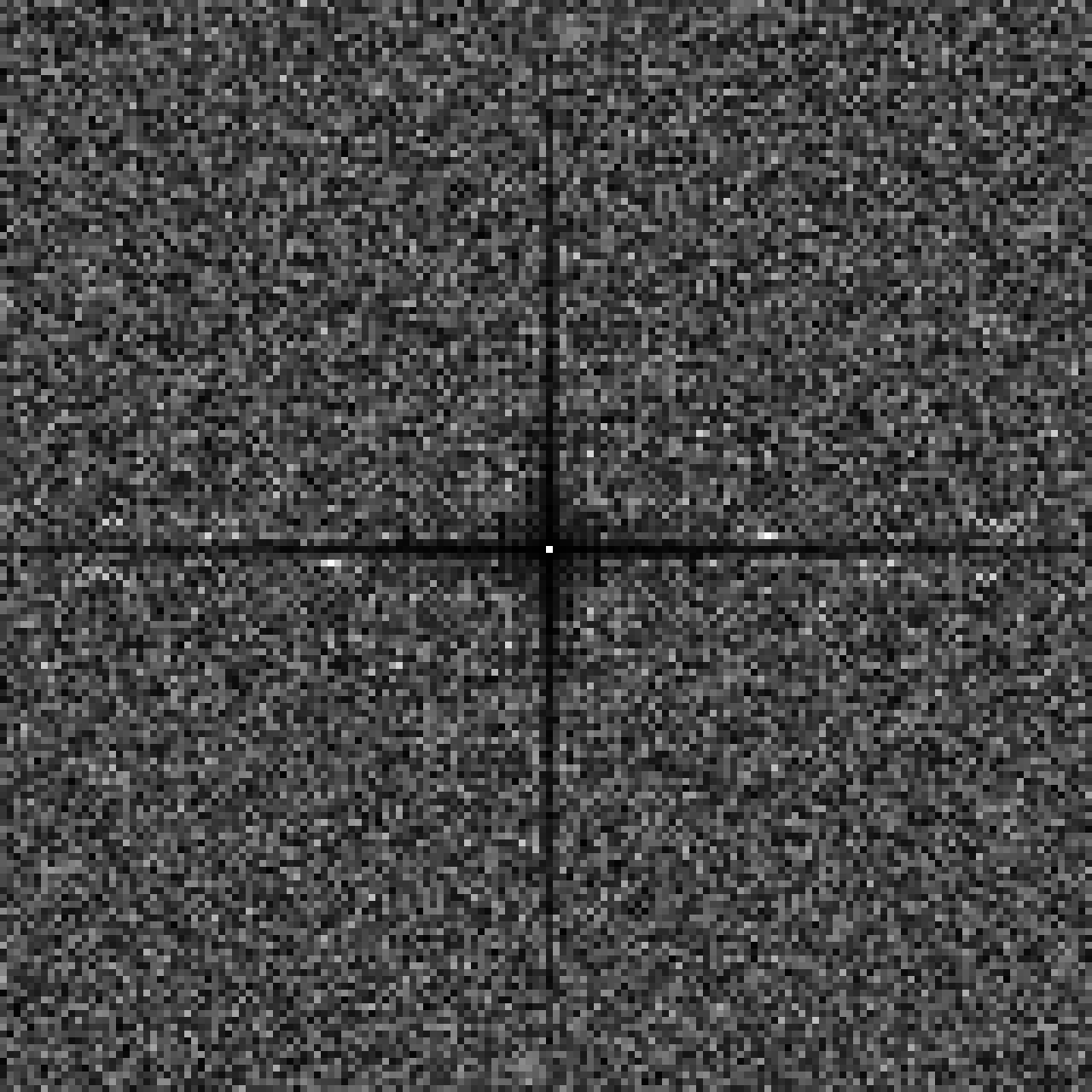}&%
	  \includegraphics[width=1\unit]{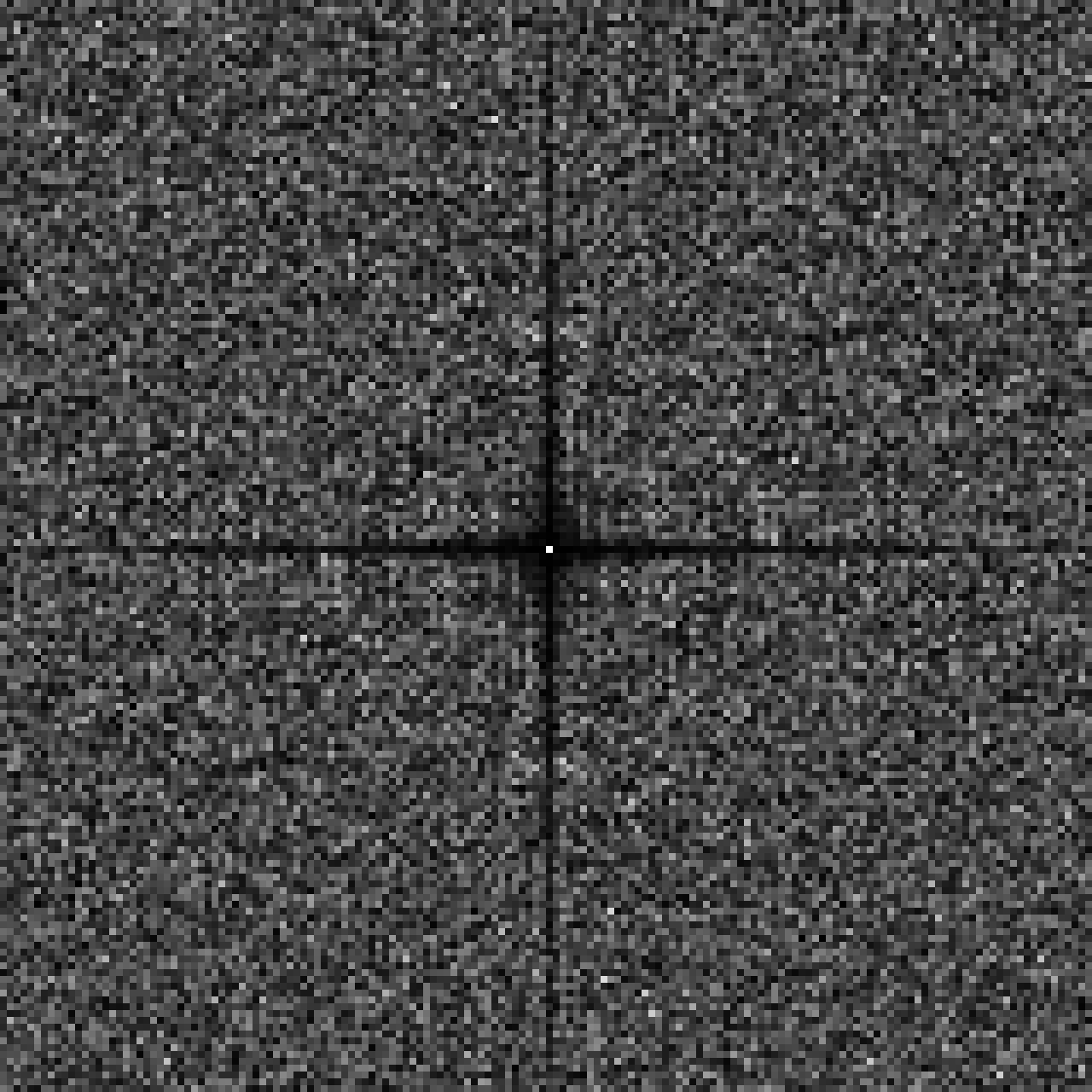}\\
	  \includegraphics[width=1\unit]{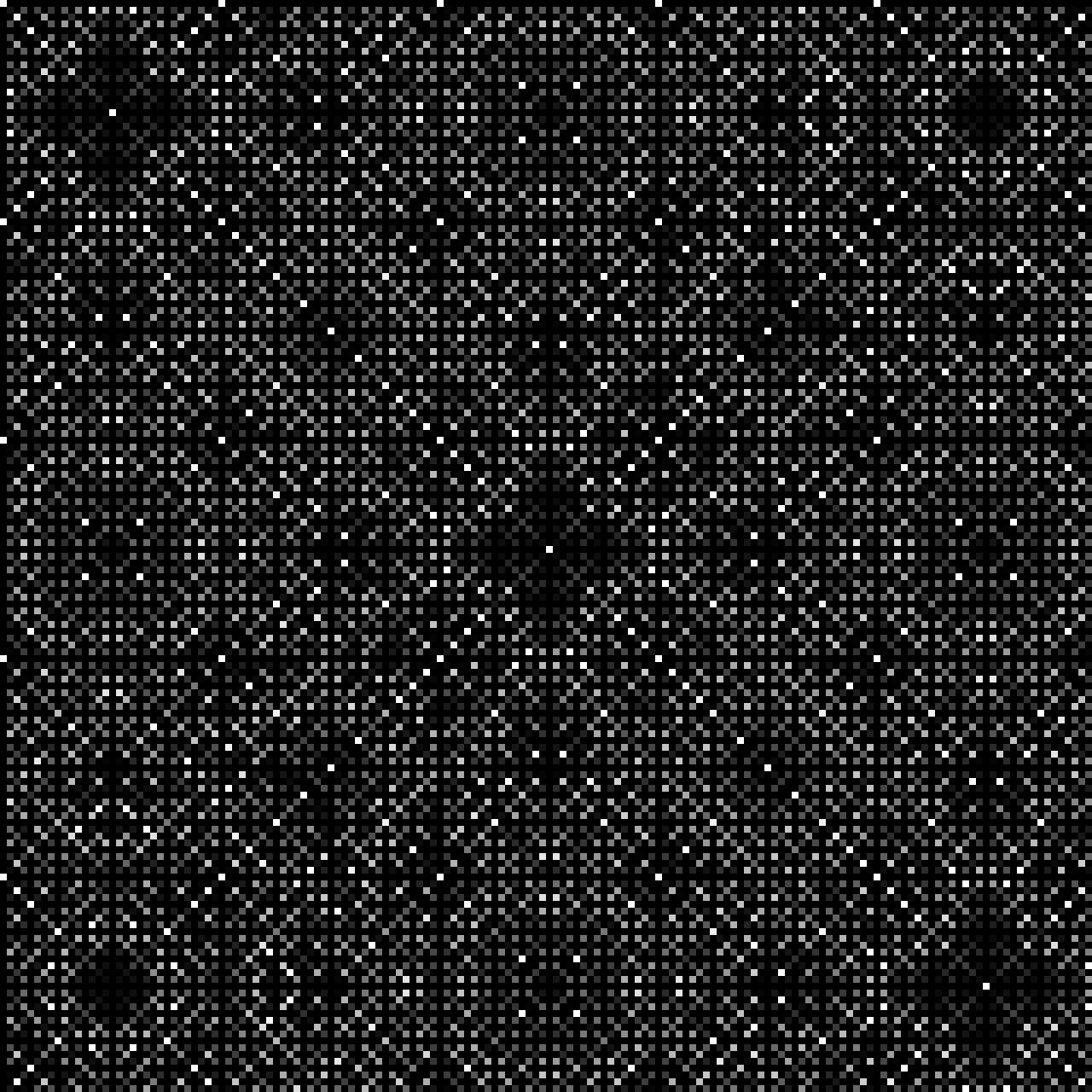}&%
	  \includegraphics[width=1\unit]{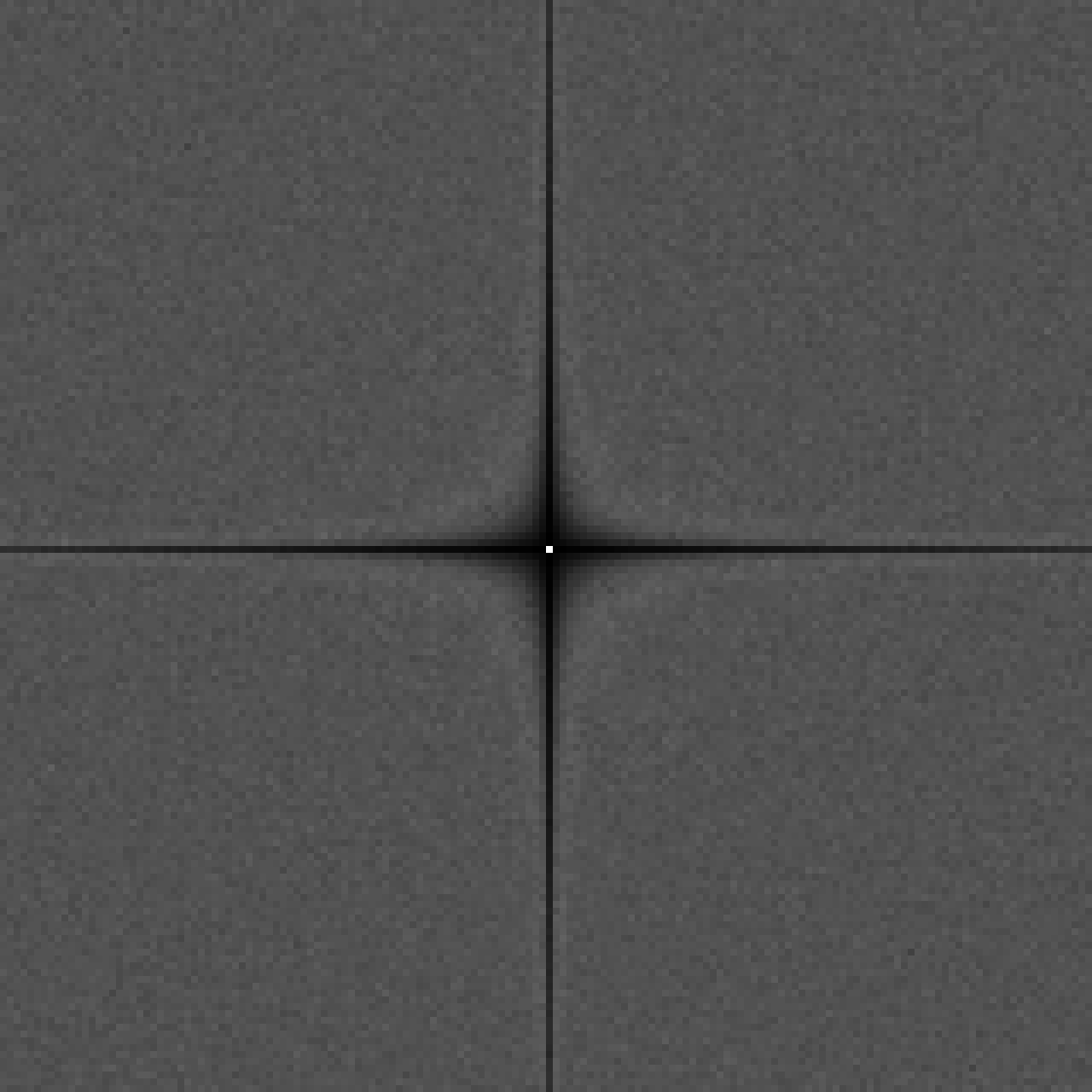}&%
	  \includegraphics[width=1\unit]{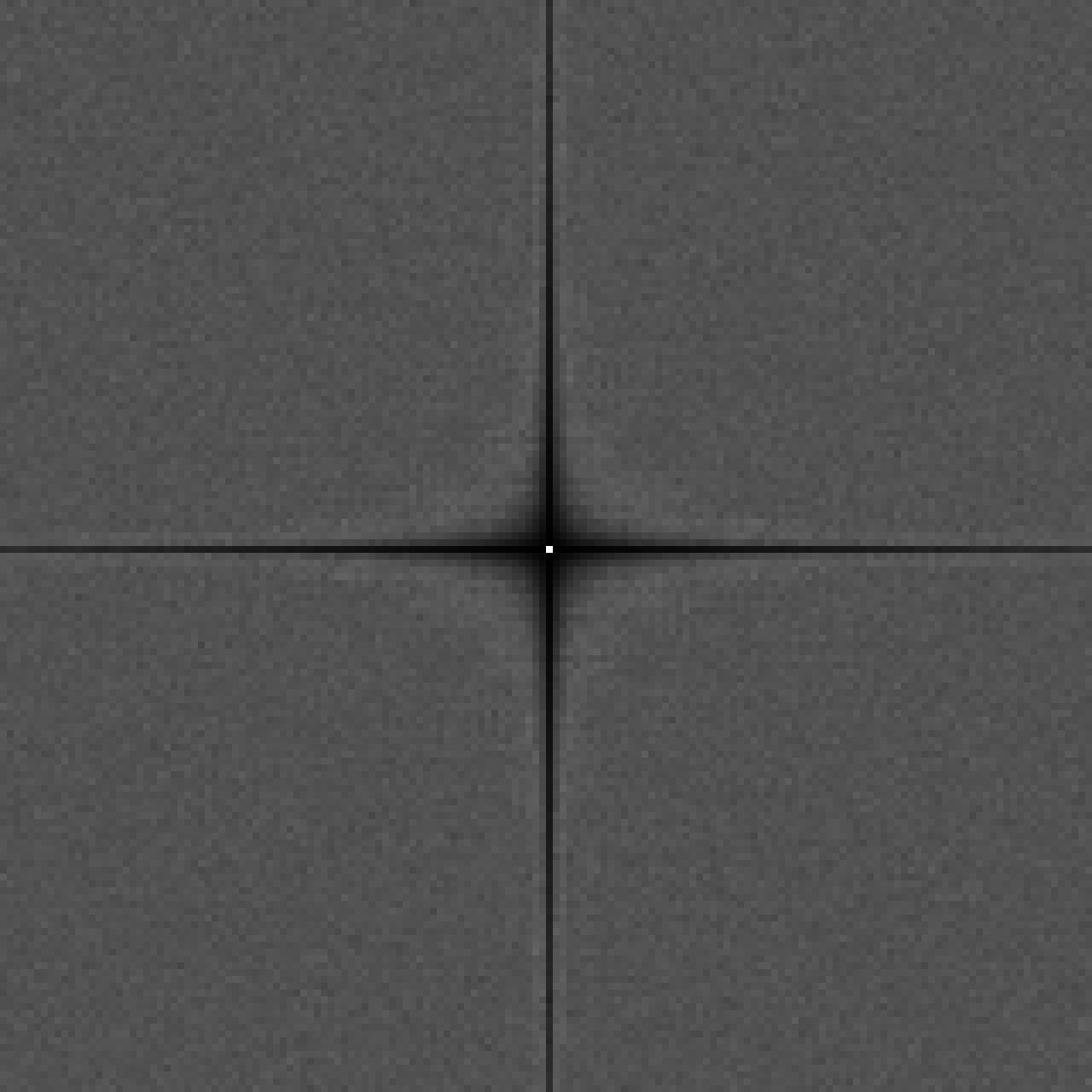}&%
	  \includegraphics[width=1\unit]{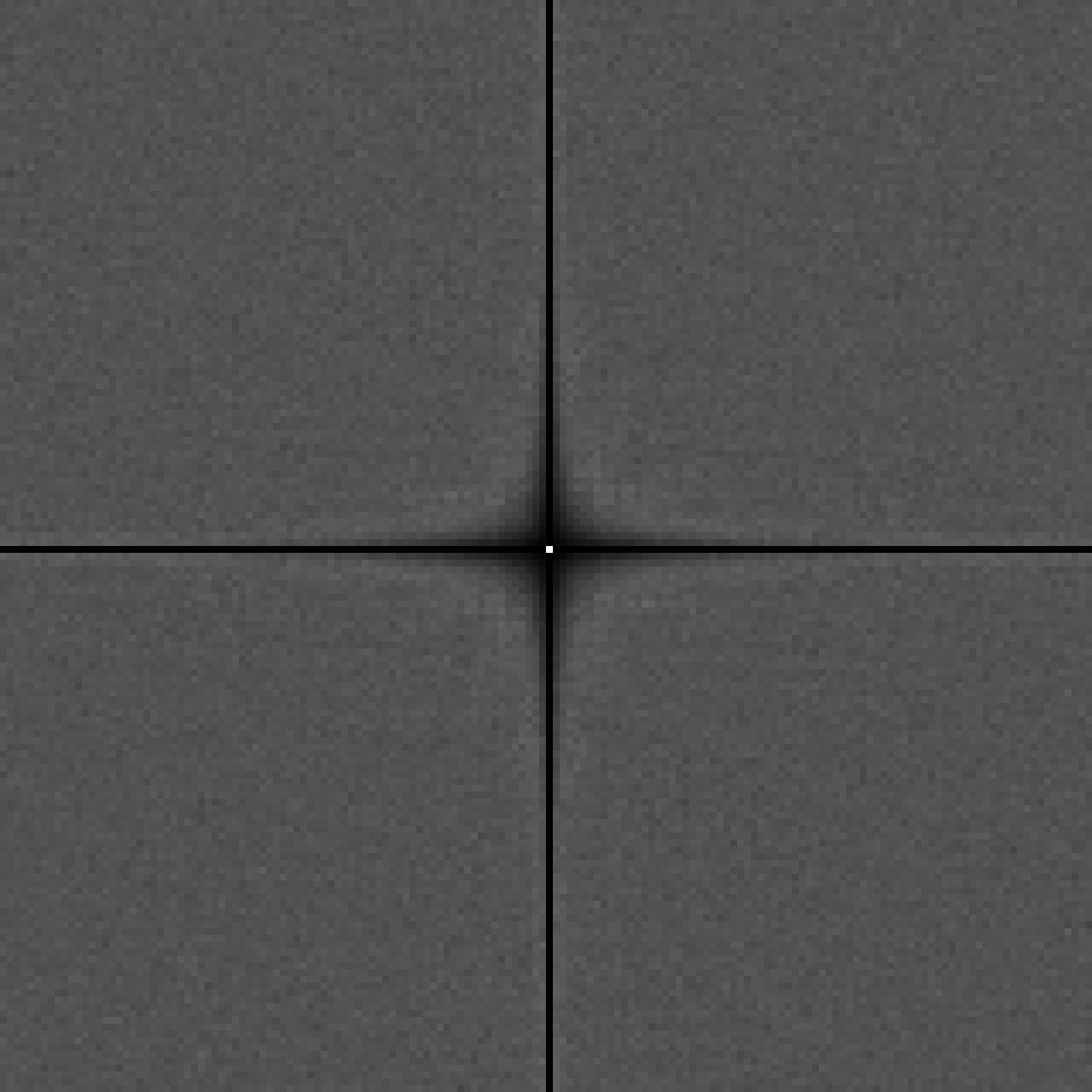}&%
	  \includegraphics[width=1\unit]{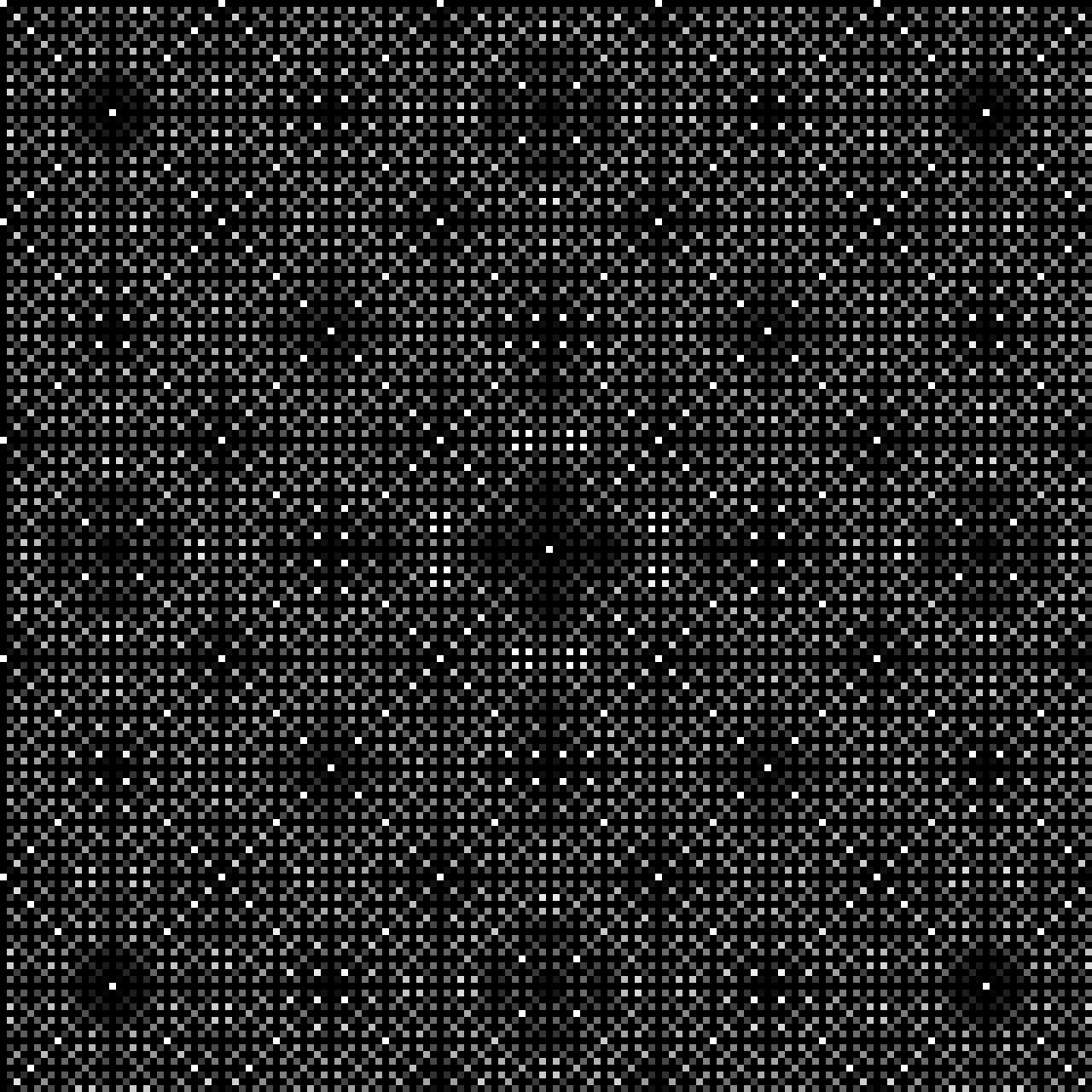}&%
	  \includegraphics[width=1\unit]{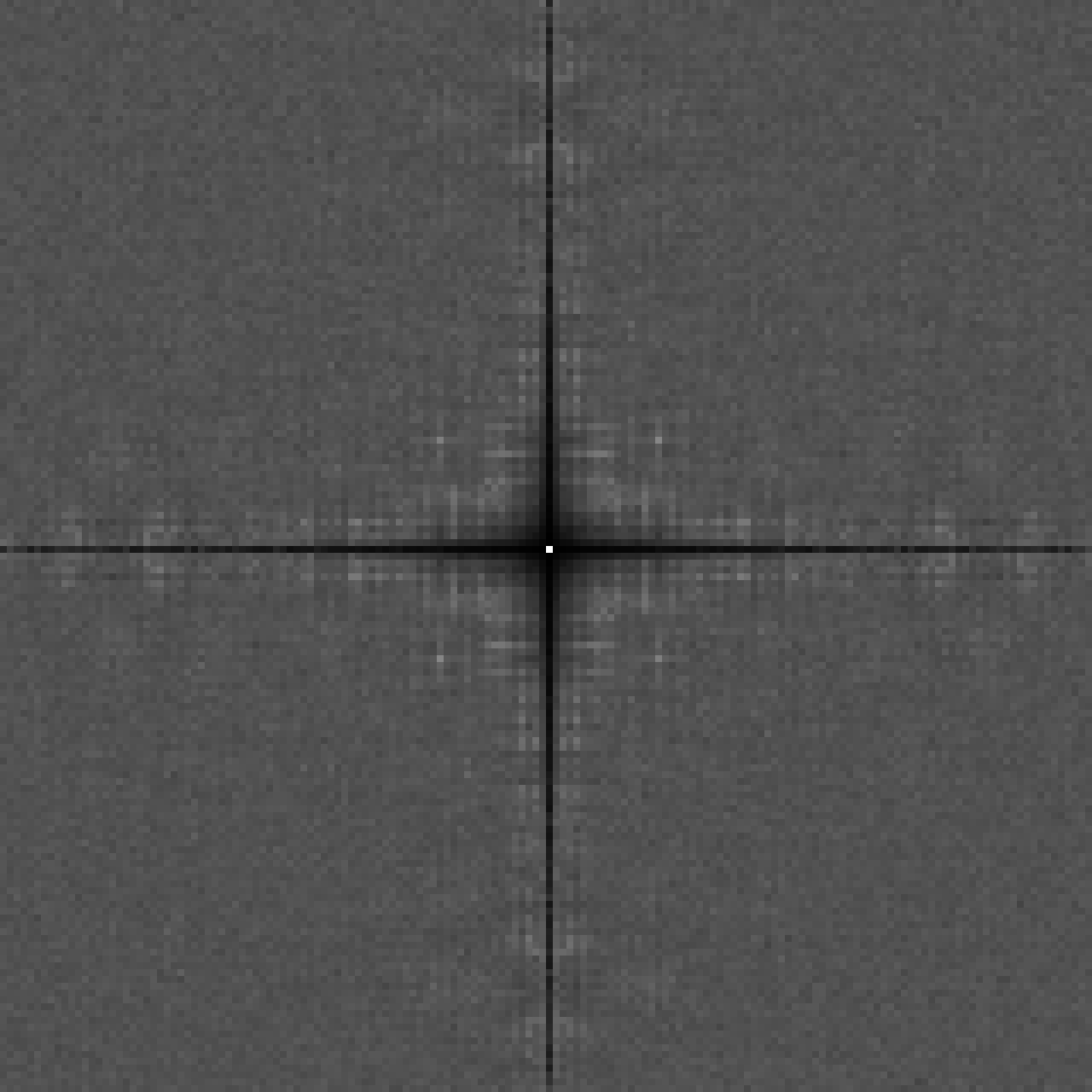}&%
	  \includegraphics[width=1\unit]{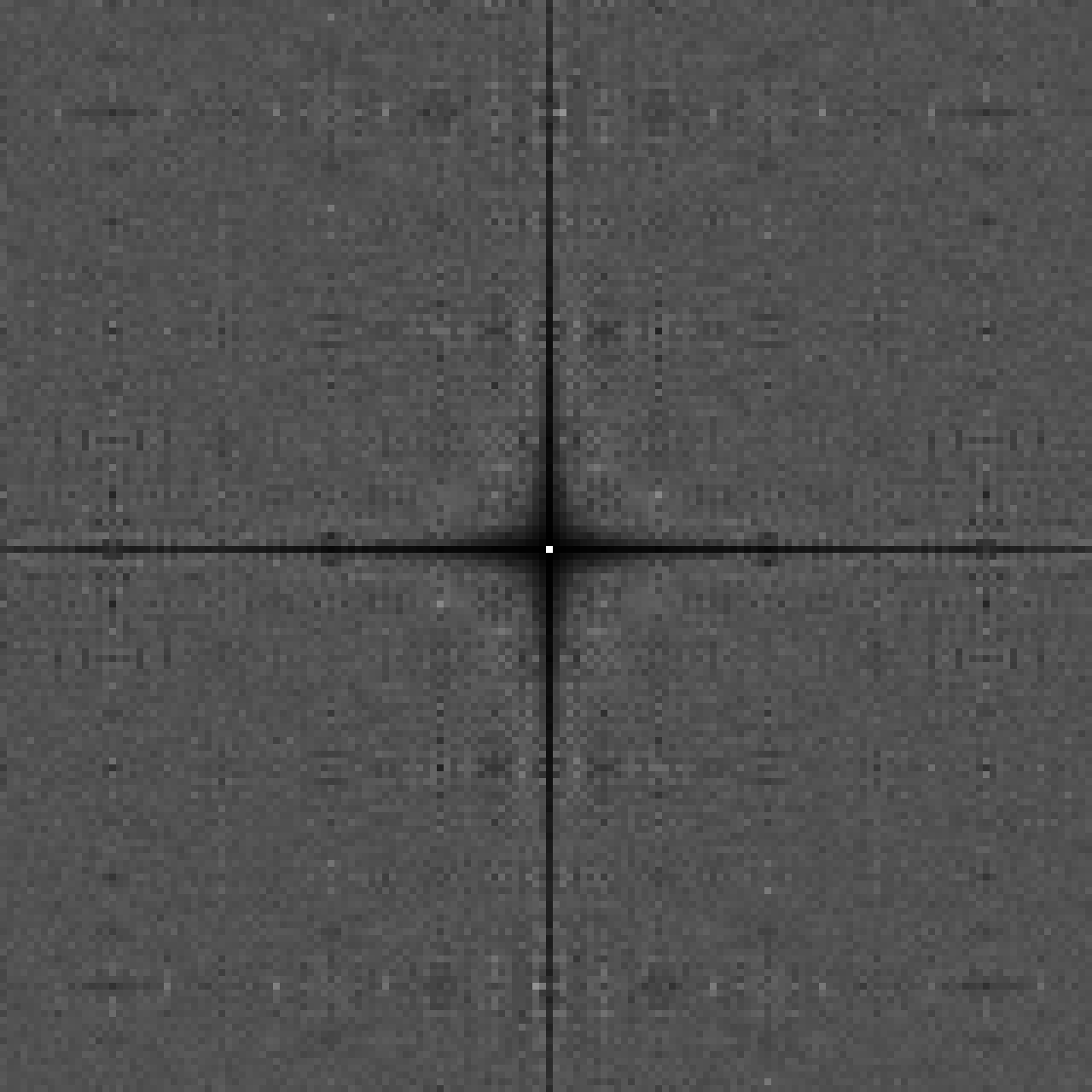}&%
	  \includegraphics[width=1\unit]{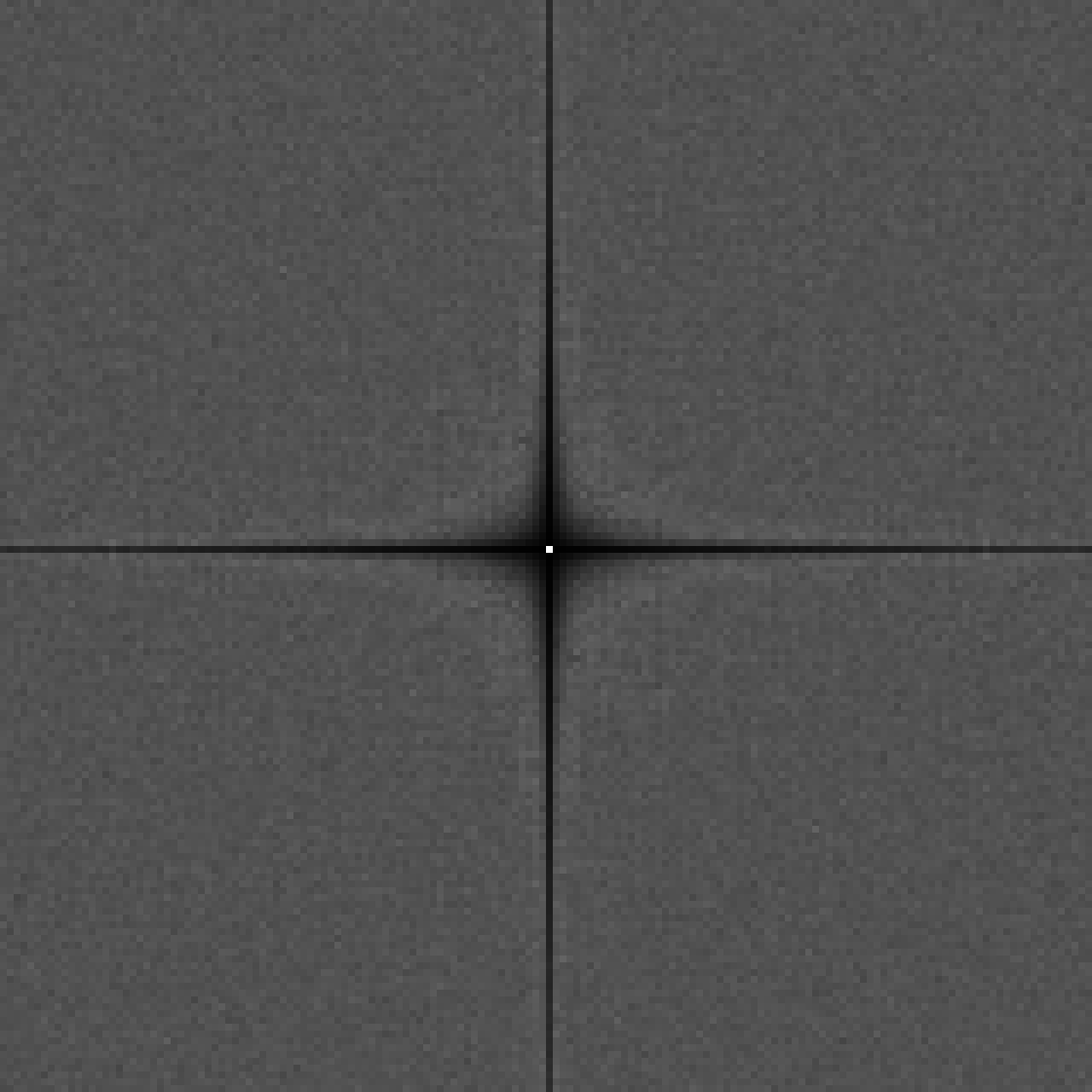}&%
	  \includegraphics[width=1\unit]{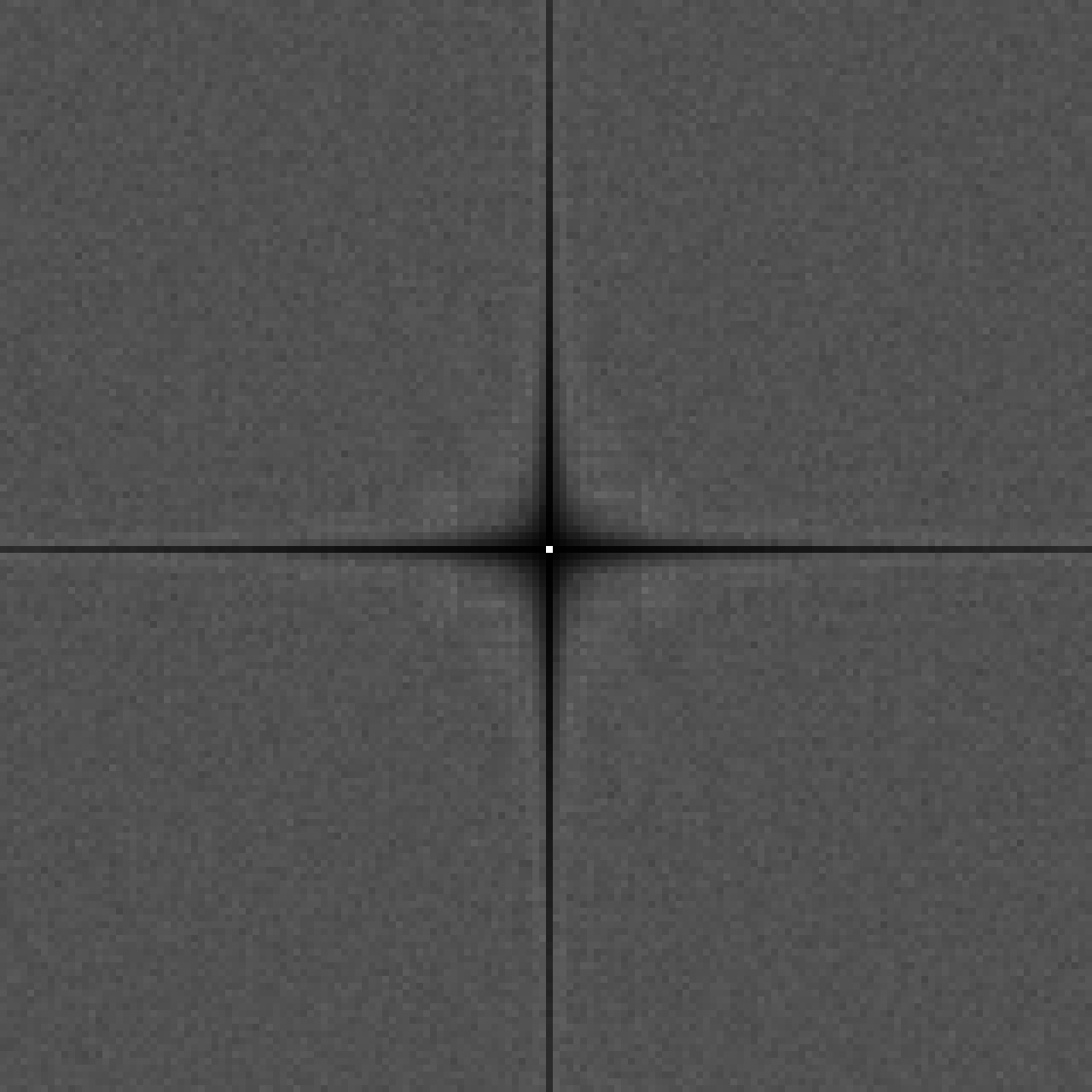}\\
	  \includegraphics[width=1\unit]{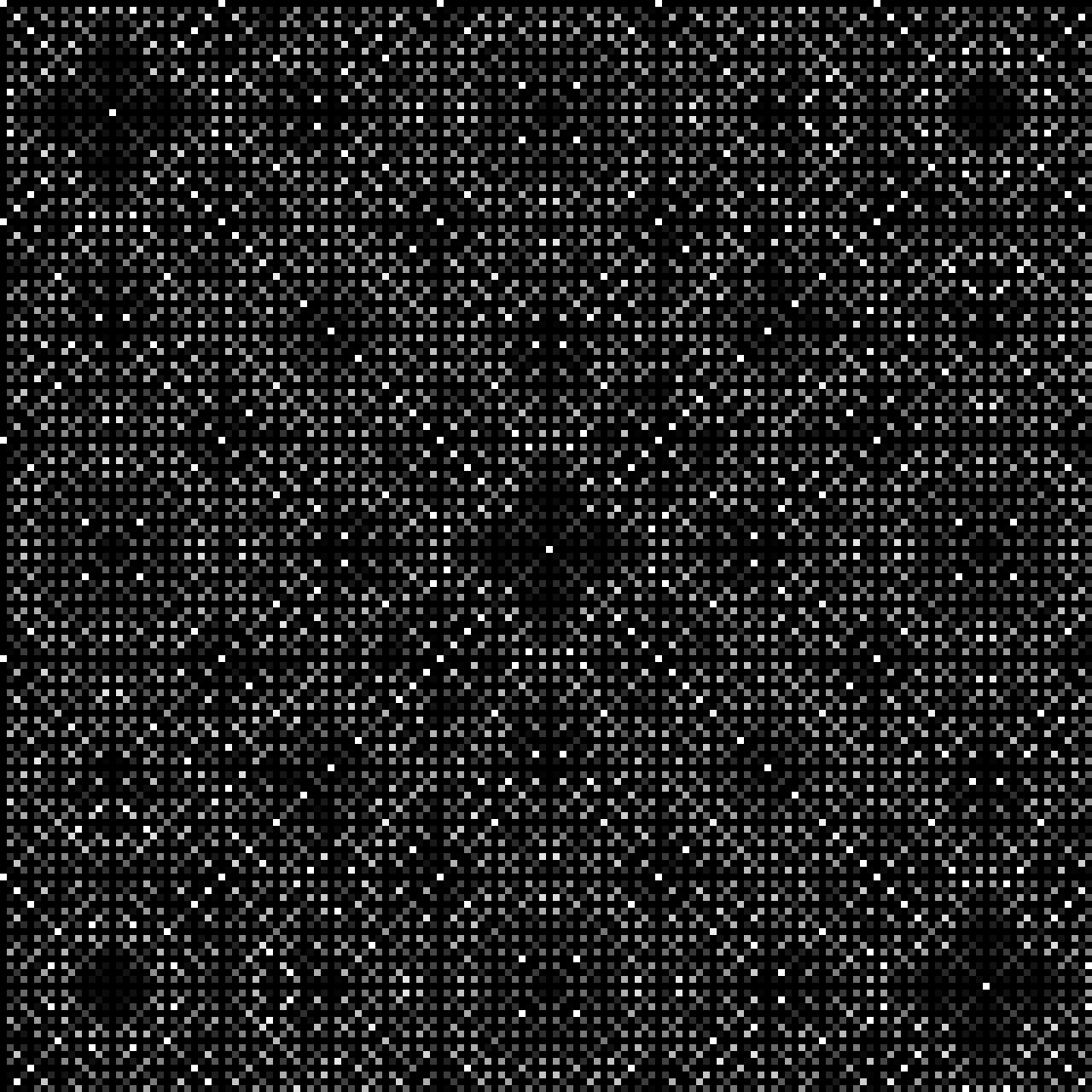}&%
	  \includegraphics[width=1\unit]{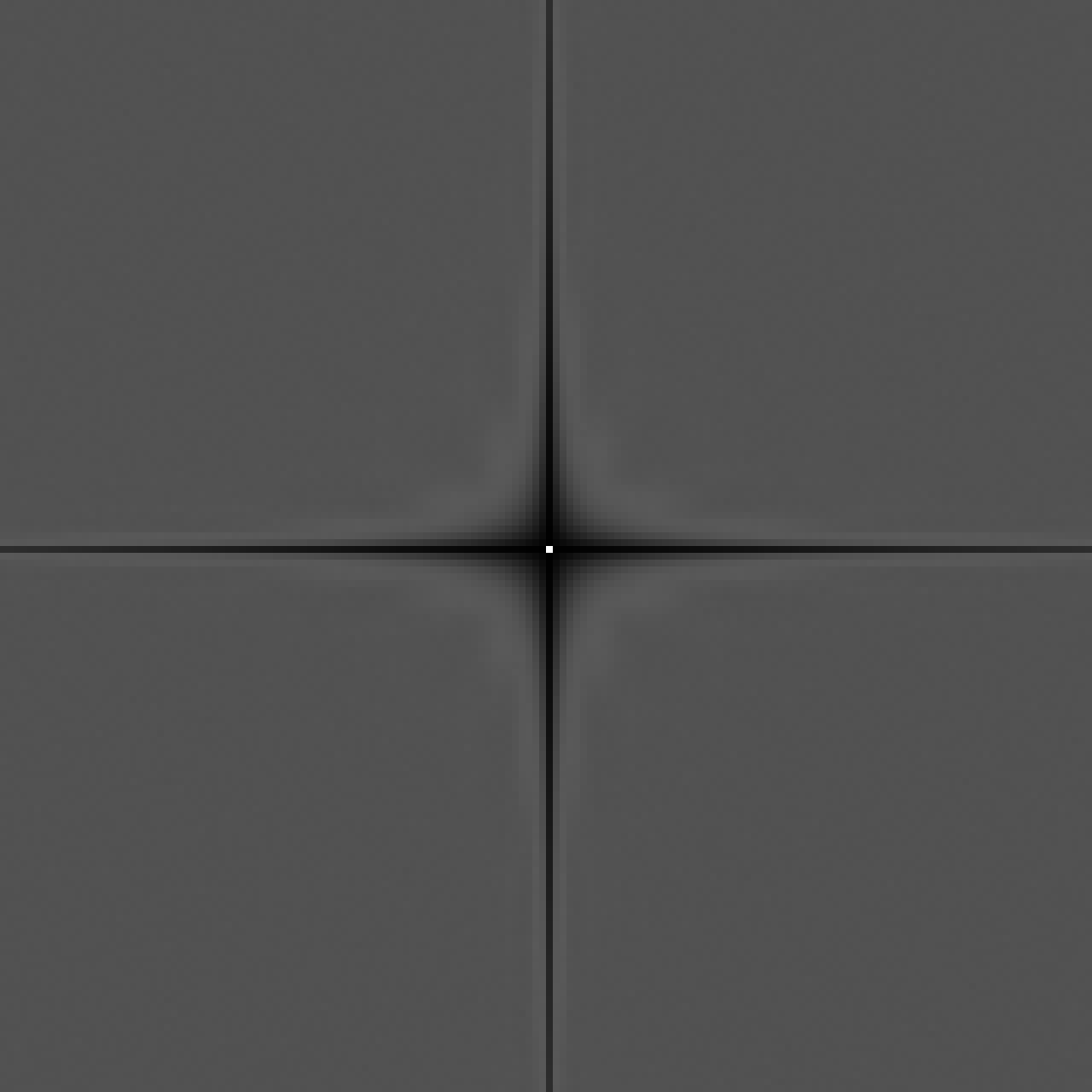}&%
	  \includegraphics[width=1\unit]{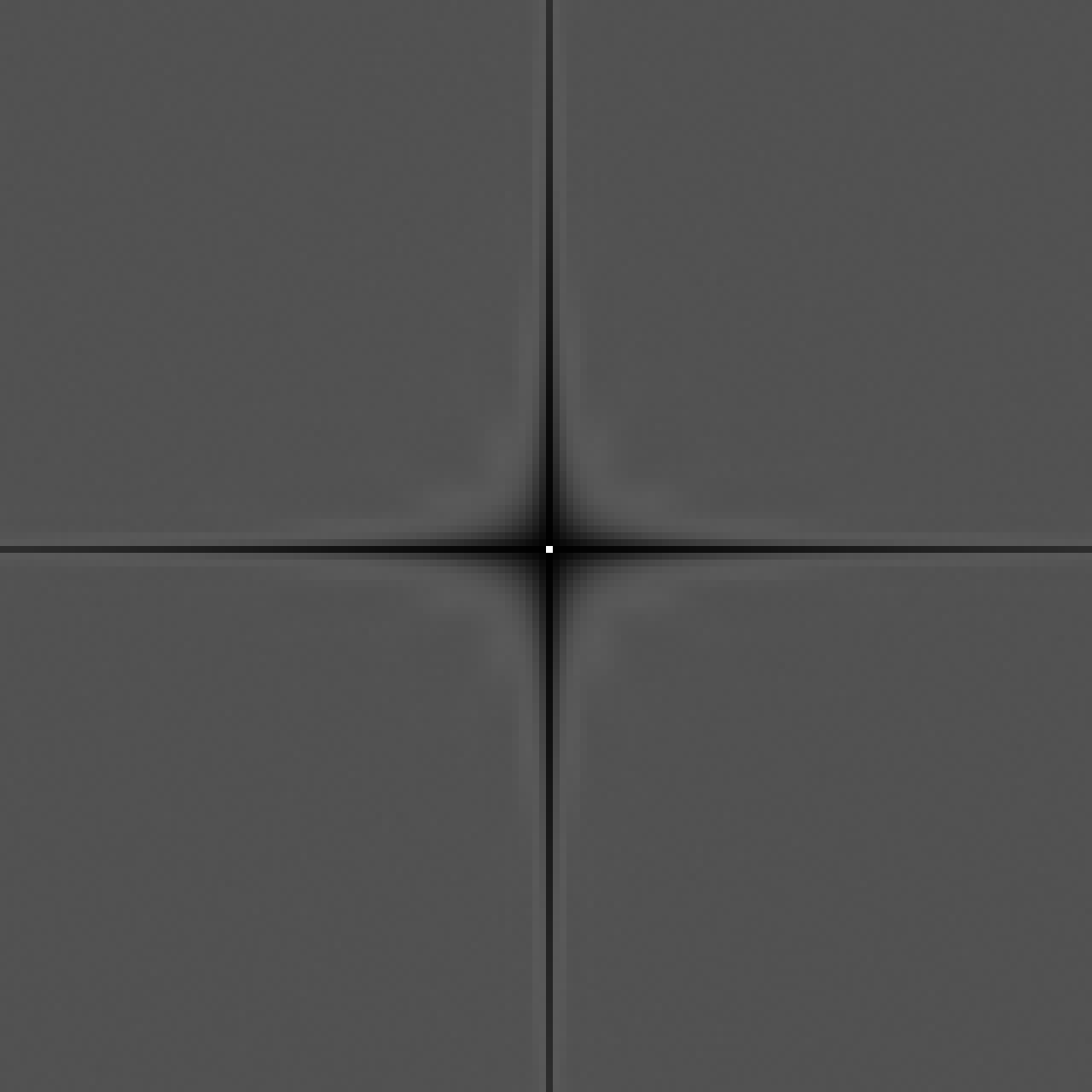}&%
	  \includegraphics[width=1\unit]{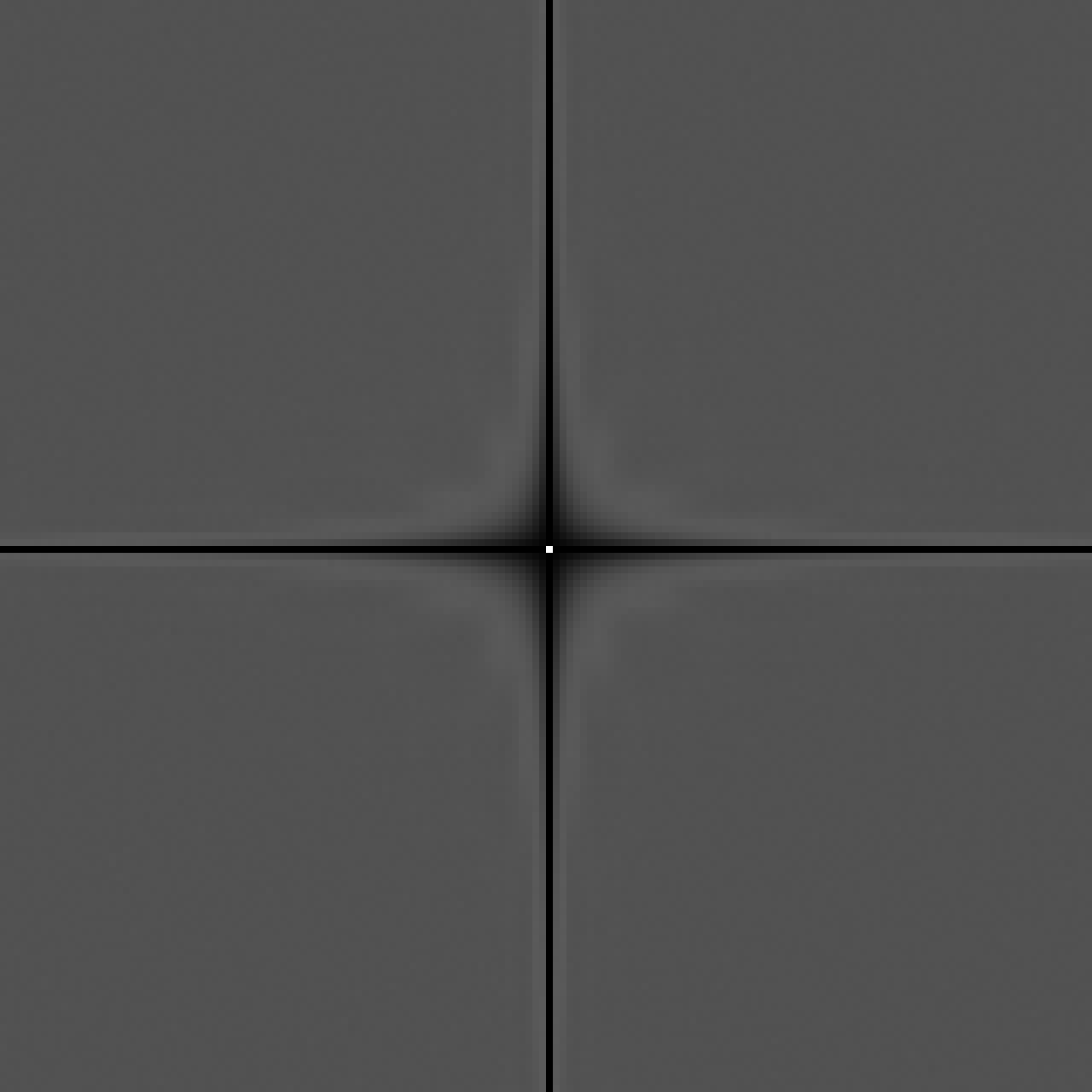}&%
	  \includegraphics[width=1\unit]{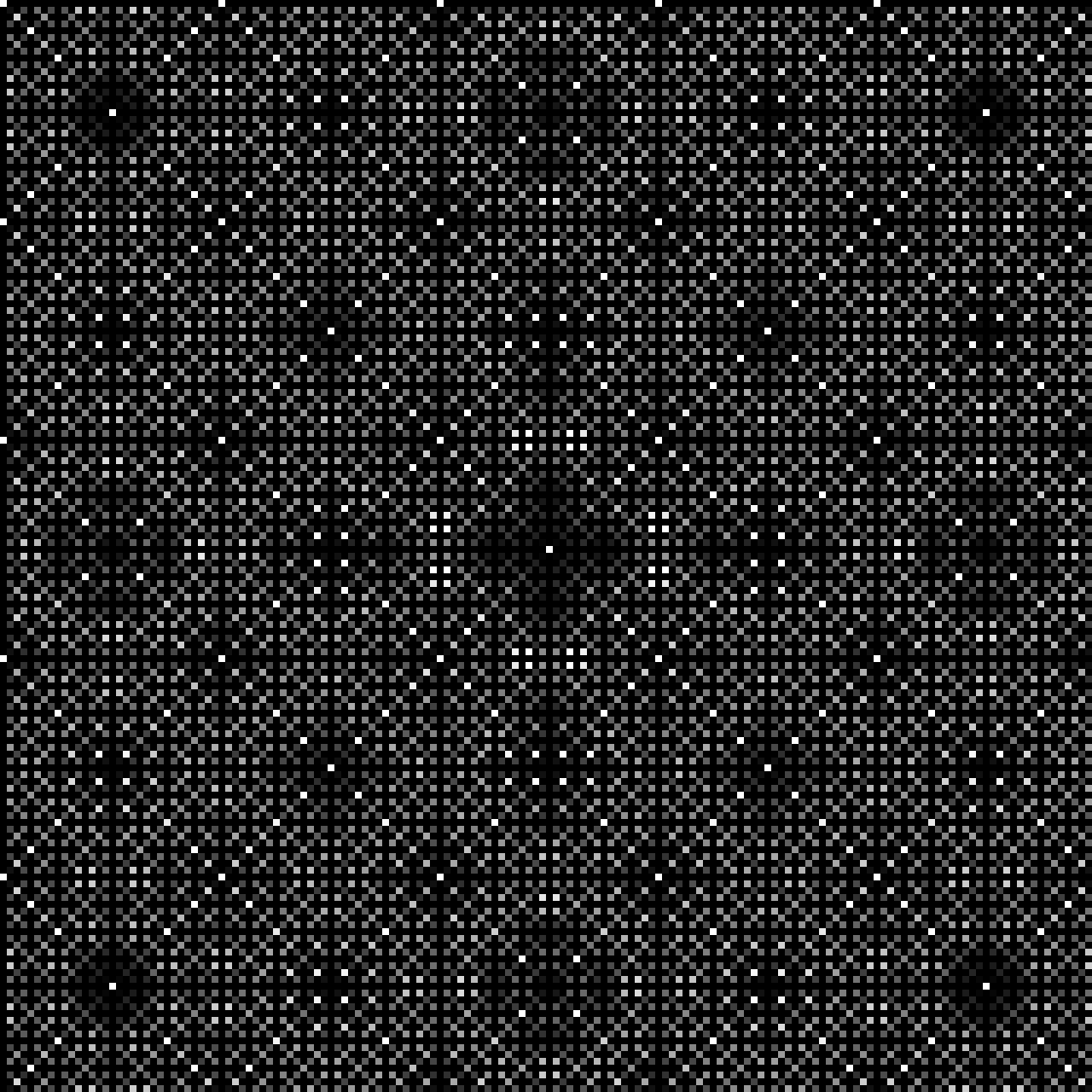}&%
	  \includegraphics[width=1\unit]{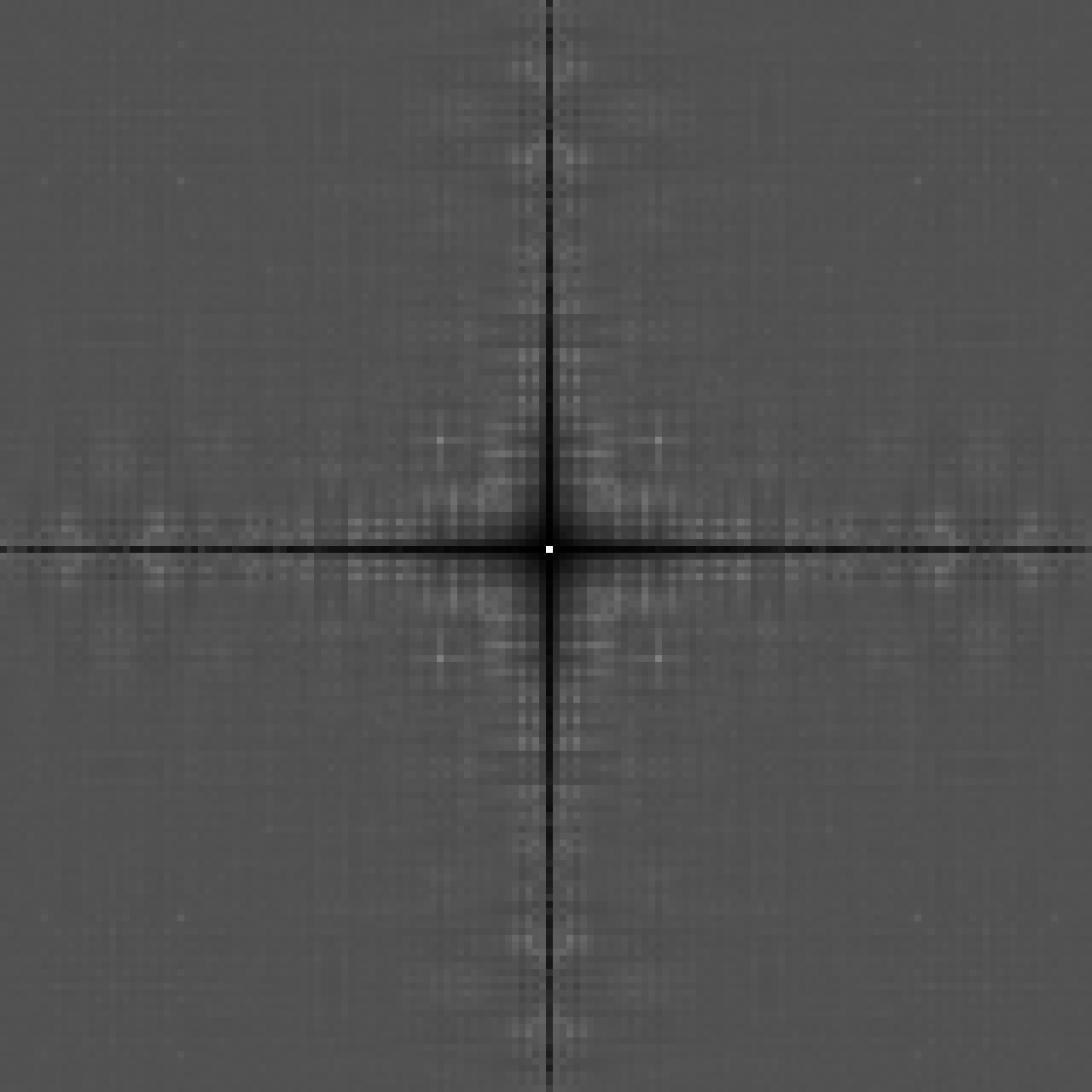}&%
	  \includegraphics[width=1\unit]{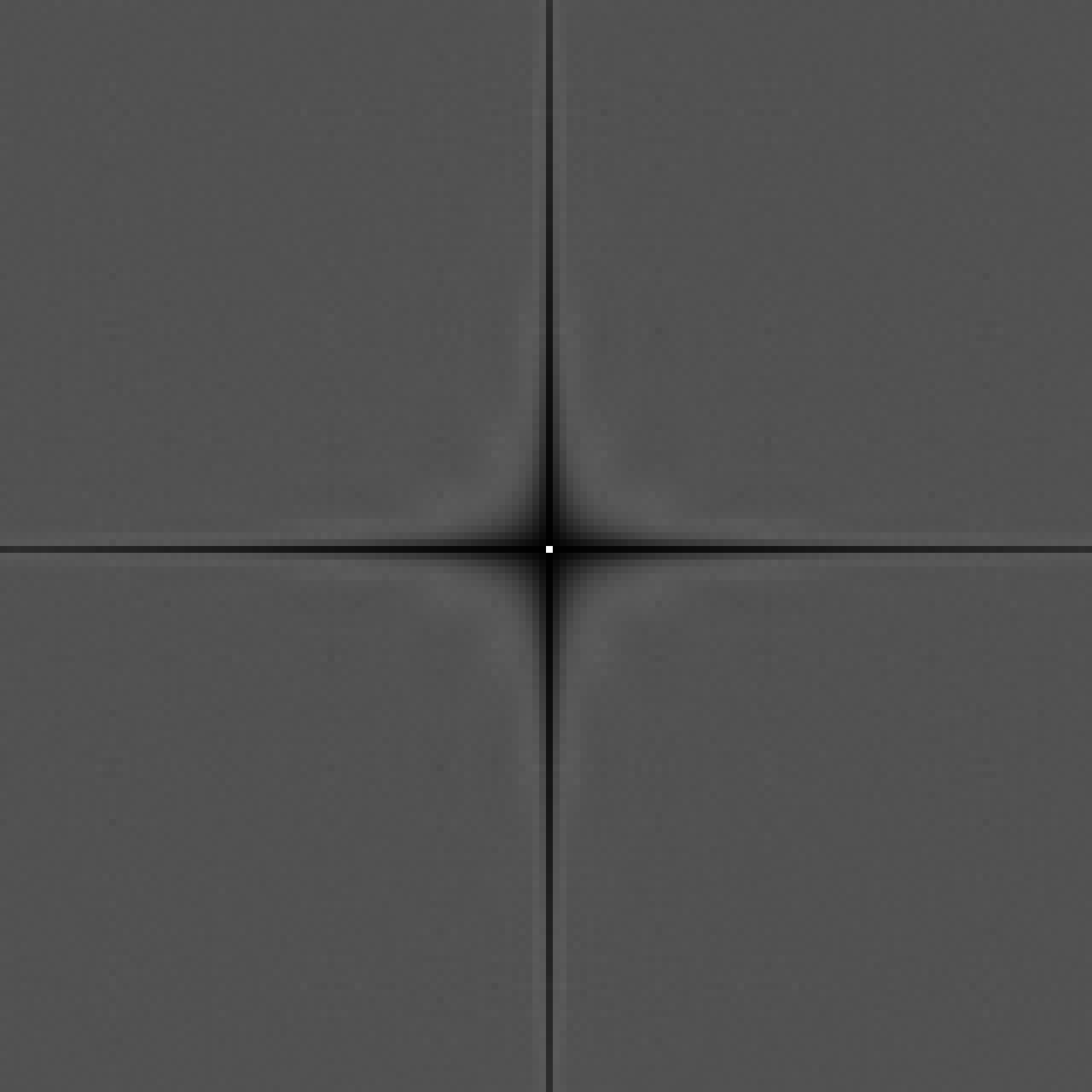}&%
	  \includegraphics[width=1\unit]{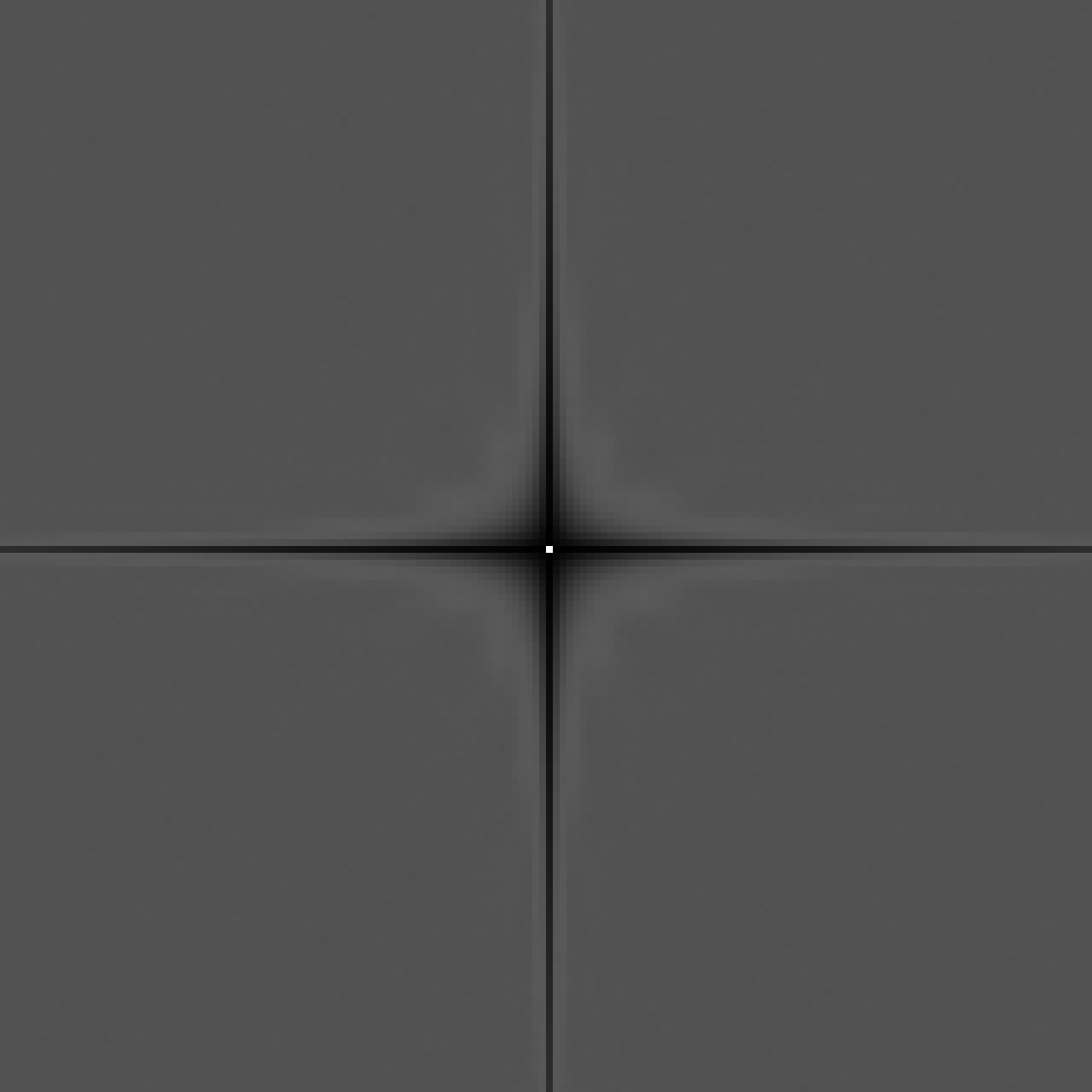}&%
	  \includegraphics[width=1\unit]{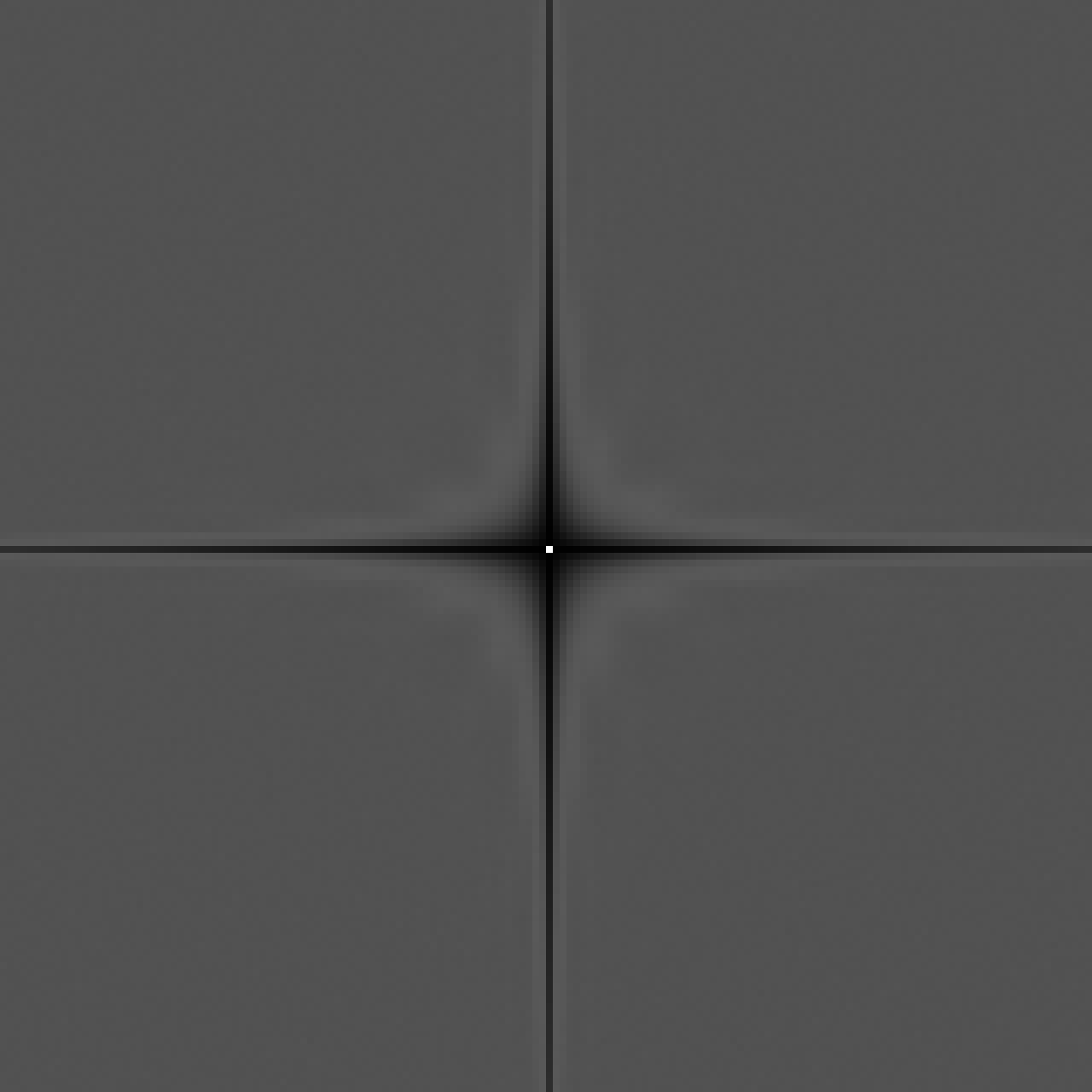}\\
	  \includegraphics[width=1\unit]{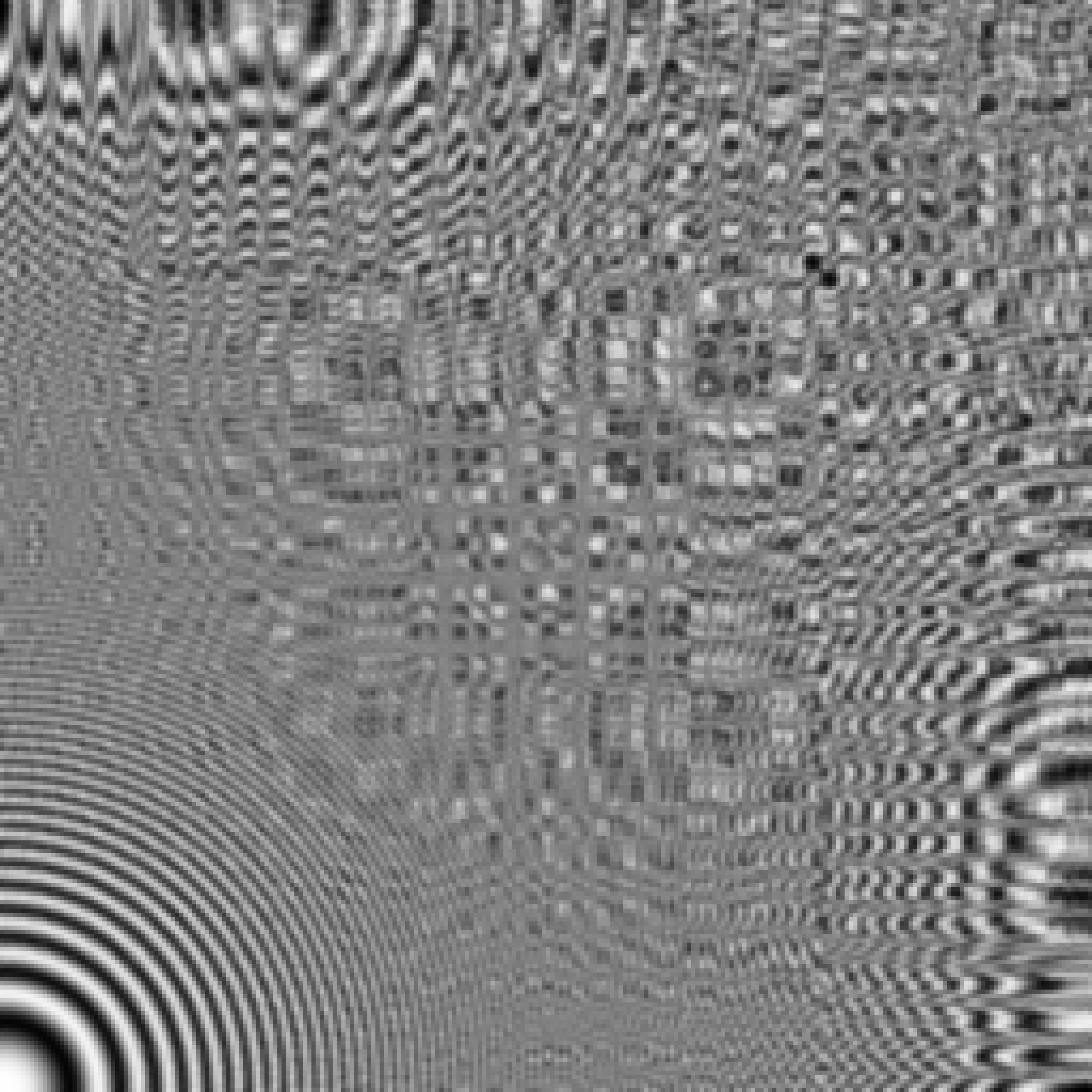}&%
	  \includegraphics[width=1\unit]{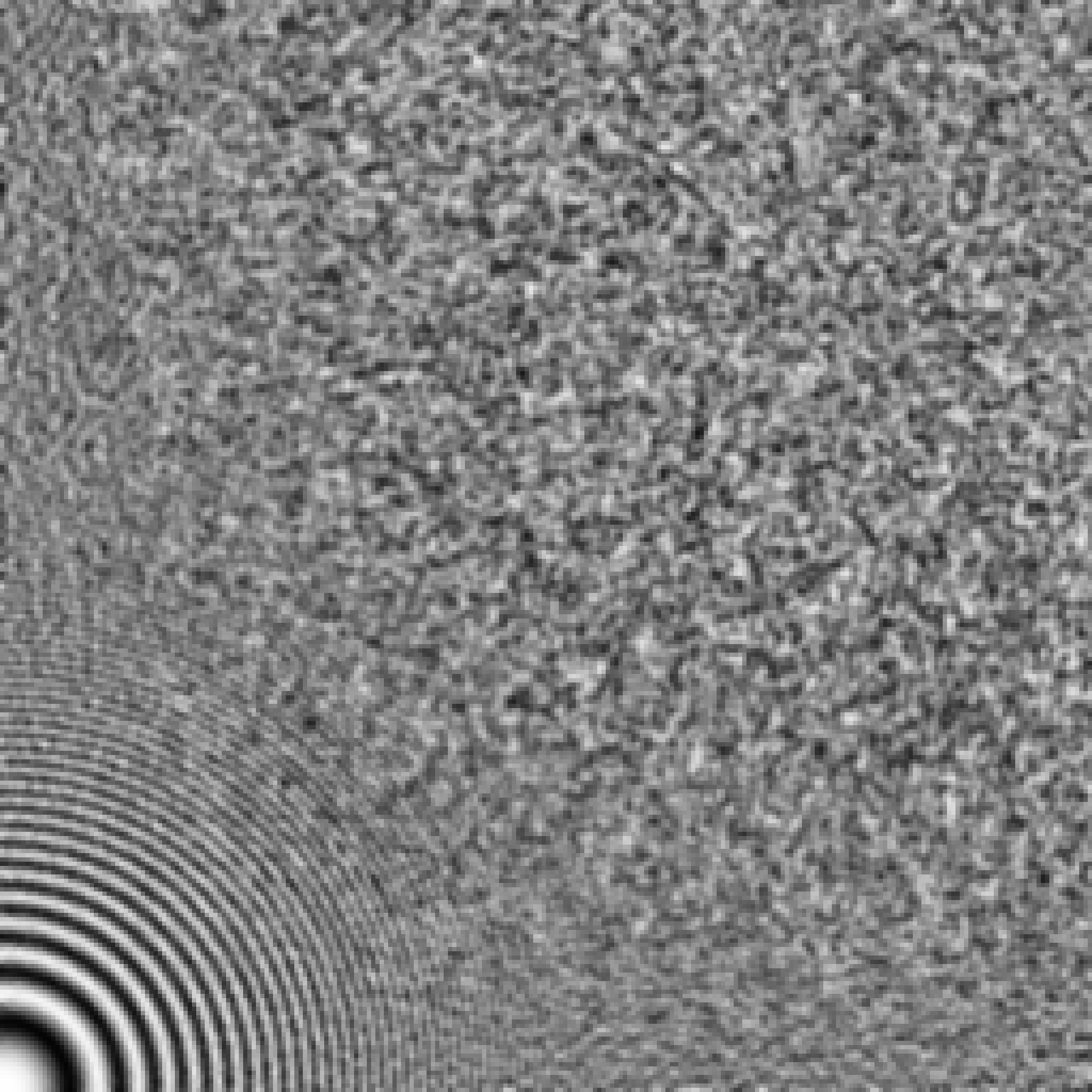}&%
	  \includegraphics[width=1\unit]{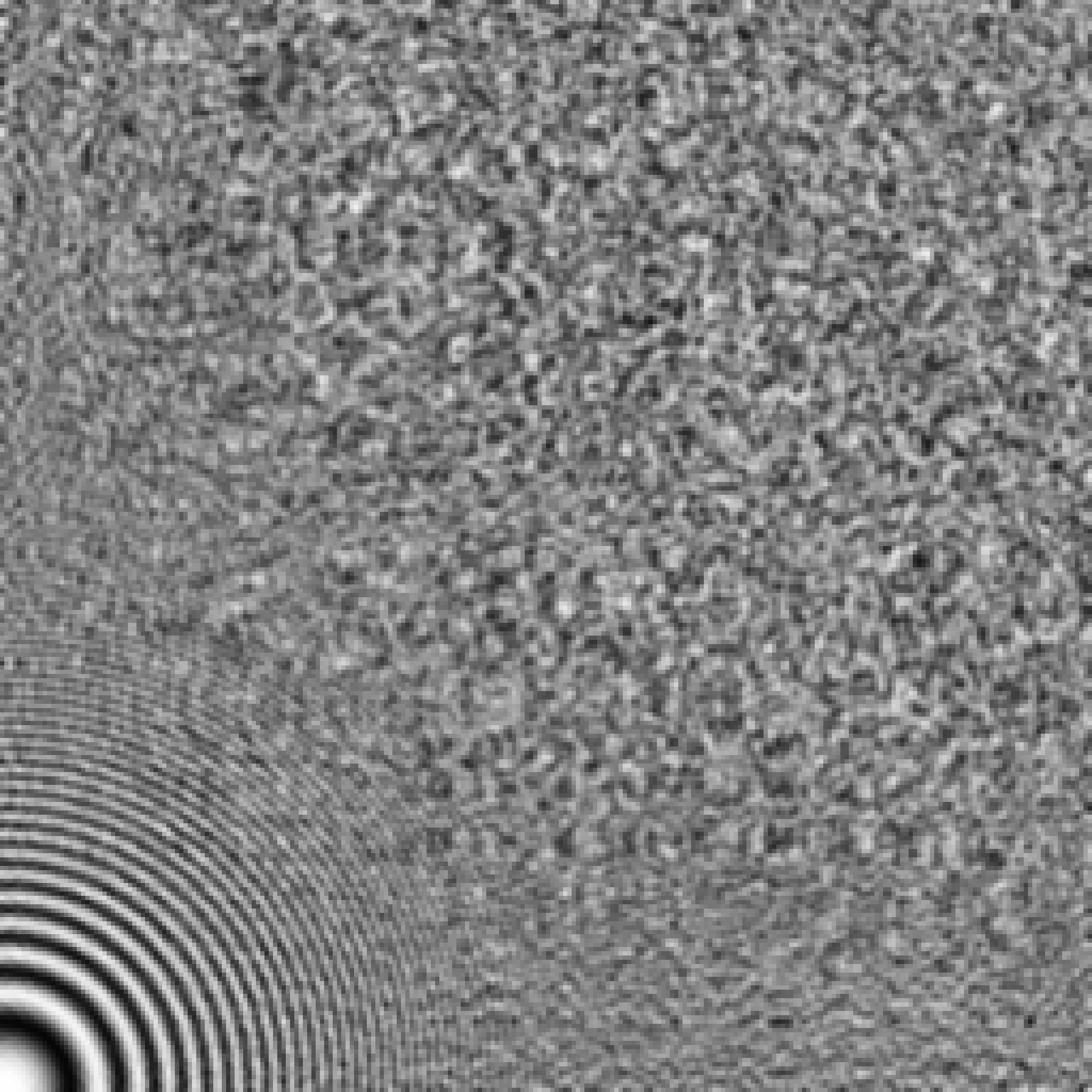}&%
	  \includegraphics[width=1\unit]{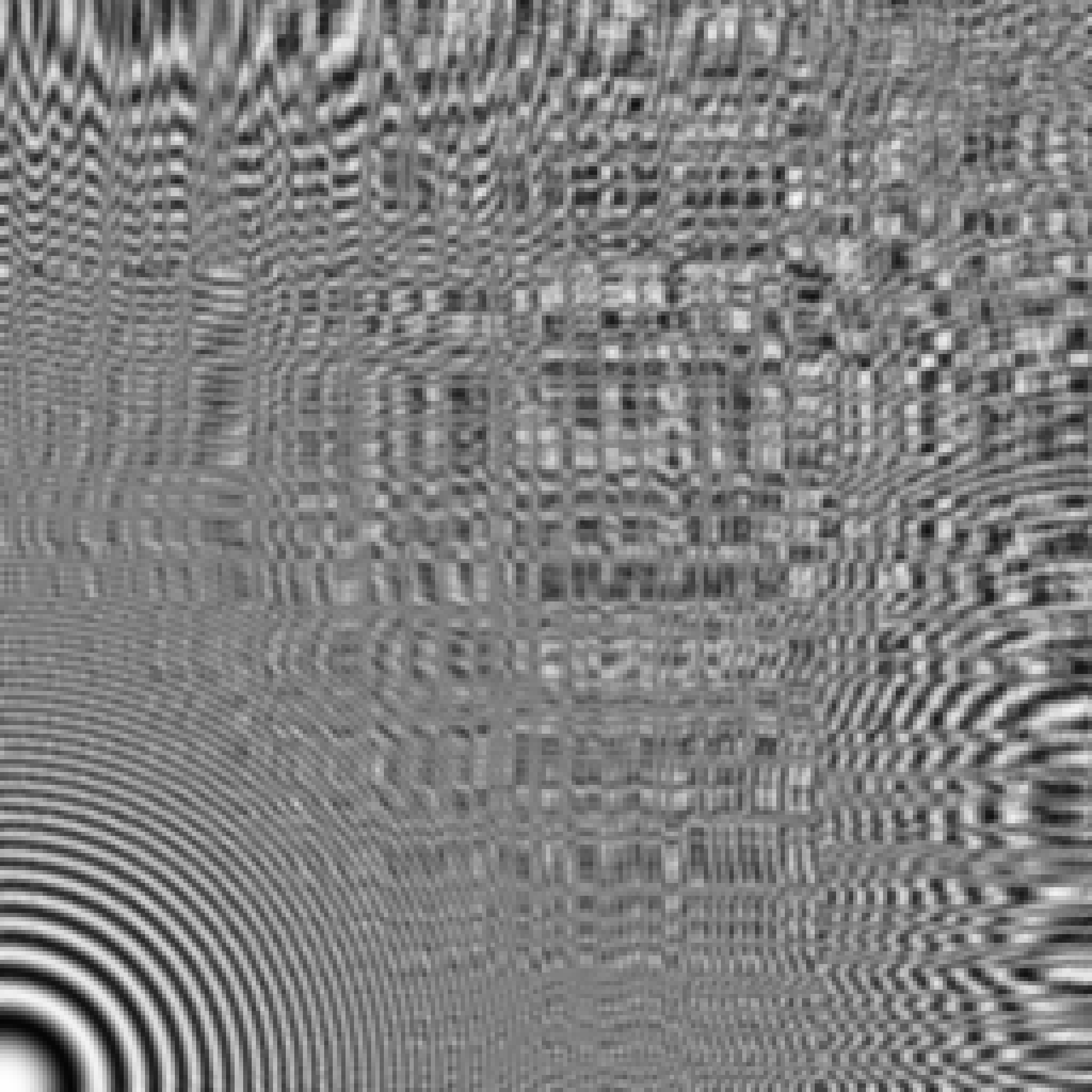}&%
	  \includegraphics[width=1\unit]{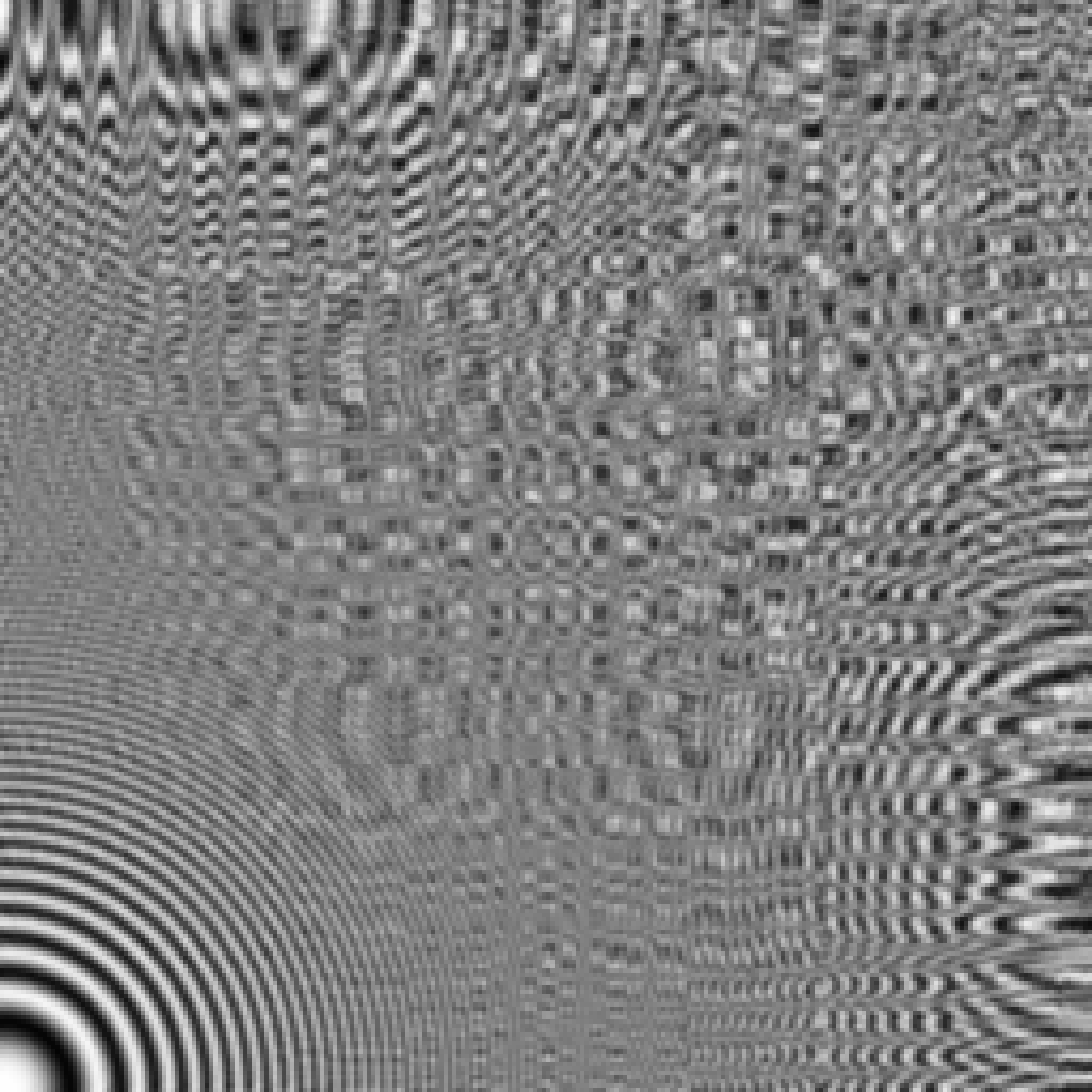}&%
	  \includegraphics[width=1\unit]{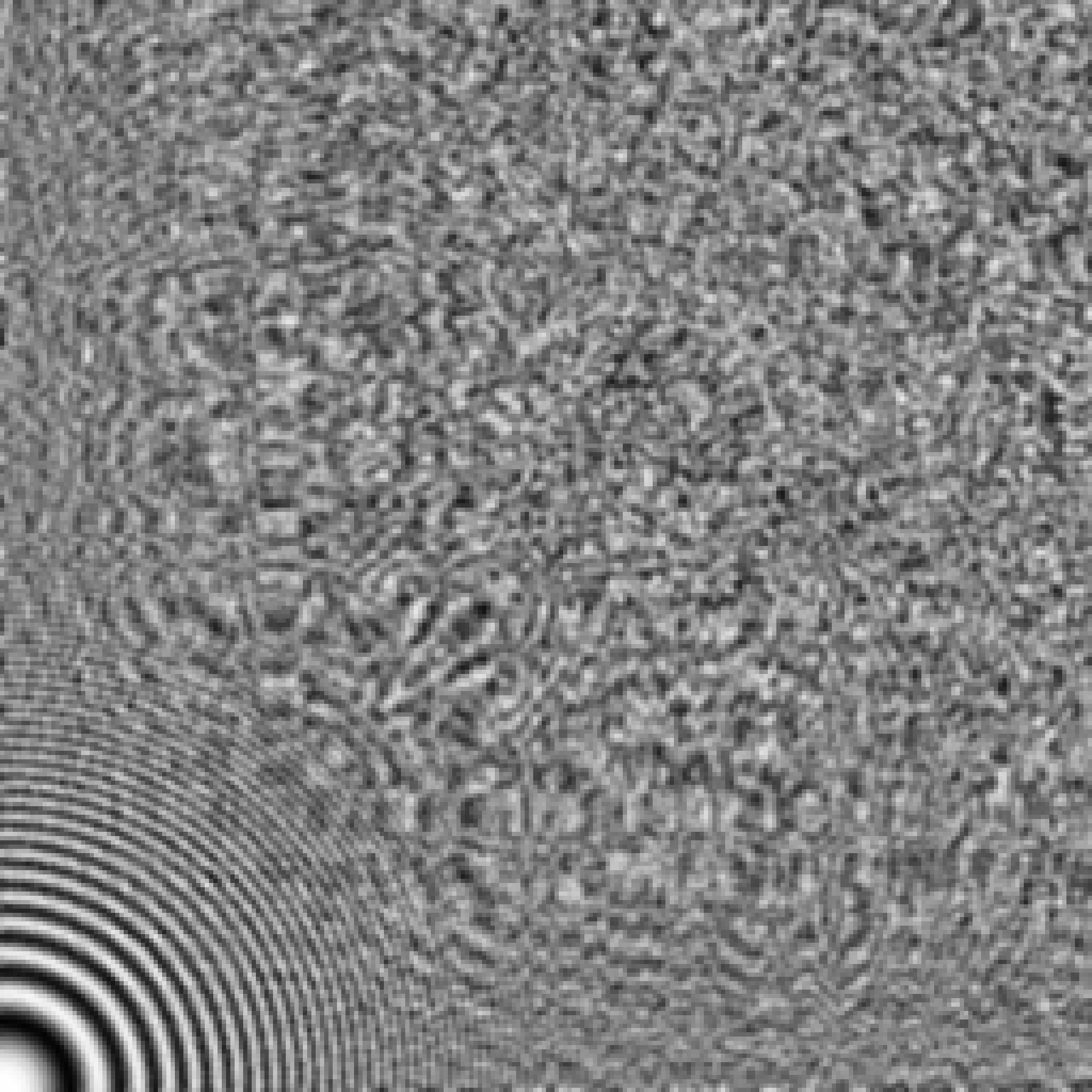}&%
	  \includegraphics[width=1\unit]{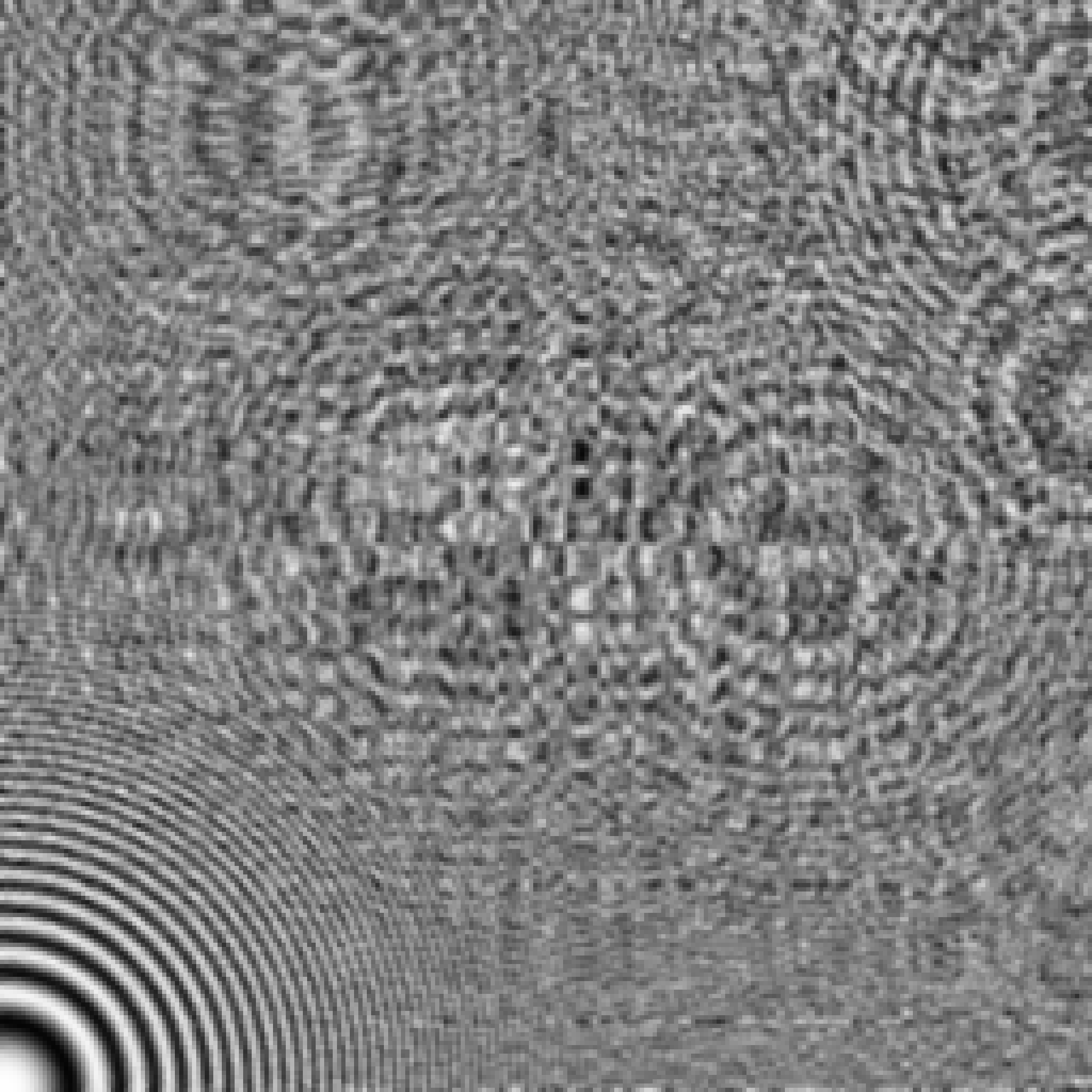}&%
	  \includegraphics[width=1\unit]{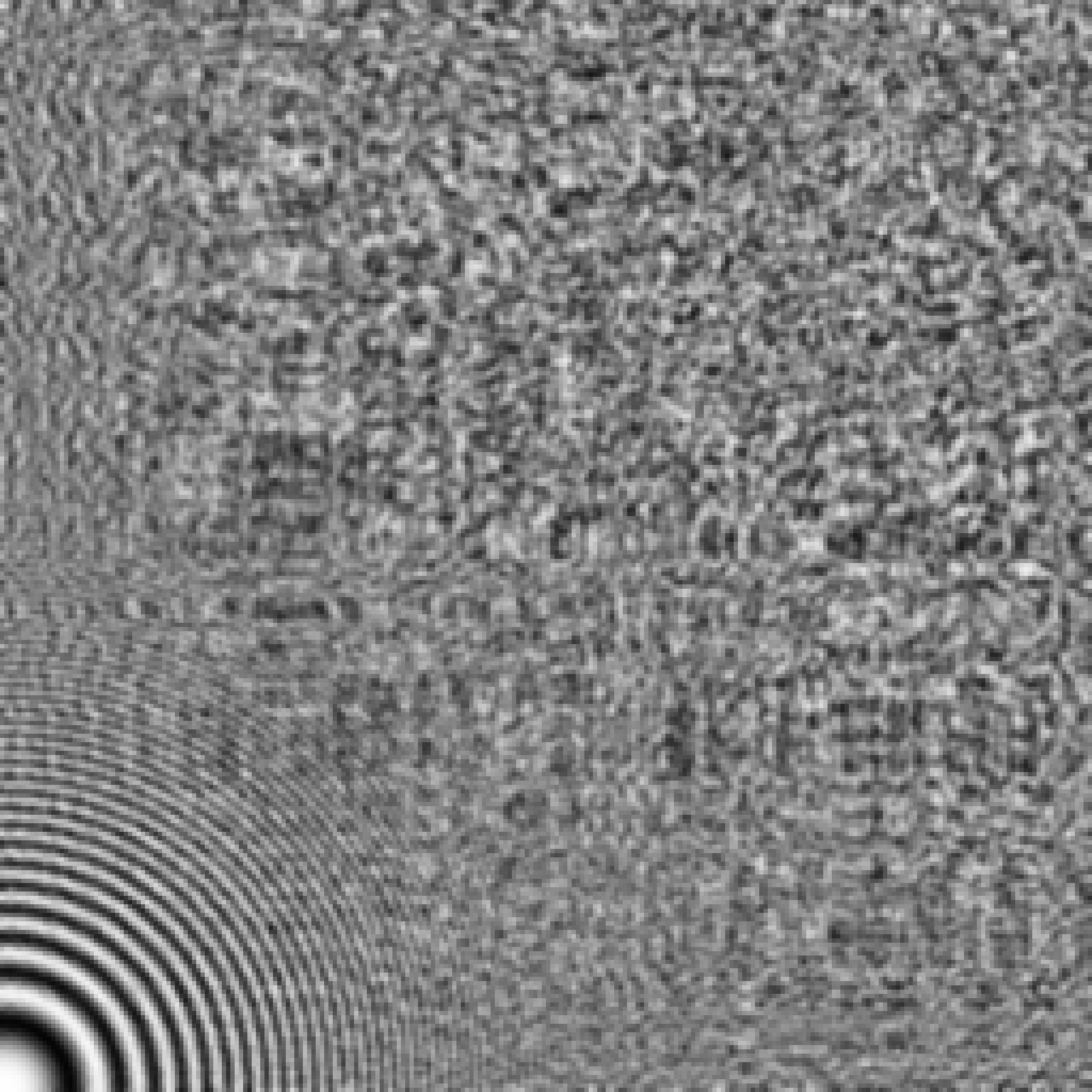}&%
	  \includegraphics[width=1\unit]{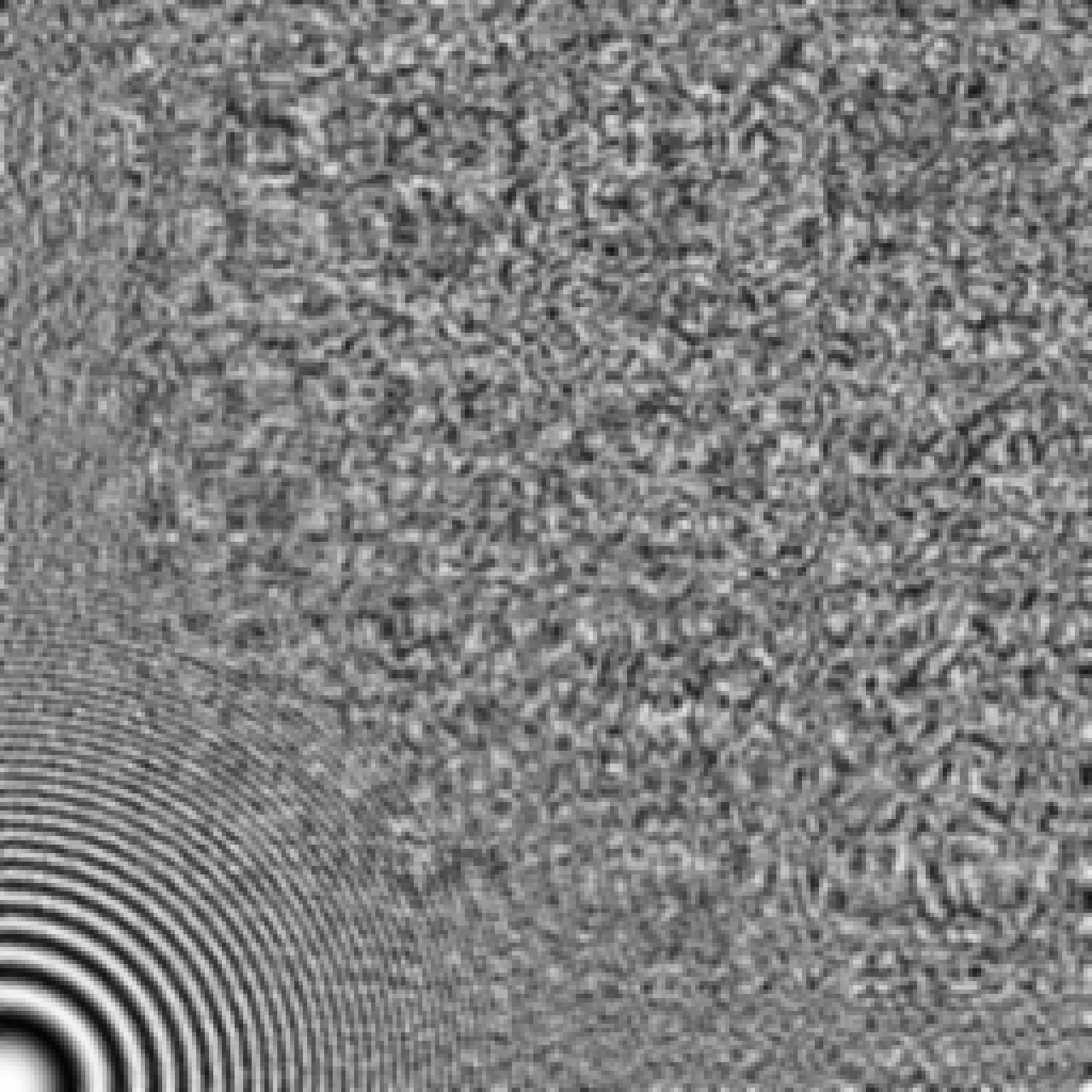}\\[2mm]
	  (a) Sobol, 0 bits &%
	  (b) Netshuffle, 13312 bits &%
	  (c) Owen, 12798 bits &%
	  (d) 8-bit Owen, 510 bits &%
	  (e) XOR, 64 bits &%
	  (f) Burley, 192 bits &%
	  (g) ART-Owen, 128 bits &%
	  (h) ART-Owen, 256 bits &%
	  (i) ART-Owen, 1024 bits
	  \end{tabular*}}
  \caption{\label{fig:periodograms}%
      Various scramblings of the first 256 points of the 2D Sobol sequence, showing along the rows
      a typical distribution of the points,
      a typical periodogram of a single set,
      an average periodogram over 100 realizations,
      an average periodogram over 10000 realizations, and
      a zoneplate plot of 256$\times$256 points; comparing:
      (a) unscrambled Sobol sequence,
      (b) Netshuffle \protect{\cite{Ahmed2021Optimizing}},
      (c) a proper Owen scrambling, using Helmer's implementation \protect{\shortcite{Helmer2021Stochastice}},
      (d) Owen scrambling to only 8-bit depth (cf.~\protect{\cite{Matousek1998L2}}),
      (e) XOR scrambling \protect{\cite{Kollig02Efficient}}: a minimal variant of Owen scrambling that uses a single word (i.e. bit-vector) per axis,
      (f) Burley's \protect{\shortcite{Burley2020Scrambling}} hash-based Owen scrambling,
      and our method, using (g) two, (h) four, and (i) 16 symbols.
      Numbers indicate the number of randomization data bits used.
      Netshuffle and Helemer's implementation of Owen scrambling work only with a prescribed size of nets, and are shown here for reference to the quality, while the remaining methods enable random access to the samples.
      Note that XOR scrambling offers almost no spectral improvement, while Burley's technique seems to bear its own frequency structure that creeps into the generated nets, imposing a persistent distortion onto the frequency spectrum.
      Our method is free of these distortions and offers a smooth trade-off between quality and memory footprint.
      The improvement over Burley's and XOR is evident in the absence of spikes from the average spectra and the reduced structures in the zoneplate plots.
      The truncated scrambling in (d) is shown to demonstrate the influence of the trailing bits. It is not noticeable for a number of points proportional to the scrambling depth, but becomes evident for a larger number of points, as may be seen in the zoneplate plot.
  }
% ------------------------------------------------------------------------------
\end{figure*}

% ==============================================================================
The power spectra of sampling patterns give additional insight into their performance.
In particular, patterns with low energy at low frequencies and uniform energy at higher frequencies are effective at converting aliasing into high-frequency noise, which is pleasing to the human visual system.
We have generated power spectra by averaging the periodograms of various sampling methods over multiple realizations.
As we can see in Fig.~\ref{fig:periodograms}, the instance performance of our method varies with the budget of the data bits, i.e. the grammar size, which makes sense.
In average, all our tested TM grammars, even with only two symbols, converged to the reference Owen-scrambling spectrum, indicating that our method performs very well in sampling the universe.
In contrast, we find persistent residual frequency spikes in Burley's scrambling, indicating a biased sampling of all possible scrambling trees.
% ------------------------------------------------------------------------------
% ------------------------------------------------------------------------------

% ==============================================================================
\subsection{Versatility\label{sec:versatility}}
% ==============================================================================

Our method presents the user with many degrees of control to tailor the model.
For example, the model is scalable to any memory budget.
Notably, even with a single-symbol grammar, the model is still able to yield.
Namely, it provably reproduces XOR scrambling.
The processing can be trimmed to any bit depth to gain some speed up; cf. \cite{Matousek1998L2}.
Finally, it can target any given scrambling tree, as discussed in Section~\ref{sec:ordered grammar}.

% ==============================================================================
\subsection{Optimization\label{sec:optimization}}
% ==============================================================================

\begin{figure}
%-------------------------------------------------------------------------------
    {\centering\scriptsize
    \settoheight{\labelHeight}{Optimized}
    \setlength{\unit}{(\columnwidth - 3\gap - \labelHeight)/4}
    \begin{tabular*}{1\columnwidth}{@{}c@{\extracolsep{\fill}}c@{\extracolsep{\fill}}c@{\extracolsep{\fill}}c@{}}
    \rotatebox{90}{\parbox{1\unit}{\centering{Raw}}}&%
    \includegraphics[height=1\unit]{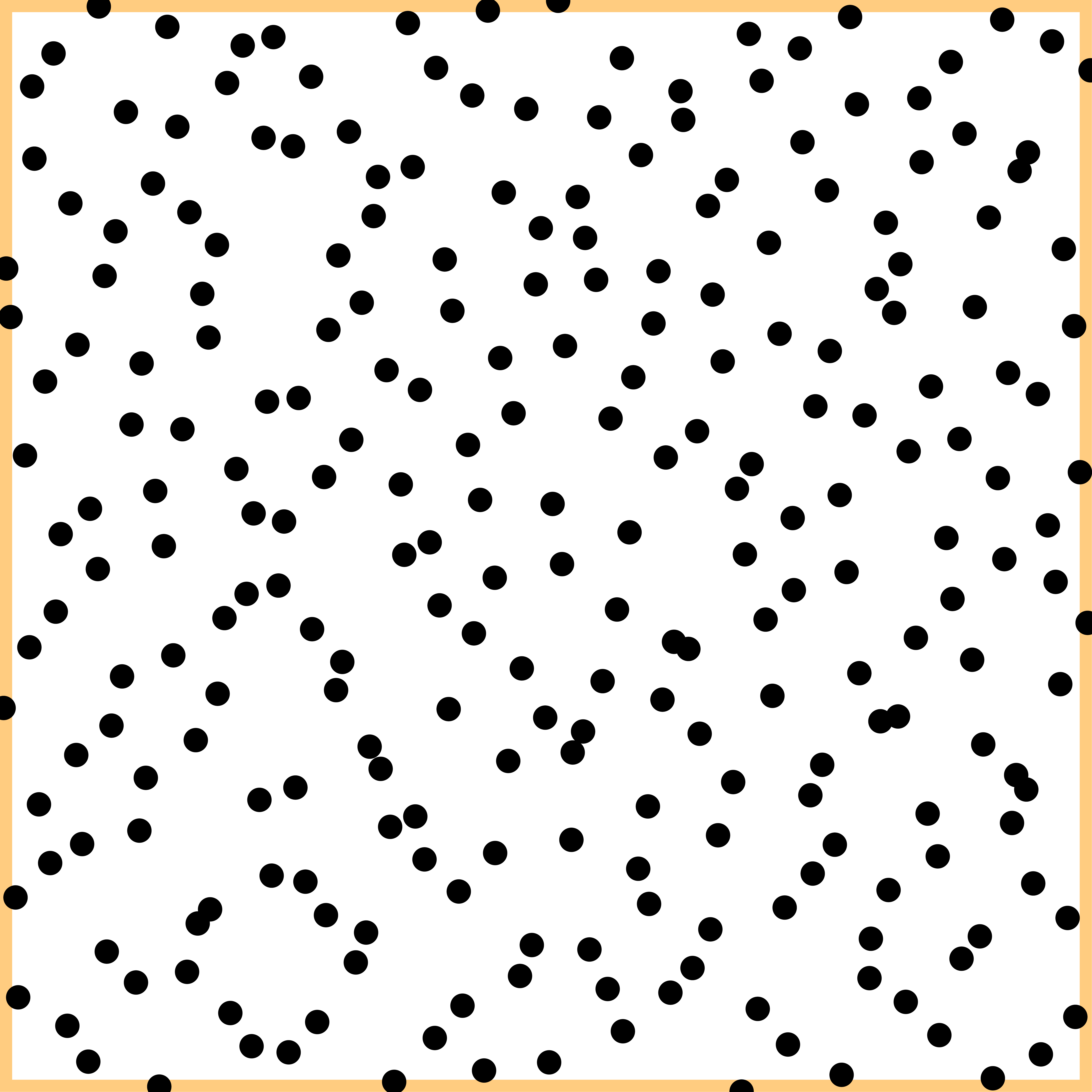}&%
    \includegraphics[height=1\unit]{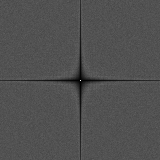}&%
    \includegraphics[height=1\unit]{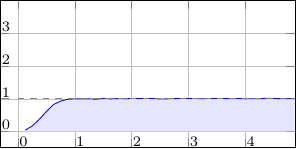}\\[1mm]
    \rotatebox{90}{\parbox{1\unit}{\centering{Optimized}}}&%
    \includegraphics[height=1\unit]{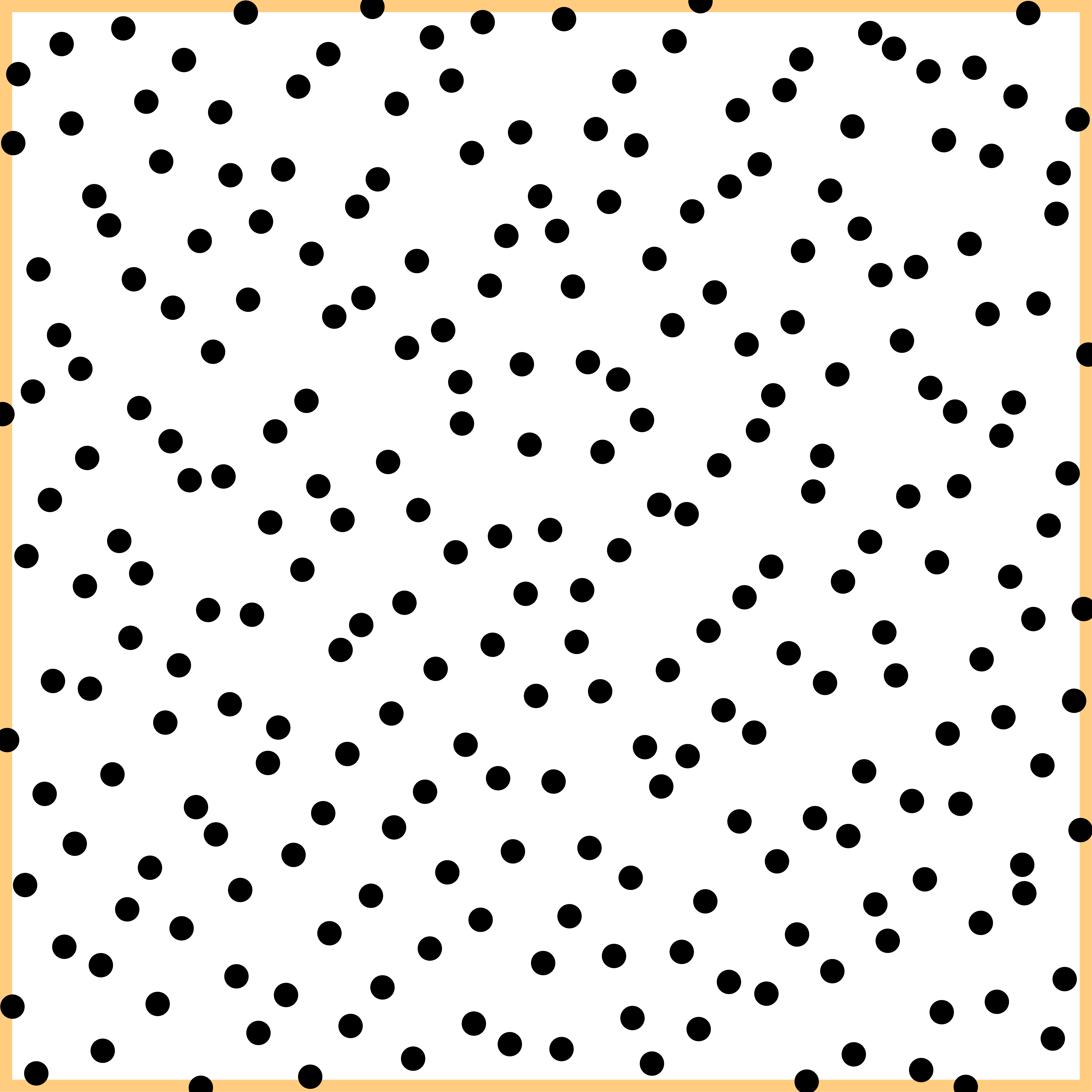}&%
    \includegraphics[height=1\unit]{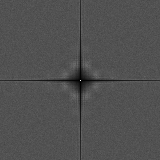}&%
    \includegraphics[height=1\unit]{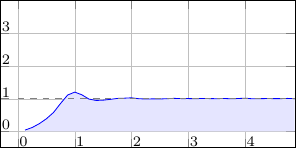}\\[1mm]
    & (a) Points & (b) Spectrum & (c) Radial Power
    \end{tabular*}}
    \caption{\label{fig:downhill}%
       Comparing raw ART-Owen scrambled  256-point sets, using random scrambling data, to obtimized sets targetting a minimum conflict radius of 0.2 and a blue noise energy $\sigma = 0.5$ \protect{\cite{Ahmed2021Optimizing}}.
       The spectral plots are averaged over 100 realizations.
    }
%-------------------------------------------------------------------------------
\end{figure}

Our framework readily suggests using a greedy descent algorithm to optimize the
scrambling data towards different targets.
The general approach is to start with a random setting, iterate through all
the data bits, flipping them one by one, and accepting the change if it
improves towards the target.
A cycle through all the bits counts as one iteration, and the algorithm stops
when no more changes are acceptable.
We made a basic implementation of this downhill idea, as listed in Algorithm~\ref{alg:downhill},
% ------------------------------------------------------------------------------
\begin{algorithm} [tb]
% ------------------------------------------------------------------------------
    \SetKwInOut{KwIn}{Input}
    \SetKwInOut{KwOut}{Output}
    \caption{
        Optimizing an ART Owen Scrambling.
    }
    \label{alg:downhill}
    \KwIn{
        (1) A production table representing an ART-Owen scrambler.\\
        (2) An initial data table.\\
        (3) A quality assessment function.\\
    }
    \KwOut{A data table optimizing the target set size towards the prescribed quality measure.}
    \Repeat{No changes accepted} {
        Reset change counter\;
        \ForEach{symbol}{
            \For{preset number of attempts} {
                randomly chose a new data vector\;
                evaluate the new scrambling\;
                \eIf{quality improves}{
                    increment change counter\;
                } {
                    restore preceding data vector\;
                }
            }
        }
    }
%-------------------------------------------------------------------------------
\end{algorithm}
%-------------------------------------------------------------------------------
and obtained the result shown in Fig.~\ref{fig:downhill}.
%-------------------------------------------------------------------------------
%-------------------------------------------------------------------------------
We do not claim any optimality in our implementation, and while the results are not exceptional, they still represent a proof of concept on the optimizability.
The point is that our model narrows down the design space of Owen scrambling, making such a search feasible. All previous optimization attempts we are aware of, e.g., [Helmer2021], try an exhaustive scan in a small depth.
There may still be considerable room for improvement.
For example, the intuitive bit-by-bit adjustment failed, but we managed to get it to work by adjusting the whole data entry at once.
We also conceived the multiple attempts idea experimentally. For the shown results we use a thousand attempts.

\begin{figure}
%-------------------------------------------------------------------------------
    {\centering\scriptsize
    \settoheight{\labelHeight}{$r_\mathrm{f}>0.615,\; \sigma = 0.5$}
    \setlength{\unit}{(\columnwidth - 3\gap - \labelHeight)/4}
    \begin{tabular*}{1\columnwidth}{@{}c@{\extracolsep{\fill}}c@{\extracolsep{\fill}}c@{\extracolsep{\fill}}c@{}}
    \rotatebox{90}{\parbox{1\unit}{\centering{$r_\mathrm{f}>0.615,\; \sigma = 0.5$}}}&%
    \includegraphics[height=1\unit]{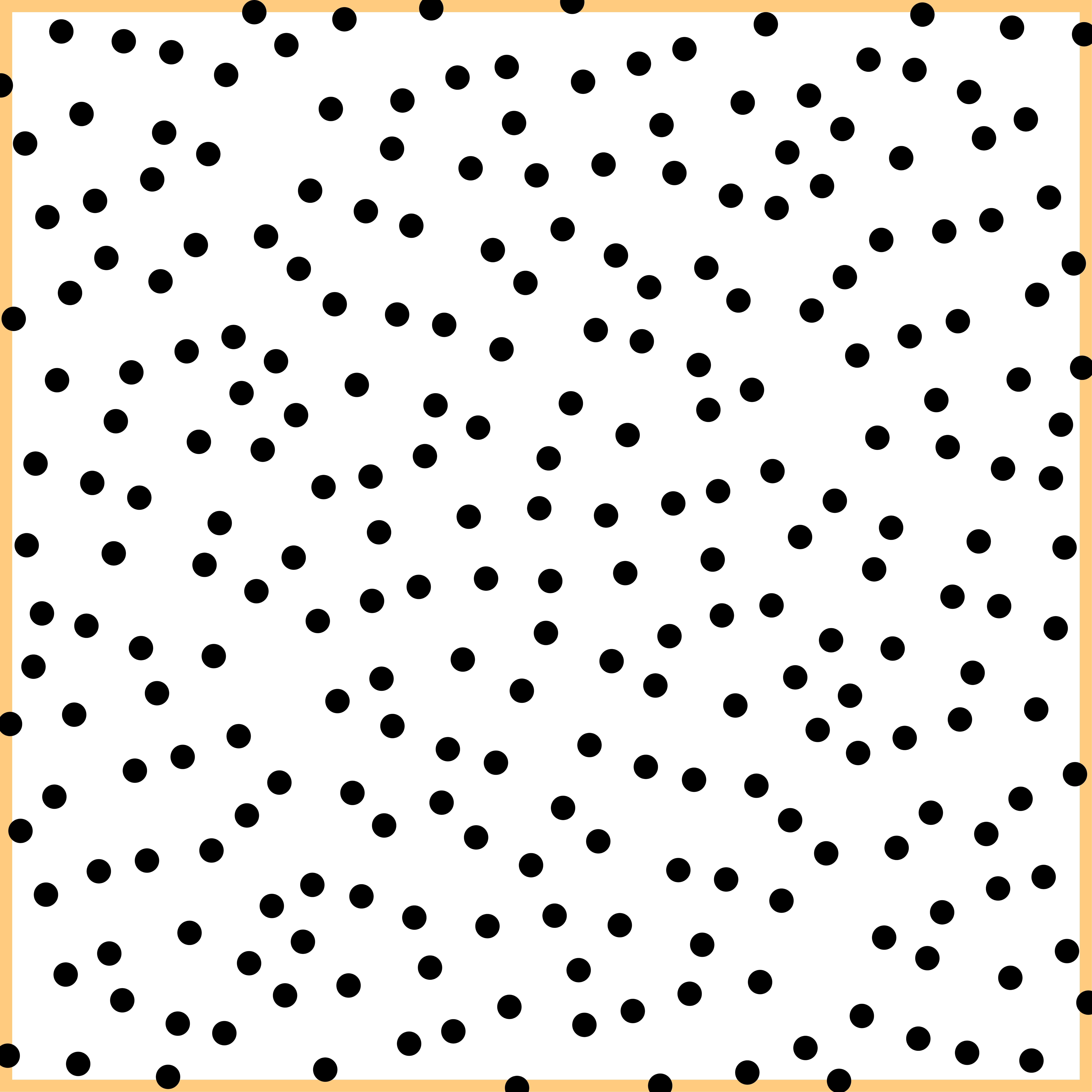}&%
    \includegraphics[height=1\unit]{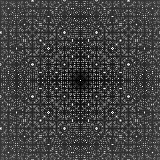}&%
    \includegraphics[height=1\unit]{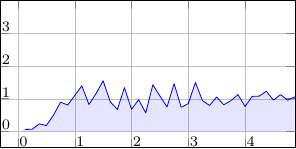}\\[1mm]
    \rotatebox{90}{\parbox{1\unit}{\centering{$r_\mathrm{f}>0.468,\; \sigma = 1$}}}&%
    \includegraphics[height=1\unit]{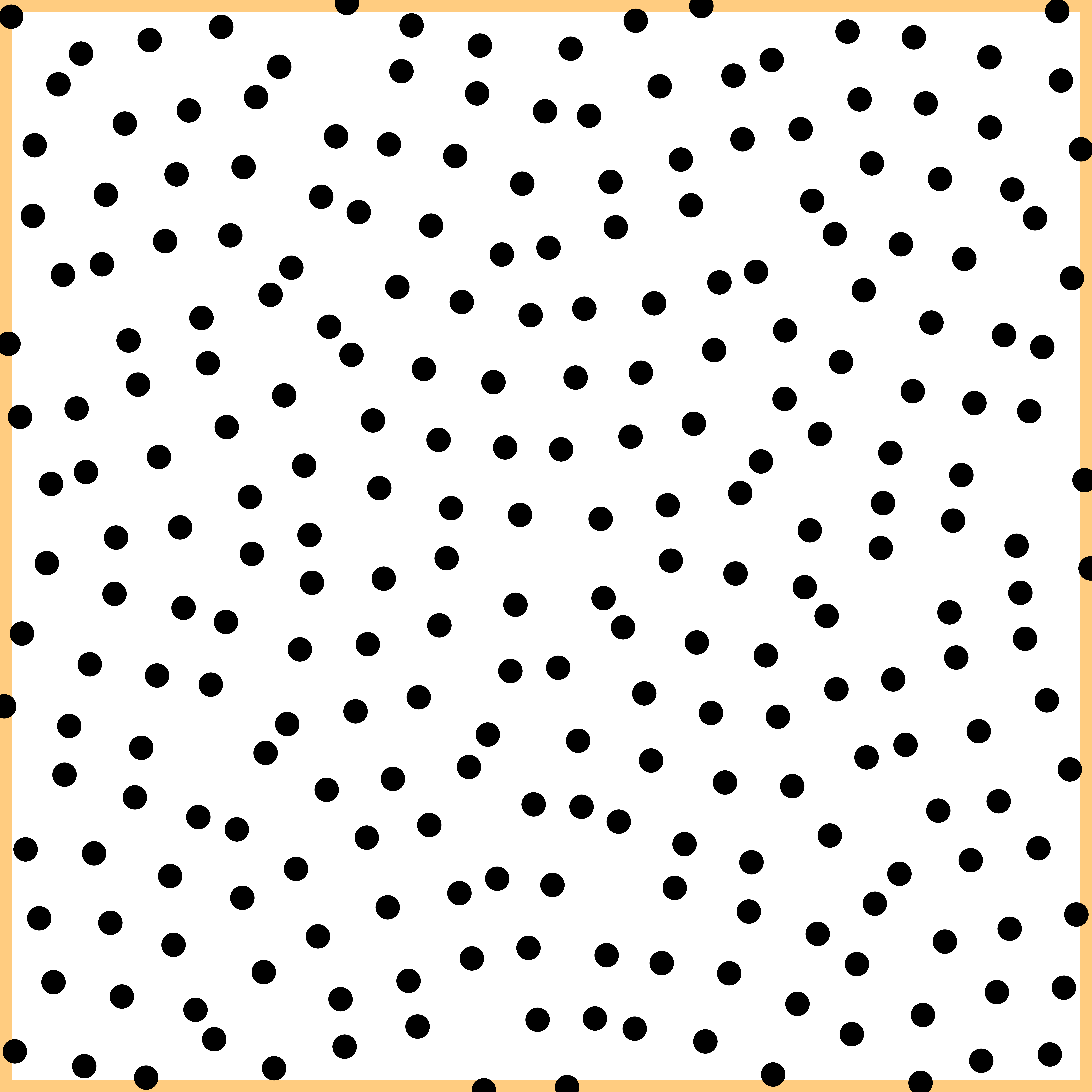}&%
    \includegraphics[height=1\unit]{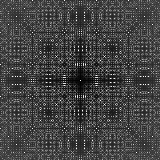}&%
    \includegraphics[height=1\unit]{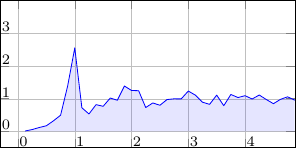}\\[1mm]
    & (a) Points & (b) Spectrum & (c) Radial Power
    \end{tabular*}}
    \caption{\label{fig:optimzaiton}%
       Best found scrambling codes for the binary TM grammar and 256 points, targeting different combinations of conflict radius and blue noise energy $\sigma$ \protect{\cite{Ahmed2021Optimizing}}.
       The points plot shows the top set and the spectral plots are averaged over the top 1000 sets.
    }
%-------------------------------------------------------------------------------
\end{figure}

Optimization targets vary widely and depend on the application scenario.
For example, for some applications, we may optimize a sequence progressively
so that all the leading power-of-two sets are optimal, while other scenarios,
e.g., Z-Sampler \cite{Ahmed20Screen}, may ask to optimize equal-sized nets
taken from the same sequence.
There are many more optimization scenarios than we can enumerate here, but to
give a non-trivial example, we considered optimizing under the two-symbol
TM grammar in Eq.~\eqref{eq:t-prodcution}, which is extremely efficient
in both memory and speed.
Indeed, the grammar may be hard-coded as a single XOR operation.
Rather than using the iterative model outlined above, we could implement an
exhaustive search over all the 256-point nets obtained by scrambling the
first points of the Sobol sequence.
With this 8-bit resolution, each scrambling data entry takes one byte, hence
the whole set of scrambling data can be stored compactly as a 32-bit word.
With the help of a GPU, we scanned the whole 4G range of choices for
scrambling codes that maximize the minimum spacing of points \cite{Gruenschlos08Nets,Gruenschlos09Nets}, also known as conflict radius, and for codes that minimize the blue-noise energy \cite{Ahmed2021Optimizing}.
We obtained even better qualities by combining the two, as demonstrated in Fig.~\ref{fig:optimzaiton}.
%-------------------------------------------------------------------------------
%-------------------------------------------------------------------------------

% ==============================================================================
% ==============================================================================
% ==============================================================================
\section{Conclusion\label{sec:conclusion}}
% ==============================================================================
% ==============================================================================
% ==============================================================================

In this paper, we presented a simple and elegant solution to the long-standing problem of efficiently implementing Owen scrambling.
Our algorithm readily integrates in rendering engines with negligible effort, and provides many degrees of freedom for users to control the distribution of points.

Because we have transformed the scrambling problem into another, possibly more interesting problem of finding a good grammar, we believe that our work will lead to further research and investigation.
Well-specified, intriguing, and challenging, we expect this problem to be quite appealing to researchers of all career stages, and we appeal to the community to develop the method even further.
For example, a tournament over graduate students to suggest the best grammars and demonstrate them may return better solutions than what we, the authors, can do.

% ==============================================================================
% Bibliography
% ==============================================================================

\bibliographystyle{ACM-Reference-Format}
\bibliography{main.bib}

% ==============================================================================
% ==============================================================================
% ==============================================================================
\appendix

% ==============================================================================
% ==============================================================================
% ==============================================================================
\section{Caching the Permutations}
% ==============================================================================
% ==============================================================================
% ==============================================================================

\edited{Noting the implementation problem of Owen scrambling discussed in this paper, Matou\v{s}ek \shortcite{Matousek1998L2} early proposed a ``trick'' to trade speed for memory by caching the permutation at the top of the tree.
The key insight is that the volume $2^m$ of data is relatively small for the few leading bits, and the underlying assumption is that the required number of points is not large.

Following that suggestion, it is tempting to stop the scrambling at a small depth.
While there are many feasible applications of this idea, it falls short in terms of scalability.
To demonstrate this, we discuss two implementation choices of caching 8 bits.
The fastest implementation is to precompute and tabulate the scrambled bytes themselves.
This model immediately loses invertability. But even if that is not needed, this model does not scale well with dimension, and the 256-byte table starts to become problematic.
Alternatively, we may consider storing the data bits.
The  $\mathcal{O}(m)$ time complexity is similar to our model, though slightly faster.
The model is also readily invertable.
It fails, however, to scale with \emph{depth} this time.
Indeed, at the time Matou\v{s}ek published his paper there might have not been practical applications imaginable that required scrambling deep down beyond 8 or 12 bits. Gr{\"u}nschlo{\ss} et al. \shortcite{Gruenschlos12Enumerating}, however, later introduced the concept of global samplers,where the whole image plane is treated as a 2D projection of a unit hypercube, hence the proposed 8-bit depth, for example, is far from reaching the subpixel samples.
This is where the advantage of our model becomes evident, since we reuse the stored data bits to recursively scramble to any depth.
Finally, we note that the concept is \emph{orthogonal} to our model, and may well be combined with it. For example, an interesting combination would be to reduce our data storage to 8 bits rather than 32.
}

\end{document}